\documentclass[aip,jcp,floatfix]{revtex4-2}

\usepackage{graphicx}
\usepackage{amsmath,amssymb}
\usepackage{mathtools}
\usepackage{dsfont}
\usepackage{booktabs}
\usepackage{threeparttablex}
\usepackage[version=4]{mhchem} 

\def\H{\hat{H}}
\newcommand{\tn}{\tnote}
\newcommand{\etal}{\emph{et al.}}
\newcommand\ket[1]{|#1\rangle}

\newcommand\overl[2]{\langle#1|#2\rangle}
\newcommand{\mbr}{\ensuremath{\mathbf r}}
\newcommand{\bR}{\ensuremath{\mathbb R}}
\newcommand{\cC}{\ensuremath{\cal C}}

\begin{document}

\title{Chemical bonding in three-membered ring systems}

\author{Nina Strasser}
\affiliation{Institute of Solid State Physics, NAWI Graz, Graz University of Technology, Graz, Austria}
\affiliation{Department of Chemistry, University of Graz,  Graz, Austria}
\author{Alexander F. Sax}
\email{alexander.sax@uni-graz.at}
\affiliation{Department of Chemistry, University of Graz,  Graz, Austria}

\begin{abstract}
The formation of the four three-ring systems \ce{c-(CH2)_{3-k}(SiH2)_{k}}, ($k=0$: cyclopropane, $k=1$: silirane, $k=2$: disilirane, $k=3$: cyclotrisilane) by addition of methylene and silylene to the double bond in ethene, disilene, and silaethene, as well as the elimination of the carbene analogs from the three-rings, was studied with CAS(4,4) wave functions in both $C_{2v}$ and $C_s$ symmetry. To reveal charge and spin redistribution during these reactions the CAS(4,4) wave functions were analyzed using the orthogonal valence bond method (OVB). The potential energy curves, different internal coordinates, and the results of the OVB analysis show, that frequently the addition and elimination reactions follow different minimum energy paths, because they are indeed diabatic reactions. In these cases, there are no energy barriers corresponding to saddle points on the potential energy surfaces but the energy increases during one diabatic reaction until, at a certain point, the system jumps to the other diabatic state and, in the following, the energy decreases. This happens for reactions in $C_{2v}$ symmetry; as soon as the system can change to the lower symmetry, the diabatic states combine to an adiabatic one and the reaction follows a single minimum energy path.\end{abstract}

\maketitle

\section{Introduction}
In 1981, two manuscripts were submitted for publication, in which the synthesis of disilenes, hitherto unknown molecules with silicon-silicon double bonds, were described. West \etal{} reported in their paper\cite{West1981} the dimerization of silylenes, whereas  Masamune \etal\cite{Masamune1982} prepared disilene by photofragmentation  of a cyclotrisilane, another hitherto unknown molecule. That hexakis(2,6-dimethylphenyl)cyclotrisilane was  ``stable to oxygen, moisture, and heat to its melting point'' indicated unexpected thermodynamic stability. Only when the cyclotrisilane was irradiated with ultraviolet light for 5 minutes, it was nearly quantitatively converted into  disilene and silylene.\cite{Masamune1982} This suggests that large aryl substituents increase the kinetic stabilization of the three-membered ring against oxidation and fragmentation, at least in the electronic ground state. When the aryl substituents were replaced by alkyl substituents\cite{Masamune1983}, the cyclotrisilanes were found to be rather sensitive to exposure to oxygen; but the light induced fragmentation took several hours. These experimental results did not explain why the cyclotrisilanes were that stable; after all, according to chemical knowledge, both the smallest cyclic silanes and unsaturated silanes were regarded as nonexistent compounds or compounds of low stability\cite{Dasent1965}. To explain the lower stability of the silicon species when compared with the carbon analogs, the overlap of silicon valence orbitals was claimed to be much poorer. In case of disilenes, the side-on overlap of p$_{\pi}$ AOs (atomic orbital) was considered to be too small, in case of small ring systems the small overlap of sp$^3$ AOs was claimed to be the cause of low stability and of high reactivity. The high stability caused by suitable substituents was surprising and demanded a correction of the explanation.

Bonding is the stabilizing process in a system composed of several subsystems,\cite{System} in molecular systems the subsystems are either atoms or stable atom groups; stable means that the atom group does not spontaneously disintegrate into smaller atom groups or atoms. Stable atom groups, also termed molecular species or molecular compounds, can be radicals or molecules; molecules are molecular species that do not react violently with other species of its own kind.\cite{Preuss1963} In this paper, we will call any subsystem of a molecular system a fragment. An important criterion for any kind of bonding is the total energy: if the total energy of a system  decreases during some process the system is stabilized and thus bonded. For chemists, the following structural criterion must be fulfilled for chemical or covalent bonding in a system: at least one atom group with a significant short interatomic distance, which  did not exist in the system before the stabilization process started, must be created. This  criterion together with the claimed role of the Lewis electron pair and the valence concept is the basis of the traditional chemical explanation of covalent bonding.
According to Lewis, the formation of the  atom groups is caused by electron pairs shared between the atoms, meaning that each atom in the atom group contributes one electron to the electron pair (Lewis' rule of two). The valence of an atom is equal to the number of unpaired electrons it has and to the number of  electron pairs the atom can share with other atoms. In chemistry parlance, one speaks of a covalent single AB bond if the AB atom group is the result of the formation of one shared electron pair between atoms A and B. If not all of the N unpaired valence electrons are used to form single bonds to different atoms, two or more electron pairs can stabilize an AB group and one speaks then of the existence of double bonds, triple bonds etc. in the AB group. Pairs of valence electrons that are located at one atom but are not shared electron pairs are called lone pairs. A serious deficiency of Lewis' electron pair model is that it does not say when electrons will pair.

As soon as the fermionic character of electrons was recognized, it became clear that a shared electron pair is an electron distribution in a singlet spin state, called a low-spin state. That means, the two electrons must be in spin projection states with different quantum numbers, in chemical parlance, one electron has $\alpha$ spin, the other has $\beta$ spin or, simply said, they are singlet coupled. If the spins of both electrons are in the same spin projection state, either  both are $alpha$ or both are $\beta$ electrons, the electron distribution is in a high-spin triplet state, the electrons are unpaired. Singlet coupled valence electrons are always paired and cannot immediately participate in the formation of shared electron pairs, whereas triplet coupled valence electrons are unpaired and therefore immediately ready for bonding. The difference in the dimerization of methylene and of silylene demonstrates this nicely. The ground state of methylene is a triplet state, the two valence electrons are triplet coupled lone pair electrons, which are ready for the formation of a double bond. The ground state of silylene, however, is a low spin singlet state, the lone pair electrons are nor ready for bonding but must be prepared by exciting the lone pair from the low-spin state to the high-spin state.\cite{Sax2015}

However, pairing of unpaired electrons is not sufficient for covalent bonding. Since the first quantum theoretical studies of bonding in the hydrogen molecule, it is known that charge shifts during bonding must be accounted for.\cite{Weinbaum1933} And the more valence electrons the atoms connected by covalent bonds have the more inter- and intra-fragment interactions can occur and must be properly treated. These interactions may cause dramatic changes in the distribution of charges and spins and, accordingly, in the structure of the molecules involved; MCSCF (multi-configurational self-consistent field) wave functions, especially the CASSCF (complete active space self-consistent field) variant, allow the sound description of reactions in such molecular systems. But local effects remain hidden if the wave functions are built with delocalized molecular orbitals (MO), and they can be revealed when valence bond (VB) methods based on localized orbitals are used. The OVB (orthogonal valence bond) analysis of CASSCF wave functions\cite{Sax2015} was developed to reveal the hidden local features by transforming delocalized CASSCF MOs into localized fragment MOs (FMO).

To find out which processes are constitutive for creating or breaking of covalent bonds, comparative studies of reactions with homologous reactants are helpful. The insertion of methylene\cite{Kollmar1979} and silylene\cite{Gordon1984,Sosa1984,Sax1985} into the hydrogen molecule were the first reactions that stimulated a re-investigation with the OVB method, followed by the dimerization of methylene and silylene to ethene and disilene, respectively,\cite{Sax2015,Sax2017} as well as the formation of silaethene from methylene and silylene.\cite{Sax2022}
The insertion reaction into hydrogen is a prototype reaction for insertions of all carbene analogs into sigma single bonds. In these reactions the bonding electron pair of the sigma bond and the singlet coupled lone pair of the carbene analogs must simultaneously be excited from low- to high-spin states, and, since all reactions are studied in systems with singlet ground states, these spin rearrangements in the reactants must occur simultaneously. Depending on the system's symmetry, polarization of the charge distribution can enhance the decoupling, and make symmetry-forbidden reactions symmetry-allowed, as found for the dimerization of carbene analogs.\cite{Sax2017}

In this paper, the elimination of methylene and silylene from cyclic three-rings and the reverse addition of methylene and silylene to the $\pi$ bonds in ethene, disilene, and silaethene are studied in both high $C_{2v}$ and low $C_s$ symmetry. In high symmetry, elimination of carbene analogs is costly, in low symmetry elimination proceeds without any barriers. This was found for all three-rings, and explains, why three-rings with large, bulky substituents, which prohibit a deformation to $C_s$ symmetry, are rather stable.  Comparison of the reactions in the cyclopropane and cyclotrisilane systems show intrinsic similarities and differences of the two small ring systems. The reactions of silacyclopropane and disilacyclopropane show the differences in the electron structure between a molecular system with a C-C double bond and silylene compared with the system with an Si-Si double bond and methylene.

\section{The role of the Pauli exclusion principle}
That electrons are fermions is fundamental for the electron distribution in many-electron systems. Both fermions and bosons are quantum particles with a spin, fermion spins have half-integer quantum numbers $s$, boson spins have integer spin quantum numbers. To each quantum number $s$ there are $2s+1$ different projections of the spin onto an external axis, and thus $2s+1$ different spin states with spin projection quantum numbers running from $-s$ to $+s$ in steps of one. Electrons are fermions with spin quantum number $s=1/2$, having two spin projections, called spin-up and spin-down; in chemistry, electrons in the spin-up state are called $\alpha$-electrons and electrons in the spin-down state are called $\beta$-electrons. Electrons in the same spin projection state are identical electrons. Many-boson and many-fermion systems have very different physical properties, which are the consequence of the different symmetries of the quantum states $\ket{\Psi}$.\cite{Tensor}  The most simple way to describe an $N$-particle systems is by looking at the probability to find one particle in a certain one-particle state $\ket{\phi_1}$, a second in one-particle state $\ket{\phi_2}$, and so on. This probability $P(1,2,\cdots,N)$ is the square modulus of the, in general complex, quantum amplitude $\overl{\phi_1,\phi_2\cdots \phi_N}{\Psi}$ or simply $\overl{1,2\cdots N}{\Psi}$. Quantum amplitudes of many-boson systems are  totally symmetric, meaning that a permutation of the individual states leaves the quantum amplitude unchanged, $\overl{1,2,\cdots,j,\cdots,i,\cdots N}{\Psi} = \overl{1,2,\cdots,i,\cdots,j,\cdots N}{\Psi}$, whereas wave functions of many-fermion systems are totally antisymmetric, $\overl{1,2,\cdots,j,\cdots,i,\cdots N}{\Psi} = -\overl{1,2,\cdots,i,\cdots,j,\cdots N}{\Psi}$. This implies that a fermion-wave function is zero whenever two individual states are identical or, in other words, a system of fermions can never be in a state in which two individual states are identical. This is the Pauli exclusion principle (PEP). If the individual states are characterized by quantum numbers, the PEP can be expressed as follows: no two individual states can agree in all quantum numbers; if the individual states are position states, the PEP says that two identical fermions can never be at the same position in space. Two identical bosons can do. Considering that the square of the modulus of the quantum amplitude can be interpreted as a probability, one can also say that, for $\mbr_i\to \mbr_j$ with $i\ne j$, the probability to find two identical electrons at $\mbr_i$ and $\mbr_j$ goes to zero. That two identical electrons avoid coming close in space
resembles a repulsive interaction, but one that is much more effective than any other physical interaction like, e.g., the Coulomb interaction.  Identical electrons avoid each other because of the PEP and the Coulomb repulsion, non-identical electrons experience only the Coulomb repulsion. This has an important effect on the spatial extension of orbitals: if two electrons occupy the same orbital they must be non-identical, thus they can come arbitrarily near and experience the strong Coulomb repulsion. To diminish the repulsion the spatial region where two non-identical electrons are located must increase, that means the orbital must expand. The same is true if two non-identical electrons occupy two different orbitals that occupy regions in space that are very close, for example the atomic orbitals (AO) representing the subshells of an electron shell in a many-electron atom. But if the AOs are occupied by identical electrons, the Coulomb repulsion is strongly reduced due to the PEP and the AOs contract.
Given incompletely filled subshells of the valence shell, the degenerate AOs can be occupied with non-identical or with identical electrons. The total spin quantum number $S$ is proportional to the difference of the number of $\alpha$-electrons, $n_{\alpha}$,  and $\beta$-electrons, $n_{\beta}$, with $N=n_{\alpha}+ n_{\beta}$. $S=(2n_{\alpha}-N)/2$ ranges from zero (for even $N$) or 1/2 (for odd $N$), to $S=N/2$, spin states corresponding to small total spin quantum numbers are low-spin states, those corresponding to large spin quantum numbers are high-spin states. The number $2S+1$ is called the multiplicity of the state. If the degenerate AOs are occupied with identical electrons, the system is in the state with the highest spin. In the carbon and the silicon atom, only two electrons occupy  the three AOs of the p-subshell. If the two electrons are identical, the state of the total spin is a high-spin state and the electrons must occupy different AOs; if the electrons are non-identical the state of the total spin is a low-spin state, and the electrons can either occupy the same AO or different AOs. In both cases, the low-spin state has a higher total energy than the high-spin state. This is expressed in one of Hund's rules, which states that for a given electron configuration (distribution of electrons in AOs) the state with the highest multiplicity has the lowest energy (rule of maximum multiplicity). Hund's rules were originally formulated for many-electron atoms, but they apply also to molecular states.\cite{PCbook}

When atoms from different periods in the periodic table are studied, one must also consider the different extensions of the AOs in the valence shell. The silicon atom has a larger extension than the carbon atom, the same is true for the valence AOs, consequently,  the Coulomb repulsion of two non-identical electrons in a single silicon valence AO is smaller than in a carbon valence AO.  Hund's rule of maximum multiplicity can be used to predict the ground states of molecules with partially occupied degenerate molecular orbitals (MO) such as the O$_2$ molecule with the $\pi_g^2$ configuration, or linear methylene or linear silylene both with the $\pi_u^2$ configuration.
All three molecules have triplet (high-spin) ground states. If the triatomic molecules bend, the degeneracy is lifted and the $\pi$ MOs develop into atom centered lone pair MOs: one becomes an sp-hybrid with a large s-contribution and a second MO that is essentially a p AO; the energy of the sp-hybrid is lower than that of the p AO. Now the different extensions decide on the orbital occupation: in methylene the extensions of both lone pair MOs is much smaller than the extension of the respective MOs in silylene, single occupation of both orbitals with identical electrons is preferable to a double occupation of one AO, and the ground state of bent methylene is indeed a high-spin triplet. The large extension of the MOs in silylene on the other hand allows a double occupation of the energetically lower sp-hybrid and the ground state of silylene is a low-spin singlet.

The PEP governs also the electron and spin distribution in larger molecules.
The existence of two identical electrons in triplet methylene allows immediately the formation of two covalent bonds when two ground state methylenes combine to singlet ethene; the two lone pair electrons at each carbon atom remain also in the ethene molecule in a local high-spin arrangement. In the reaction of two singlet silylenes to singlet disilene, the two non-identical lone pair electrons, which occupy the energetically low lone pair sp-hybrid, are not able to form  covalent bonds between the silicon atoms, they must be  transformed into a pair of identical electrons by a spin flip, by which the singlet coupled spins are decoupled. This means that the lone pair electrons in
methylene and silylene fragments in ethene, disilene, or silaethene are identical electrons. Speaking of lone pair electrons in fragments assumes that the lone pair MOs of the isolated fragments retain their characteristics also in the molecule and that it makes sense to say that in the molecule the lone pair electrons occupy fragment MOs (FMO) that resemble the lone pair MOs of the isolated fragments. And only then one can say, identical electrons located in lone-pair  FMOs constitute a local high-spin state; similarly constitute non-identical electrons in lone pair FMOs local low-spin states. One must not forget, however, that these local spin states are not fragment states, which are always mixed states\cite{Cohen1977}; the justification for speaking of local spin states is the possibility to construct spin eigenfunctions of the molecular system from spin eigenfunction of the fragments.\cite{Pauncz1979}

\section{The OVB analysis of CASSCF wave functions}
Configuration state functions (CSF) of an $M$ electron system are antisymmetrized products of a spatial wave function, in most cases a product of spatial MOs, and an eigenfunction of the spin operator $\hat{S}^2$.\cite{Pauncz1979} The most simple CSF of a system with $M=2N$ electrons is a closed-shell CSF, it consists of the $N$ lowest MOs, all occupied with two electrons, combined with a singlet spin eigenfunction. In general, the MOs are calculated with the SCF (self consistent field) method. A  wave function that consists of only one closed-shell CSF cannot  properly describe the change of the electron distribution of a molecular system during chemical reactions, but a linear combination of CSFs, called a multi-configurational (MC) wave function, can do it. For chemical reactions only valence electrons are relevant, which are the electrons in the valence shell, that is the outermost electron shell of an atom; the sum of all atomic valence electrons is the number of valence electrons of a molecule. Valence AOs are all AOs describing the valence shell of an atom; valence MOs are all MOs that are linear combinations of valence AOs. The CSFs of an MC wave function are constructed by taking a certain number of electrons, called the active electrons, out of doubly occupied MOs of the closed shell CSF and placing them into some unoccupied MOs. Active MOs and active valence electrons are defined as those valence MOs and valence electrons, respectively, that are regarded as important for the description of a chemical reaction. The coefficients of the CSFs in MC wave functions are termed CI (configuration interaction) coefficients.
In the MCSCF method (multi-configurational SCF) all active MOs and the CI coefficients are simultaneously optimized.  An important class of MCSCF wave function are the so called CASSCF (complete active space SCF) wave functions made with the maximum number of CSF that can be constructed with $n$ active electrons and $m$ active MOs. The symbol for such a CASSCF wave functions is CAS($n$, $m$). The number of possible CSFs, given by the Robinson-Weyl formula\cite{Pauncz1979}, depends not only on $n$  and $m$ but also on the spin quantum number $S$ of the wave function, it is

\begin{displaymath}
M(n,m,S) = \frac{2S+1}{n+1} \binom{m+1}{\frac{n}{2} - S}\binom{m+1}{\frac{n}{2}+S+1}.
\end{displaymath}

The specification of the active MOs and active electrons is part of the modelling of a chemical reaction. CASSCF wave functions are designed to give  optimal descriptions of reactions, therefore they are also called FORS (fully optimized reaction space) wave functions.

Because of the variational optimization of CASSCF wave functions, all information about the interaction between reactants in a chemical reaction is stored in the orthogonal CASSCF MOs. And because the MOs are orthogonal also the CSFs made with them are orthogonal. This allows to interpret the square of the CI coefficient of a CSF as the weight of the CSF in the wave function. If CSFs of an MC wave function are constructed with non-orthogonal orbitals, the CSFs are also non-orthogonal and the square of the CI coefficient of a CSF cannot be directly  interpreted as the weight of the CSF.

As mentioned above, spin functions of composite molecular systems can be constructed from spin functions of the subsystems, that is the fragments. The antisymmetrized product of FMOs and such a spin function, gives a CSF that can be interpreted as antisymmetrized product of subsystem CSFs or fragment CSFs (pFCSF). If the FMOs are obtained from the delocalized CASSCF MOs by an orthogonal transformation,\cite{Sax2012,Sax2015} also the FMOs are orthogonal as are the pFCSFs. The FMOs store the same interaction information as do the original MOs, but now it is possible to trace the interactions in the molecular system to specific interactions between the fragments. In other words, the wave function of the composite system can be represented by $M(n,m,S)$ CSFs with delocalized MOs or by the same number of pFCSFs. This is the basis of the OVB analysis of CASSCF wave functions.\cite{Sax2015,Sax2017,Sax2022,Sax2023}

The charge distribution of $n_f$ electrons in an isolated neutral fragment is described by a neutral pFCSF. If the molecular system is composed of two fragments, the following descriptions of the fragments are possible: the system consists of a pair of neutral fragments, or of a cation/anion fragment pair, or of a dication/dianion fragment pair and so on. Accordingly, the pFCSFs are products of neutral FCSFs, or of one FCSF describing a cationic charge distribution and a second one describing an anion, and so on. $M(n,m,S)$ is, in general, much larger than the number of neutral pFCSFs, but it is equal to the sum of all neutral and all ionic pFCSFs, that are products of FCSFs describing states of cation/anion fragment pairs. A single ionic pFCSF describes the charge distribution in a cation/anion pair, which can be regarded as the result of a charge transfer between neutral fragments, accordingly, ionic pFCSF are frequently called charge transfer CSFs. However, only wave functions of the molecular system that consist of a single charge-transfer pFCSF describe a physical  electron transfer, in an MC wave function, charge-transfer pFCSFs describe charge shifts (charge polarization) in the composite system.

FMOs that are calculated for one isolated fragment with an SCF method are orthogonal to each other, but they are not, in general, orthogonal to FMOs of another fragment. pFCSFs made with such FMOs are not orthogonal to each other, except when the fragments are spatially far from each other, that is, when the system is dissociated. Because of the non-orthogonality of the FMOs, an electron occupying an FMO in the first fragment has a finite probability to be found in an FMO of the second fragment, and this interference describes also a charge shift. This means, also neutral pFCSFs constructed with non-orthogonal FMOs contribute to the description of charge shifts, which are hidden by the mathematical form of the pFCSFs. This makes a unique and sound physical interpretation of such pFCSFs impossible. When orthogonal FMOs are used, the electron distributions in the fragments are always either neutral or ionic, and charge shifts are described by linear combinations of neutral and ionic pFCSFs. As mentioned in the Introduction, the dimerization of silylene is only possible when the singlet coupled lone pair electrons in each fragment are decoupled to a triplet high spin state, and these local triplet states are coupled to a global singlet state.  Local spin-rearrangements in neutral fragments are invisible when wave functions with delocalized MOs are used, they are however revealed by an OVB analysis. In a recent paper by one of the authors,\cite{Sax2023} the charge and spin rearrangements during the dissociation of the C$_2$ molecule were investigated.

In the following sections, we will discuss only properties of pFCSF, so we replace this acronym by CSF.

\section{The description of chemical reactions}\label{sec:chemreac}
\subsection{Potential energy surface, minimum energy path, potential energy curve}
A fundamental theoretical tool for the study of processes in molecular systems is the potential energy surface (PES), which is a consequence of the Born-Oppenheimer approximation. A lucid discussion of this topic was given by Sutcliffe and Woolley.\cite{Sutcliffe2012} Separating the translational and rotational contributions from the Hamiltonian of an $N$-atomic  molecular system yields a system of moving electrons and nuclei, the latter can vibrate,  permutation of identical nuclei are also possible. In a last step one can assume that the nuclei do not change their relative position, which are given by $n$ general coordinates $\{q_1, q_2, \cdots,q_n\}$, the $n$-tuples are elements of the nuclear configuration space $X$. Any point $P\in X$, also called the system point,  represents a certain geometry of the molecular system; changing $P$ means changing the geometry of the molecular system, as it happens during chemical reactions. The system at point $P$ is described by the clamped-nuclei Hamiltonian $\H{P}$, the eigenstates $\ket{\Psi(P)}$ and the eigenvalues $E(P)$ depend also on $P$. The energy is the so called electronic energy of the system, the energy $E(P)$ as a function of $P$ is the potential energy surface (PES), which is a hypersurface in the $(n+1)$-dimensional space $X\times \bR$.  Eigenstates and eigenvalues can be indexed $I=1,2, \dots$ at any point $P$ in order of increasing energy $E$. The eigenstates of a clamped-nuclei Hamiltonian are called adiabatic, as are the wave function describing an eigenstate and the corresponding PES. Because of the way adiabatic PESs are constructed, they can have local minima connected by saddle points, troughs separated by ridges and so on; one says adiabatic PESs are not smooth. The interpretation of processes in the molecular system is therefore more difficult than when the PESs are smooth. A non-smooth adiabatic PES indicates often large changes of the molecular system accompanied by dramatic changes of the electronic structure. Atchity and Ruedenberg classified, therefore, states as adiabatic if ``..in certain regions of coordinate space, drastic changes occur in the electronic structures ...''.\cite{Ruedenberg1993,Atchity1997}
Chemists are used to represent equilibrium structures of molecular systems by conventional structural formulas, which store essential chemical informations like the connectivity and the type and number of bonds between connected atoms, and represent important parts of the characteristics of the electronic structure of the system.  In an adiabatic reaction, \ce{A -> B}, a system A with its essential characteristics is smoothly transformed into a system B having a different characteristics.

A curve (path, trajectory) $\cC(\lambda)$ in $X$ is a subset of $X$ that can be indexed (parametrized) by a single continuous parameter $\lambda$; if the arc length of the curve is used as parameter one speaks of the natural parametrization of the curve. In chemistry, the arc length is called the reaction coordinate. It may happen that, for a certain curve $\cC(\lambda)$ in $X$, one of the $n$ coordinates $q_i$ makes the largest contribution to the arc length,  or is regarded to describe the most important aspect of the structural change in the system considered. For example, when during a chemical reaction two fragments approach each other, the structure of the reactants will change as will the relative orientation of the fragments,  but the distance $R$ between the reactants is often considered to be most important. Then, in chemical parlance, the coordinate $R$ is often called the reaction coordinate, although it is indeed only an approximate reaction coordinate, describing only one aspect of the change of the molecular structure during the reaction. The reaction coordinate used in this paper is always an approximate coordinate $R$ that measures the distance between the reactants. If a reaction occurs without any symmetry restrictions for the  system's geometry, then, for a certain $\lambda$, all internal coordinates will be different; if there are symmetry restrictions, some internal coordinates will have either equal values or they will differ by the sign. Then the curve will lie in a subset of $X$, whereas, without symmetry restrictions, the curve is allowed to meander through a much larger subset of space $X$.

An adiabatic PES that describes a simple chemical reaction has typically two trough-shaped channels, the entrance and the exit channel, which are, in general, connected by a saddle point. The curve at the bottom of the troughs passing the saddle point is called the energy profile or the potential energy curve (PEC); it is a curve $E(\cC(\lambda))$ in $X\times \bR$ and can be regarded as the graph of the energy function $E(P)$ over $\cC(\lambda)$ as domain,  called the minimum energy path (MEP). A PEC lies always on a PES, PECs are often used to make important aspects of a PES more obvious, such as  reaction barriers, energy differences etc. If $P_A\in X$ and $P_B\in X$ represent the structures A and B of the molecular system, the MEP between $P_A$ and $P_B$ has a length $l(\cC)$, which is, in general, assumed to be finite. For the graphical representation of the energy profile the MEP is always rectified, meaning, that the curve $\cC(\lambda)$ is replaced by a line segment having the same length  $l(\cC)$ as the MEP. The reaction coordinate is then element of an  interval $I\subset \bR$ with $l(I)=l(\cC)$. By this, the energy profile becomes the graph of the function $E(\lambda)$, $\lambda\in I$, the graph is subset of the cartesian product $I\times \bR$. A PEC must be a curve without kinks or jumps, the PEC of a reaction going over a saddle point has always a local maximum and frequently two local minima. In practice, the true reaction coordinate $\lambda$ is mostly replaced by an approximate reaction coordinate $R$, and the true reaction profile $E(\lambda)$ is replaced by an approximate potential energy curve $E(R)$.
Because of the high dimension, $3N-6$, of the configurational space $X$, it is impossible to imagine all geometry changes along a path $\cC(\lambda)$, but the $n$ coordinate curves $q_i(\lambda)$, $i=1,2, \dots, n$, can help to do it. If the approximate reaction coordinate $R$ is used, the $n-1$ curves $q_i(R)$ are used. Too often the change of coordinates along a path is not appropriately considered, although only they can describe structural changes; changes in the energy alone, as described by a PEC, cannot do this. Only a PEC together with all coordinate curves can give a hint to the physical processes behind a chemical reaction.

\subsection{Diabatic and adiabatic states}\label{sec:diabat}
Two reactants C and D without unpaired electrons cannot form a product with covalent bonds between them; only when the reactants are prepared for bonding, \ce{C^{*}} and \ce{D^{*}}, they can form the product \ce{C^{*}D^{*}}. Both the combination reaction
\ce{C + D -> C^{*}D^{*}} and the dissociation reaction \ce{ C^{*}D^{*} -> C + D} are adiabatic reactions, in which the electron structure of the molecular system changes dramatically. On the other hand, in the  two reactions,  \ce{C + D -> CD} and  \ce{C^{*}D^{*} -> C^{*} + D^{*}}, the electronic characteristics of the initial structures are not changed. In the recombination reaction, the fragments are compressed to the structure of the product, but they are still not prepared for bonding; in the dissociation reaction, the fragments are still prepared for bonding, but the distance between them is large. According to Atchity and Ruedenberg, states, the corresponding wave functions, and the reactions are called diabatic, if ``.. [the] electronic structures maintain their essential characteristics over the entirety of such regions''.\cite{Ruedenberg1993,Atchity1997}

\begin{figure}[ht]
\includegraphics[width=0.9\textwidth]{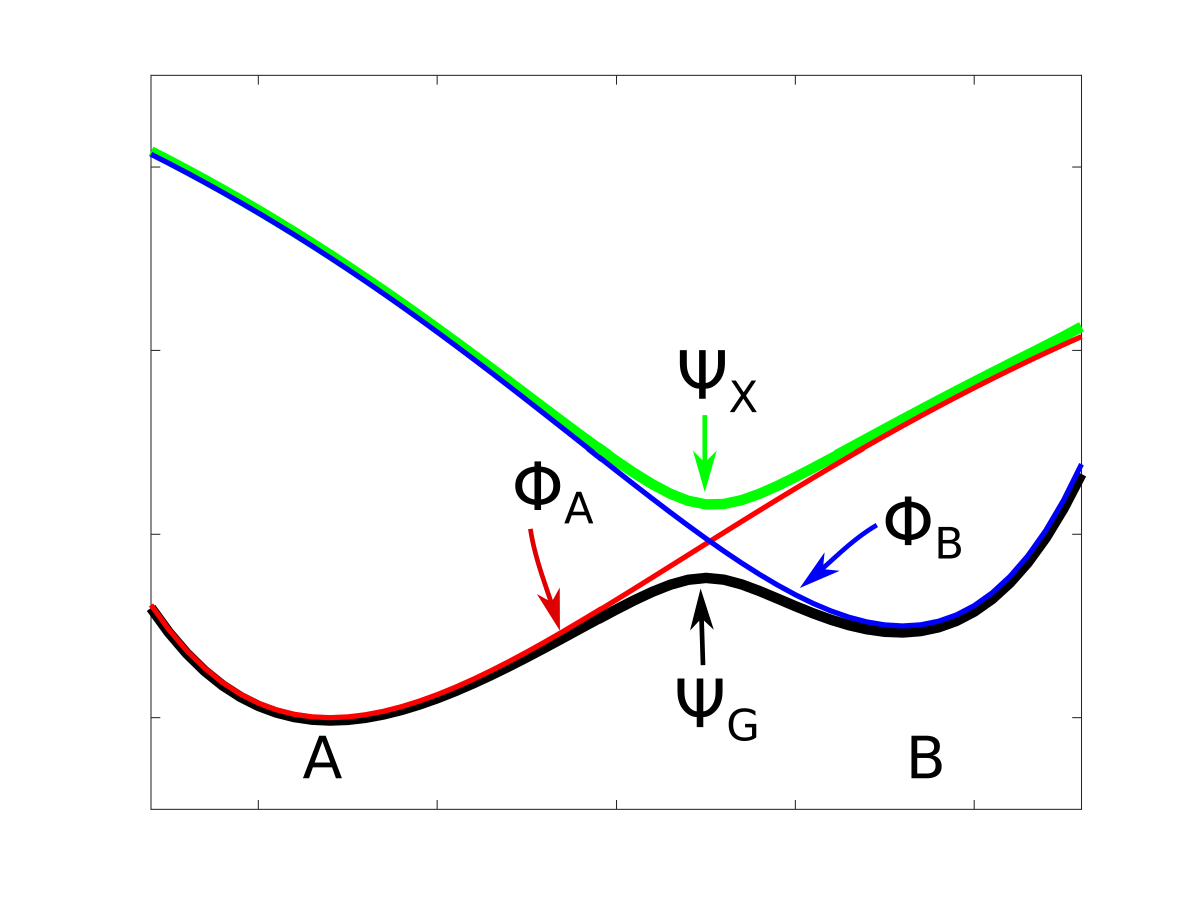}
\caption{PECs obtained with the adiabatic wave function $\Psi_G$ for the ground state and the adiabatic wave function $\Psi_X$ for an excited state, and with the diabatic wave functions $\Phi_A$ and $\Phi_B$.}\label{fig:adiadia}
\end{figure}

Adiabatic states are well defined as eigenstates of the Hamiltonian of the system, the adiabatic PECs are made with the corresponding eigenvalues. The PEC of an adiabatic state with wave function $\Psi_G$ that describes the ground state of a molecular system having two stable  structures A and B will show two local minima connected by a local maximum. See Figure \ref{fig:adiadia}. There will be an excited adiabatic state with wave function $\Psi_X$, which has the same symmetry as  $\Psi_G$, the PEC of this excited state has a local minimum where the ground state PEC has a local maximum; one speaks of an avoided crossing as a consequence of the non-crossing rule for states with  the same spin or spatial symmetry,  see for example Cohen-Tannaoudji\cite{Cohen1977}. The diabatic states, $\Phi_A$ and $\Phi_B$, are not eigenstates of a Hamiltonian, they must be constructed, for example by linear combinations of adiabatic states, but this can be done in different ways as is described by Shu \etal\cite{Shu2022}. Each of the diabatic PECs has a single local minimum at one of the two stable structures; there the shapes of the diabatic PECs are nearly identical with the shape of the PEC of the adiabatic ground state. And the PECs of the diabatic states cross where the adiabatic PECs show the avoided crossing. In Figure \ref{fig:adiadia} the two diabatic states and the two adiabatic states describe a reaction following a single MEP. Thermal reactions are nearly always adiabatic reactions; most photochemical reactions or electron transfer reactions, in general all fast reactions are  diabatic reactions, they connect system structures in the ground state with system structures in excited states.  A PEC that consists of the parts of the diabatic PECs below the crossing point is a curve with a cusp. It never describes a real chemical reaction along the MEP.

In our studies, we describe frequently forward and  reverse reactions following different, non-crossing MEPs; the reaction valleys are then two skew troughs that do not coalesce. In this case the PECs resemble those of diabatic states, but they are indeed adiabatic PECs because the energies are the lowest eigenvalues of the Hamiltonian.
As long as one does not study the potential energy surface in detail, one can only assume that the two skew troughs are separated by a ridge of unknown height. But chemical reactions never occur in such a way that the system point follows strictly a MEP; the vibrations of the atoms result in trajectories that oscillate about the MEP, similarly to the motions of a skater in a half-pipe.  If the system energy is higher than the energy at the bottom of the trough, and if the oscillations of the system point are so strong that its energy is larger than the height of the ridge, the system point can jump from one trough to the other.

If the MEPs are rectified and parametrized over the same interval $I$ by the approximate reaction coordinates $R$, and if we assume that the PEC of the forward reaction is a monotonically increasing curve and that of the backward reaction is monotonically decreasing, the two PECs will intersect at the crossing point $R=R_0$, resembling crossing diabatic PECs.  Only when all coordinate curves are considered, it becomes clear that the crossing curves are PECs, belonging to reactions that follow different MEPs, for which no non-crossing rule holds.

In the center of the treatment of diabatic states by Ruedenberg and Atchity\cite{Ruedenberg1993,Atchity1997}  is the observation that diabatic states fulfill the criterion of electronic uniformity, meaning that in certain regions of the nuclear configuration space the electronic structure does not drastically  change, whereas the electronic structure in adiabatic states does. The electronic uniformity criterion can be equivalently replaced by the configurational uniformity criterion. This criterion is based on the representation of a wave function in terms of CSFs made with some kind of MOs. If only few CSFs have non-neglectable weight, and if the weight of these CSFs remains large along a path in configuration space, one speaks of configurational uniformity and, since the CSFs represent a certain electronic structure, of electronic uniformity.
Now it is found, that diabatic CSFs can fulfill the configurational uniformity criterion only with proper MOs, called diabatic MOs. Different strategies have been proposed for constructing diabatic MOs,\cite{Shu2022} but it was found for the \ce{(H2)2} system,\cite{Nakamura2001} that the diabatic MOs are localized on the two \ce{H2} fragments. We found that the CSF, made with FMOs localized on the fragments, fulfill the configurational uniformity criterion, and, since it is easy to attribute a certain charge and spin distribution to each CSF, (see next section) also the criterion of electronic uniformity. This leads us to the conclusion that the reactions along different MEPs show the properties of diabatic reactions.

In this paper we study ground state reactions, and many of them follow indeed different MPEs. The PECs describing these reactions and shown in the figures consist of the branches of two PECs that lie below the crossing point $R_0$. The region with $R > R_0$ will be called the dissociated system, the fragments are mostly in low-spin states, and we will say that the wave function has low-spin characteristics; the region with $R < R_0$ is the bonded system, the fragments are in high-spin states and therefore the wave function has high-spin characteristics. The PEC of a reaction that follows a single MEP is a smooth curve without singular points like cusps or jump discontinuities. In some systems, jumps are caused by slow geometry optimization on an obviously not quadratic PES indicating a vanishing ridge between the troughs; in other systems, there is indeed a sudden change of the structure of a fragment.

In this paper, an OVB analysis is made only for the wave functions corresponding to the PECs shown in the figures. These curves can be either smooth curves following a single MEP or  they consist of pieces of PECs over different MEPs. In this case, the energy curves may have cusps or even jumps. Coordinate curves  have jump discontinuities where the energy curves either cross or have jumps. Nevertheless, all curves in this paper are represented as if they were graphs of continuous functions. In the Supporting Information the continuous representation is compared with the correct discontinuous one.

\section{The OVB analysis of the CAS(4,4) wave functions}
All molecular systems studied in this paper are in singlet ground states; the dissociated system consists of two fragments, an ethene analog and a carbene analog; the bonded system consists of a single fragment, the 3-ring. The chemical reactions are modeled with four active FMOs and four active electrons. In the bonded system, the four delocalized active MOs are the bonding and antibonding MOs describing the two $\sigma$ bonds between the two fragments, the four active electrons are in the two bonding MOs. In the dissociated system, the four active MOs are the bonding and the antibonding $\pi$ MOs in the ethene analog, and the s-type and the p-type lone pair AOs in the carbene analog.  The four active electrons are the electron pair occupying the $\pi$ MO and two electrons that can occupy the two lone pair MOs. The delocalized active MOs in the dissociated system are indeed already strongly localized at the two fragments, and onto the MOs of these fragments the delocalized MOs will be localized.  This choice of active orbitals assumes that the $\sigma$ MO in the double bond of the ethene analog is not involved in the reaction, and a CAS(6,6) wave function is not necessary to describe the reactions; two reactions were treated with both CAS(4,4) and CAS(6,6) wave functions to check whether this assumption is reliable. See the Supporting Information.

\begin{figure}
\includegraphics[width=0.7\textwidth]{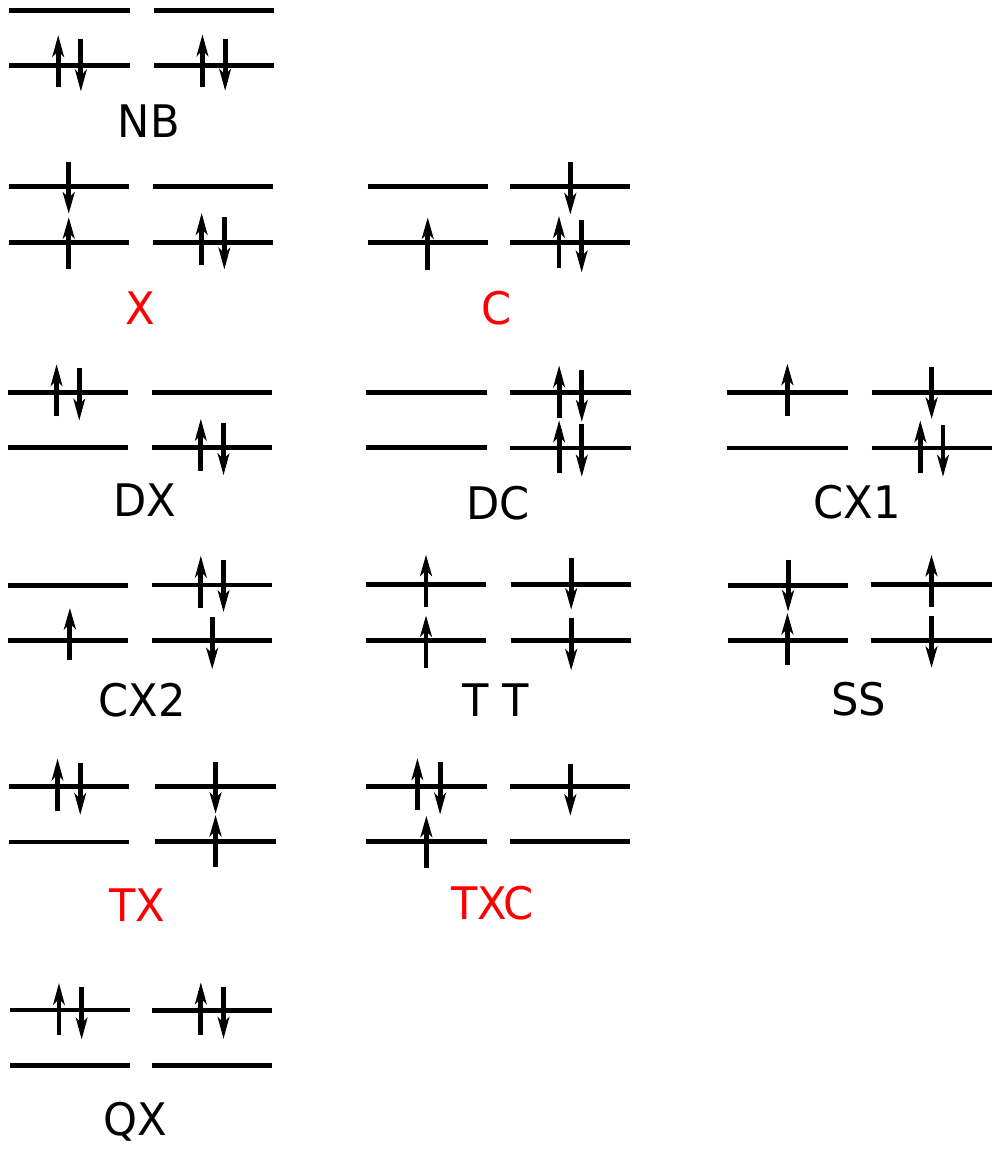}
\caption{The types of product FCSFs used to interpret a CAS(4,4) singlet wave function. CSF labelled in red contribute only in $C_s$ symmetry.}\label{fig:CSFs}
\end{figure}

For an OVB analysis of a system with geometry $P\in X$, the delocalized CASSCF MOs are localized on the FMOs calculated for the non-interacting fragments at geometry $P\in X$. The four orthogonal FMOs of the CAS(4,4) wave function are, in the order as they occur in the CSFs,  the  $\pi$ and $\pi^*$ FMOs of the ethene analog, and the  $s$, and $p$ lone pair FMOs of the carbene analog.
A CAS(4,4) wave function for a singlet state is a linear combination of at most 20 CSFs, the electron and spin distribution and the labels characterizing it can be seen in Figure~\ref{fig:CSFs}. The CSFs  NB, TT, SS, and QX describe  a unique distribution, all other labels describe two  distributions, which differ by switching the two fragment,in Figure~\ref{fig:CSFs}  only one electron distribution is shown. In Table \ref{tbl:CSFs},  the occupation types of all 20 CSFs are listed together with character strings describing the orbital occupation. Note, that the graphical representations of the CSFs do not contain any energetic information.

\begin{table}
\caption{All CSFs contributing to a singlet CAS(4,4) wave function. In the strings, $a$ and $b$ indicate $\alpha$ and $\beta$ spins, respectively, 2 indicates double occupation of a FMO, that means a singlet coupled spin pair.}\label{tbl:CSFs}
 \begin{tabular}{l|llllllllllll}
  \toprule
label   & NB & X & C & DX & DC & CX1 & CX2 & TT & SS & TX & TXC & QX\\ \midrule
occ & $|2020|$ & $|ab20|$ & $|a02b|$ & $|0220|$ & $|0022|$ & $|0a2b|$ & $|a0b2|$ & $|aabb|$ & $|abba|$ & $|02ab|$ & $|a20b|$ & $|0202|$ \\ &          & $|20ab|$ & $|2ab0|$ & $|2002|$ & $|2200|$ & $|2a0b|$ & $|a2b0|$ &          &          & $|ab02|$ & $|0ab2|$   \\
  \bottomrule
 \end{tabular}
\end{table}

\begin{itemize}
\item[1)]NB, is the single No-Bond CSF; the two low-energy FMOs are occupied by electron pairs, both fragments are in singlet states, therefore no unpaired electrons are available for bond making. This CSF describes spin distributions in the fragments for which, during the reaction, the PEP will be most important. In each fragment the lowest FMO is doubly occupied by an $(a,b)$ spin pair. Therefore, there are two inter-fragment spin pairs $(a,a)$ and $(b,b)$, which will strongly resist the approach of the fragments (no-bond). This CSF dominates the dissociated system. All other CSFs can be described as excitations or charge transfers with respect to the NB CSF.
\item[2)]X labels the two CSFs with a single excitation in one fragment without spin flip.
\item[3)]C labels the two CSFs with a single electron transfer from one fragment to the other.
\item[4)]DX labels the two CSFs with a double excitation of a singlet coupled electron pair in one fragment.
\item[5)]DC labels the two CSFs describing the transfer of an electron pair from one fragment to the other.
\item[6)]CX1 labels the two CSFs describing a single charge transfer and a single local excitation without spin flip from one doubly occupied MO.
\item[7)]CX2 labels the two CSFs describing a local double excitation in one fragment and a  single charge transfer from the other.
\item[8)]TT labels the single CSF describing two fragments in local triplet states coupled to a singlet. In both fragments two unpaired electrons are ready for bond making. There are no inter-fragment pairs of identical electrons, the PEP is not effective. This CSF is important for the description of the bonded system. With respect to NB, this CSF describes in each fragment a singe excitation with spin flip.
\item[9)]SS labels the single CSF describing two fragments in local excited singlet states. As in CSF NB there are two inter-fragment pairs of identical electrons and the PEP effectively resists the approach of the fragments. The electron configuration described by SS is never  important.
\item[10)]TX labels the two CSFs describing triple excitations, that is a double excitation in one and a single excitation without spin flip in the other fragment.
\item[11)]TXC labels the two CSFs describing a  triple excitation plus a single charge transfer.
\item[12)]QX labels the single CSF describing a quadruple excitation, that is a double excitation in each fragment. As in NB, there are two inter-fragment spin pairs $(a,a)$ and $(b,b)$, which will strongly resist the approach of the fragments. This CSF is never important.
\end{itemize}
The CSFs NB, X, DX, TT, SS, TX and QX describe neutral electron structures, in each fragment is the same number of active electrons;  all CSF containing a C in the label are ionic.

Covalent bonding needs essentially mutual charge shifts from one bonding partner towards the other, as was shown by Ruedenberg and coworkers.\cite{Ruedenberg2007,Bitter2007,Ruedenberg2009,Bitter2010,Schmidt2014} This charge shift is a consequence of the constructive interference of fragment wave functions. When orthogonal FMOs are used, ionic CSFs are necessary to describe the charge shift.

All reactions are studied in either the high $C_{2v}$ or in the low $C_s$ symmetry group. In $C_{2v}$, the low-energy FMOs $\pi$ and $s$ belong to irreducible representation (IRREP) $a_1$ and the high-energy FMOs $\pi^*$ and $p$ belong to IRREP $b_2$, so the eight CSFs X, C, TX, and TXC do not belong to the totally symmetric IRREP $A_1$ but to IRREP $B_2$. The totally symmetric wave function in $C_{2v}$ is therefore a linear combination of only twelve CSFs. The bonded system with two covalent bonds between the fragments is described by a superposition of the neutral CSF TT and some ionic CSFs. CSFs CX1 describe charge shifts between the low-energy FMOs; CSFs CX2 describe charge shifts between the high-energy FMOs. The DX CSFs are necessary to describe left-right correlation in the ethene analogs or angular correlation in the carbene analogs.

In $C_s$, all active orbitals belong to IRREP $a'$, and therefore the totally symmetric singlet $A'$ ground state wave function is a linear combination of all 20 CSFs. The eight CSF that did not contribute in $C_{2v}$ describe asymmetric local charge deformations and charge shifts, only C and X are found to be important.

The important data obtained in an OVB analysis are the diagonal elements of the CI matrix, that is the energy expectation values calculated with the CSFs, and the eigenvector of the CI matrix describing the investigated system state. The squares of the CI coefficients are the weights of the CSFs in the wave function. In the figures, the weight and energy curves of only large CSFs are shown, that are those CSFs for which the weight was larger than 0.1 (10 percent) somewhere along the reaction coordinate. To show how much the large CSFs contribute to the molecular wave function, the sum of the weights of the large and of the small CSFs  are shown for all reactions.

\begin{figure}
\includegraphics[width=0.45\textwidth]{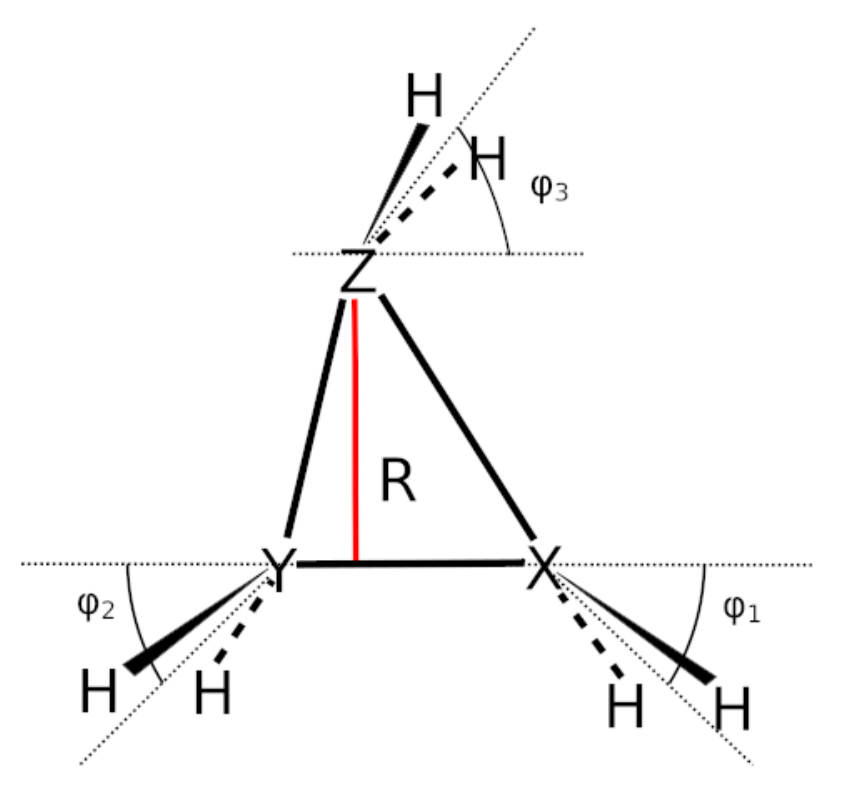}
\caption{A sketch of the molecular geometry of the investigated 3-rings and of important geometry parameters.}\label{fig:molgeom}
\end{figure}

Figure \ref{fig:molgeom} shows a sketch of the molecular geometry of the 3-rings. The three heavy atoms X, Y, and Z lie in the x-z plane; atoms X and Y lie on the x axis, the x-coordinate of atom X is always larger than that of atom Y; the normal distance from atom Z to the x axis is the approximate reaction coordinate $R$. Of all internal coordinates, only the three out-of-plane angles $\varphi_1$, $\varphi_2$, and $\varphi_3$ are indicated.
In $C_{2v}$ symmetry, atoms X and Y are symmetry equivalent, 7 of all 21 internal coordinates are unique, the four bond lengths r(X-Z), r(X-Y), r(X-H), and r(Z-H), three angles),  and the three bond angles HXH and HZH, and the out-of-plane angle $\varphi_1$. Out-of-plane angle $\varphi_3$ has always a value of 90 degrees.
In low $C_s$ symmetry, all three atoms are non-equivalent, 12 internal coordinates are unique, the six bond lengths r(X-Z),  r(Y-Z), r(X-Y), r(X-H), r(Y-H), and r(Z-H), the six bond angles HXH, HYH,  HZH, and the out-of-plane angles $\varphi_1$, $\varphi_2$, and $\varphi_3$.

The out-of-plane angles $\varphi_1$ and $\varphi_2$ have positive values if the hydrogen atoms lie below the x-y plane.  The positive values of angle $\varphi_3$ are counted counter-clockwise from 0 degrees to 180 degrees in the first and second quadrants, the negative values are counted clockwise from 0 degrees to $-180$ degrees in the fourth and third quadrants. That means, if the hydrogen atoms of the ZH$_2$ moiety  are closer to atom X than to atom Y, the $\varphi_3$ values are in the first or the fourth quadrant. If the ZH$_2$ moiety is parallel to the x-y plane, $\varphi_3$ is either 0 degrees or 180 degrees (equal to -180 degrees); $\varphi_3$ 90 degrees means that ZH$_2$  is parallel to the y-z plane with atom Z closer to the X-Y moiety than the hydrogen atoms.

The following reactions are investigated, all systems are in their respective totally symmetric singlet ground states.
\begin{itemize}
\item[\textbf{R1v}]\ce{c-C3H6 <=> C2H4 + CH2} \qquad in $C_{2v}$.
\item[\textbf{R1s}]\ce{c-C3H6 <=> C2H4 + CH2} \qquad in $C_s$.
\item[\textbf{R2v}]\ce{c-Si3H6 <=> Si2H4 + SiH2} \qquad in $C_{2v}$.
\item[\textbf{R2s}]\ce{c-Si3H6 <=> Si2H4 + SiH2} \qquad in $C_s$.
\item[\textbf{R3v}]\ce{c-CSi2H6 <=> Si2H4 + CH2} \qquad in $C_{2v}$.
\item[\textbf{R3s}]\ce{c-CSi2H6 <=> Si2H4 + CH2} \qquad in $C_s$.
\item[\textbf{R4v}]\ce{c-SiC2H6 <=> C2H4 + SiH2} \qquad in $C_{2v}$.
\item[\textbf{R4s}]\ce{c-SiC2H6 <=> C2H4 + SiH2} \qquad in $C_s$.
\item[\textbf{R5s}]\ce{c-CSi2H6 <=> CSiH4 + SiH2} \qquad in $C_s$.
\item[\textbf{R6s}]\ce{c-SiC2H6 <=> CSiH4 + CH2} \qquad in $C_s$.
\end{itemize}

With the  CAS(4,4) wave function only the bonding pairs of the two $\sigma$ bonds in the 3-ring are correlated, the $\sigma$ bond of the double bond is never correlated. Therefore, the equilibrium geometries of cyclopropane and cyclotrisilane do not have the correct $D_{3h}$ but only $C_{2v}$ symmetry. In reactions \textbf{R5s} and \textbf{R6s}, the system has always $C_s$ symmetry; the C-Si $\sigma$ MO in silaethene is always doubly occupied, but for the other C-Si $\sigma$ bond a correlated description is used; therefore the two C-Si bonds in \ce{c-SiC2H6} and in \ce{c-SiC2H6} are not equivalent.

Geometry data of ground and excited states calculated for the isolated fragments can be found in the Supporting Information.

\newpage
\subsection{The reactions \ce{c-C3H6 <=> C2H4 + CH2}}

All three atoms X, Y, and Z from the sketch in Figure \ref{fig:molgeom} are carbon atoms. We investigated the elimination of CH$_2$ from the 3-ring as well as the recombination of CH$_2$ and ethene (addition reaction); both reactions were performed in $C_{2v}$ (\textbf{R1v}) and  $C_s$ symmetry (\textbf{R1s}). The approximate reaction coordinate $R$ for both reactions varies between $R_m = 1.2$\,\AA{} (``molecule'') and $R_d=5.0$\,\AA{} (``dissociated system''). At the equilibrium geometry of \ce{c-C3H6} the approximate reaction coordinate is $R=1.332$\,\AA.

During the addition of methylene to the $\pi$ bond in ethene, the geometry around the C atoms in ethene changes from planar (sp$^2$ hybridization) to pyramidal (sp$^3$ hybridization), and this affects also the C-C $\sigma$ bond. One could assume that adding the C-C $\sigma$ MO to the set of active MOs will change the electron structure considerably and will strongly influence the shape of the curves describing the total energy and the geometry parameters. The results of the CAS(6,6) optimization can be found in the Supporting Information; the agreement with the CAS(4,4) results shows, however, that a CAS(4,4) ansatz is indeed sufficient for the description of  this reaction.

\begin{figure}
\includegraphics[width=0.46\textwidth]{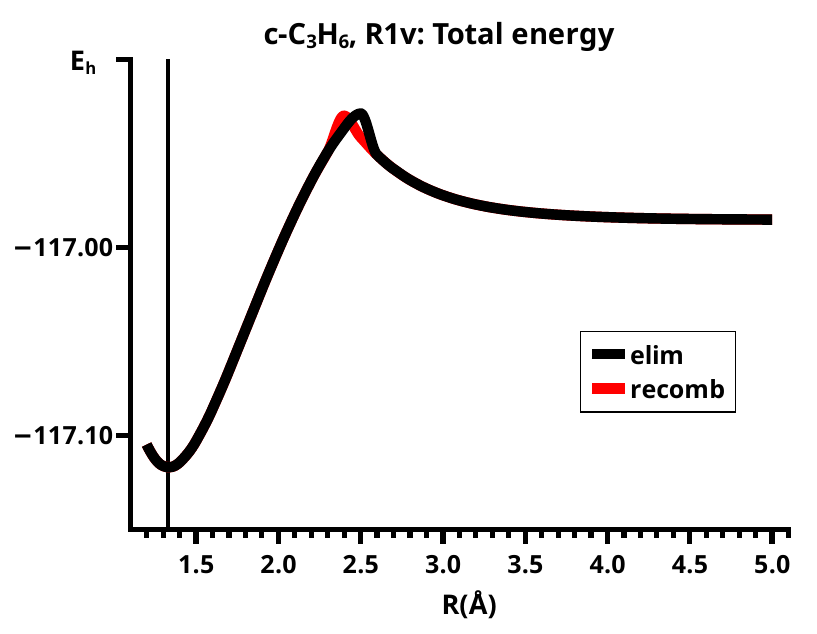}
\includegraphics[width=0.46\textwidth]{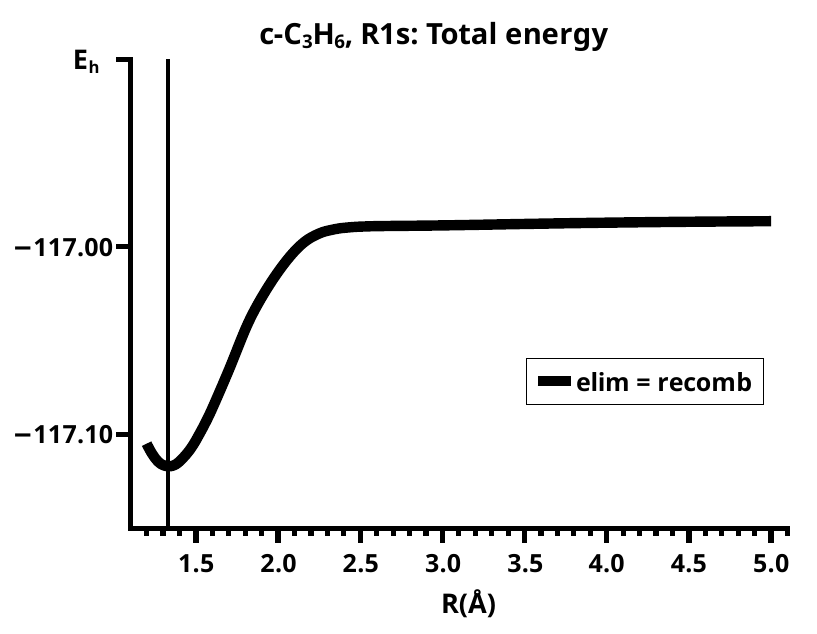}
\caption{Total energies for the reactions \ce{c-C3H6 <=> C2H4 + CH2} in $C_{2v}$ and $C_s$.}\label{fig:cPro_allE}
\end{figure}

\subsubsection{\textbf{R1v}: The elimination and recombination reactions  in $C_{2v}$}
\begin{figure}
\includegraphics[width=0.8\textheight]{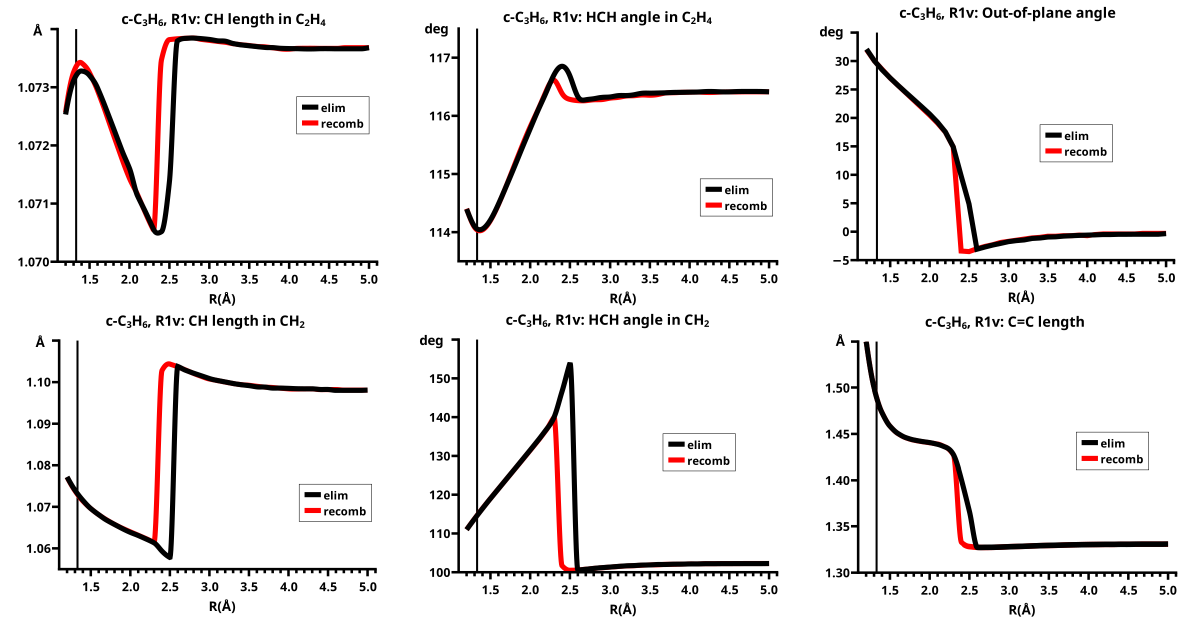}
\caption{Geometry parameters of the methylene and the ethene fragment in reactions \textbf{R1v}.}\label{fig:cProp_C2v_geo}
\end{figure}

The  PECs for both reactions, see Figure \ref{fig:cPro_allE}, have a prominent cusp at the crossing point (see comments on curve representation in \ref{sec:diabat}) of two different energy profiles (see the discussion in section \ref{sec:chemreac}), from which only the branches with energies below the crossing point are depicted. These branches belong to different energy profiles with different MEPs, this becomes obvious when all internal coordinates describing the structure of the molecular system are considered (see Figure \ref{fig:cProp_C2v_geo}). For $R \le 2.3$\,\AA{} and $R \ge 2.6$\,\AA{} the PECs of the elimination and the recombination reactions are identical, the crossing points are at $R=2.5$\,\AA{} and $2.4$\,\AA, respectively. This means that the jump between the different reaction channels occurs in the elimination reaction at larger separation of the fragments than in the recombination reaction.

\begin{figure}
\includegraphics[width=0.8\textheight]{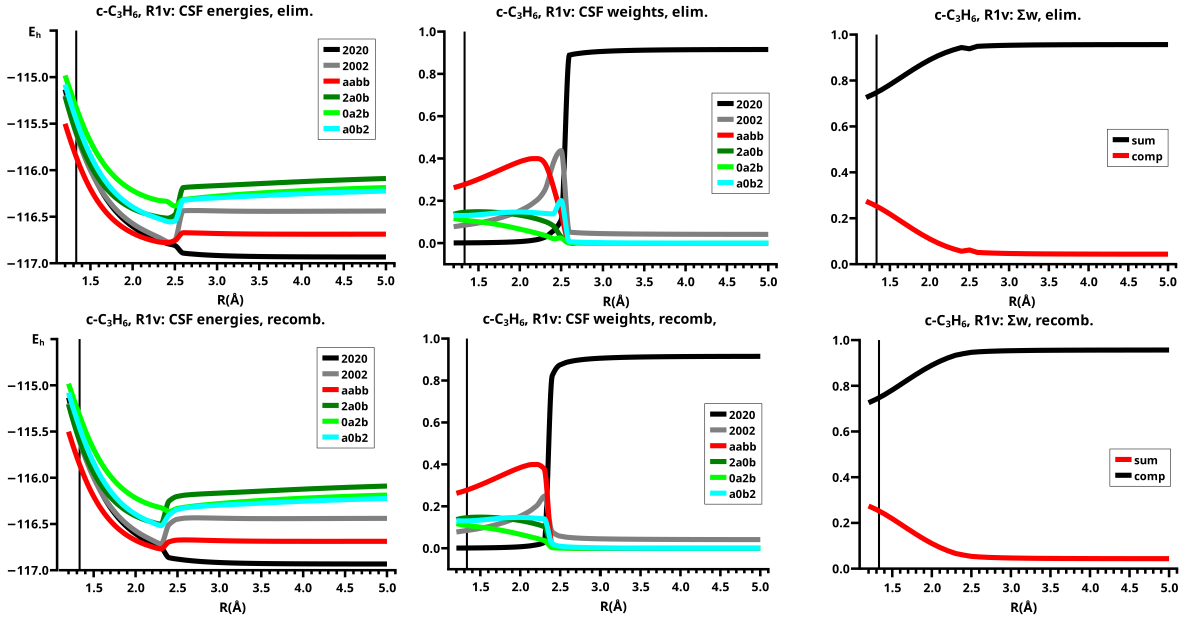}
\caption{CSF energies and weights, and sum of weights of the largest contributions to the wave functions in reactions \textbf{R1v}.}\label{fig:cProp_C2v_Ew}
\end{figure}
\newpage

As the geometry parameters of the fragments show, the dissociated system consists of ethene and methylene in their lowest singlet states; in the bonded system, the fragment geometries resemble those of the fragments in their lowest triplet states, this justifies to speak of the low-spin and the  high-spin part of the reaction. In case of ethene, we do not mean the fully relaxed triplet geometry, which has $D_{2d}$ symmetry, but the $C_{2v}$ geometry of the triplet state. Similarly, we will speak of  low-spin and high-spin characteristics, respectively, of the wave function. During the recombination reaction, the electronic structure of the dissociated system has low-spin characteristic up to $R=2.4$\,\AA, the total energy increases monotonically; at $R=2.3$\,\AA{} is the energy below that at the crossing point, and the fragment geometries are those of the high-spin states in the bonded system. The elimination reaction starts at a geometry close to the relaxed 3-ring geometry of the bonded system; the energy increases monotonically and the fragments remain in high-spin structures. At $R=2.6$\,\AA, the characteristics of the electronic structure  has  changed from high-spin to  low-spin; the total energy is lower than at the crossing point, and the fragment have adopted low-spin geometries. The different crossing points of the energy curves of the recombination and elimination reactions show that for each reaction the system stays in the entrance channel and follows the MEP to geometries where the energy is  higher than the energy of the entrance channel of the reverse reaction; and only then the system jumps to the lower lying channel. Every jump of the system to structures with lower total energy is accompanied by sudden changes of the fragment geometries that indicate characteristic changes in the electron structures of the fragments; frequently the changes can be traced to spin-rearrangements such as spin flips.

In the recombination reactions, both fragments must change from low-spin to high-spin states to create unpaired electrons in the fragments making them ready for bonding; in  elimination reactions, the spin flip must change the high-spin states of the fragments to low-spin, if the  ground state of the dissociated system is a low-spin state.
This is supported by the geometry data. In the recombination reaction, the methylene fragment has,  in the dissociated system, the characteristic geometry of the lowest $^1A_1$ state (1-$^1A_1$): the CH bond length is 1.098\,\AA{} and the HCH angle is 102.2 degrees; the geometry parameters of the ethene fragment are very close to the equilibrium values of the free fragment: the C=C bond length in 1.331\,\AA, the CH bond lengths are 1.073\,\AA{} and the HCH bond angles are 116.4 degrees.  During the approach of the fragments the geometry parameters change very little: at $R=2.4$\,\AA{} the  CH bond length in methylene is about 1.103\,\AA{} and the bond angle is 101.7 degrees; the C=C length in ethene is 1.332\,\AA{}, the CH bond lengths are 1.073\,\AA, and the HCH bond angles are 116.3 degrees, there is a slight deviation from planarity, the out-of-plane angles are -3.5 degrees with the CH bonds bent towards the methylene fragment. At $R=2.3$\,\AA, the system has already the high-spin geometry: the C=C bond length in ethene (1.427\,\AA), has increased by 0.1\,\AA, the CH bonds in ethene have shrinked to 1.071\,\AA, the HCH bond angles (116.7 degrees) are nearly unchanged, that means, the CH moieties in ethene are nearly unchanged; only the out-of-plane angles have increased by about 18 degrees, but they are still much smaller than in the 3-ring. The CH bond length in methylene (1.063\,\AA) has decreased by 0.04\,\AA, and the bond angle (139.9 degrees) has increased by roughly 40 degrees. The geometry parameters of methylene are markedly different from either the 1-$^1A_1$ state or the $^3B_1$ state; indeed, these values are typical for the second $^1A_1$ state (2-$^1A_1$) of methylene (see Supporting Information). The geometry parameters of the methylene fragment indicate, that at $R=2.4$\,\AA{} the fragments are still in low-spin states; the change of the 1-$^1A_1$ methylene state to the $^3B_1$ ground state proceeds via an intermediate excited singlet state, in which the lone pair changes from the s-type AO to the p-type AO and only at about 1.9\,\AA{} follows the spin flip to the $^3B_1$ state with singly occupied lone pair FMOs. As long as the methylene is in a singlet state also the ethene fragment must be in some excited singlet state, as the increase of the C=C bond length and the deviation from planarity show.
Further decrease of the distance between the two fragments causes a linear change of all geometry parameters, except the C=C bond length, towards the values of the equilibrium structure of the 3-ring with CH bond lengths of about 1.074\,\AA, HCH bond angles of about 114 degrees, out-of-plane angles are about  30 degrees, and the length of the former C=C bond is now about 1.48\,\AA.

In the elimination reaction, the characteristics of the bonded system changes without any sudden geometry changes, at $R=2.3$\,\AA{} the methylene fragment has already adopted the 2-$^1A_1$ geometry, but, with increasing reaction coordinate, the geometry parameters of methylene do not show expected the change to the 1-$^1A_1$ geometry. Instead, the CH bond length decreases and the HCH angle increases further, at $R=2.5$\,\AA, the CH bond length is only 1.058\,\AA, and the HCH bond angle is about 155 degrees. The change of the geometry parameters of ethene is rather slow. See Figures \ref{fig:cProp_C2v_geo}. At $R=2.6$\,\AA, the system is in the dissociated system, and all geometry parameters have the expected values.

The unexpected change of the methylene geometry during both reactions needs some further considerations. The two lowest $^1A_1$ states of bent methylene, the 1-$^1A_1$ and the 2-$A_1$ state, correlate with one $^1\Sigma_g^+$ state and one component of the lowest $^1\Delta_g$ state of linear methylene (CH bond length 1.056\,\AA). Both states are made with the two degenerate $\pi$ MOs, traditionally labelled $\pi_x$ and $\pi_y$, which are centered at the carbon atom. The $^1\Sigma_g^+$ is proportional to $\pi_x^2+\pi_y^2$; the two $\Delta$ components are proportional to $\pi_x^2-\pi_y^2$ and $^1(\pi_x\pi_y)$.

In $C_{2v}$ symmetry, that is when  methylene bends, one $\pi$ MO becomes the s-type lone pair AO of $a_1$ symmetry and the other $\pi$ MO becomes the p-type AO of $b_1$ symmetry, the former $^1\Sigma_g^+$ state and the $\pi_x^2-\pi_y^2$ component of the $\Delta$ state have $A_1$ symmetry; the second $\Delta$ component has $B_1$ symmetry. To get a totally symmetric ground state of the molecular system, the $1^B_1$ state must couple with an excited singlet state of ethylene of the same symmetry, which is described by the CSF SS but never plays a role in the composition of the total wave functions. The two $^1A_1$ states of methylene are dominated by doubly occupied lone pair AOs, the 1-$^1A_1$ state is dominated by $s^2$ and the 2-$^1A_1$ state is dominated by $p^2$. During the approach of methylene and ethene, the lone pair in the s-type AO point directly to the doubly occupied $\pi$ MO of ethene, which is resisted by the PEP. An angular change of the lone pair density from the s-type to the p-type AO will certainly minimize the so called Pauli repulsion but increase the total energy considerably. A spin-flip from the local 2-$^1A_1$ to the $^3B_1$ state yields an energetic stabilization and prepares the methylene fragment for bonding.
The OVB analysis of the CASSCF wave functions can shed more light on the bonding processes.

A final note: The geometry parameters of methylene change much more strongly than the corresponding parameters of ethene: the maximal variation of the CH length in methylene is 0.046\,\AA{}, in ethene it is only 0.003\,\AA{}; the maximal variation of the HCH angle in methylene is 39.4 degrees, in ethene it is 2.9 degrees. The variations differ roughly by a factor of 10.

In the dissociated system the neutral low-spin NB CSF has the largest weight in the wave function, the second largest weight has the DX CSF $|2002|$, which  describes the angular correlation of the lone pair electrons in the 1-$^1A_1$ state of methylene (see Figure \ref{fig:cProp_C2v_Ew}. In the bonded system the neutral CSF TT has highest weight, the contributions of the two CX1 CSFs $\rm |0a2b|$ and $\rm |2a0b|$, and of one CX2 CSF, namely $\rm |a0b2|$, are important between $R=1.2$\,\AA{} and $R= 2.3$\,\AA. These ionic CSFs describe the charge shift during covalent bonding. CSF DX  $|2002|$, which corresponds to the intermediate 2-$^1A_1$,  has second highest weight only between $R=2.1$\,\AA{} and $R= 2.3$\,\AA, where it supports the change from 1-$^1A_1$ to $^3B_1$. At the equilibrium geometry the sum of the ionic CSFs is larger than the weight of CSF TT.
The only differences between elimination and recombination reaction are found between $R=2.4$\,\AA{} and $R=2.5$\,\AA. Since the crossing point of the energy profile for the elimination reaction is at $R_0=2.5$\,\AA, the weight of CSF $|2002|$ becomes even larger than that of CSF TT.

The CSF energy curves show the effect of the contraction of the lowest FMOs: in the dissociated system, the lowest FMO in each fragment is doubly occupied, the Coulomb repulsion of non-identical electrons causes an extension of at least these two FMOs; in the bonded system the FMOs of each fragment are singly occupied by identical electrons and therefore all FMOs contract. When during the recombination reaction the extended FMOs contract, the energy of CSF NB increases because of the increase of the Coulomb repulsion in the less extended FMOs, the energy of all other CSF, not only of TT, decreases. See Figures \ref{fig:cProp_C2v_Ew}.

Since the descriptions of both reactions are nearly identical also the sums of the weights of the large CSFs and the respective complementary contributions of the small CSFs are nearly identical. The complementary contributions show that the small CSFs, which are mostly ionic CSF,  become important when the new covalent bonds are formed, the various ionic CSF describe all possible deviations of the electron distribution during the reaction. See Figure \ref{fig:cProp_C2v_Ew}.

It is important to remember that also the curves of the CSF energies and weights are composed of contributions of two reactions following different MEPs, even when the continuous curves suggest a reaction following a single MEP.

\subsubsection{\textbf{R1s}: The elimination and recombination reactions in  $C_s$}
\begin{figure}[ht]
\includegraphics[width=0.8\textheight]{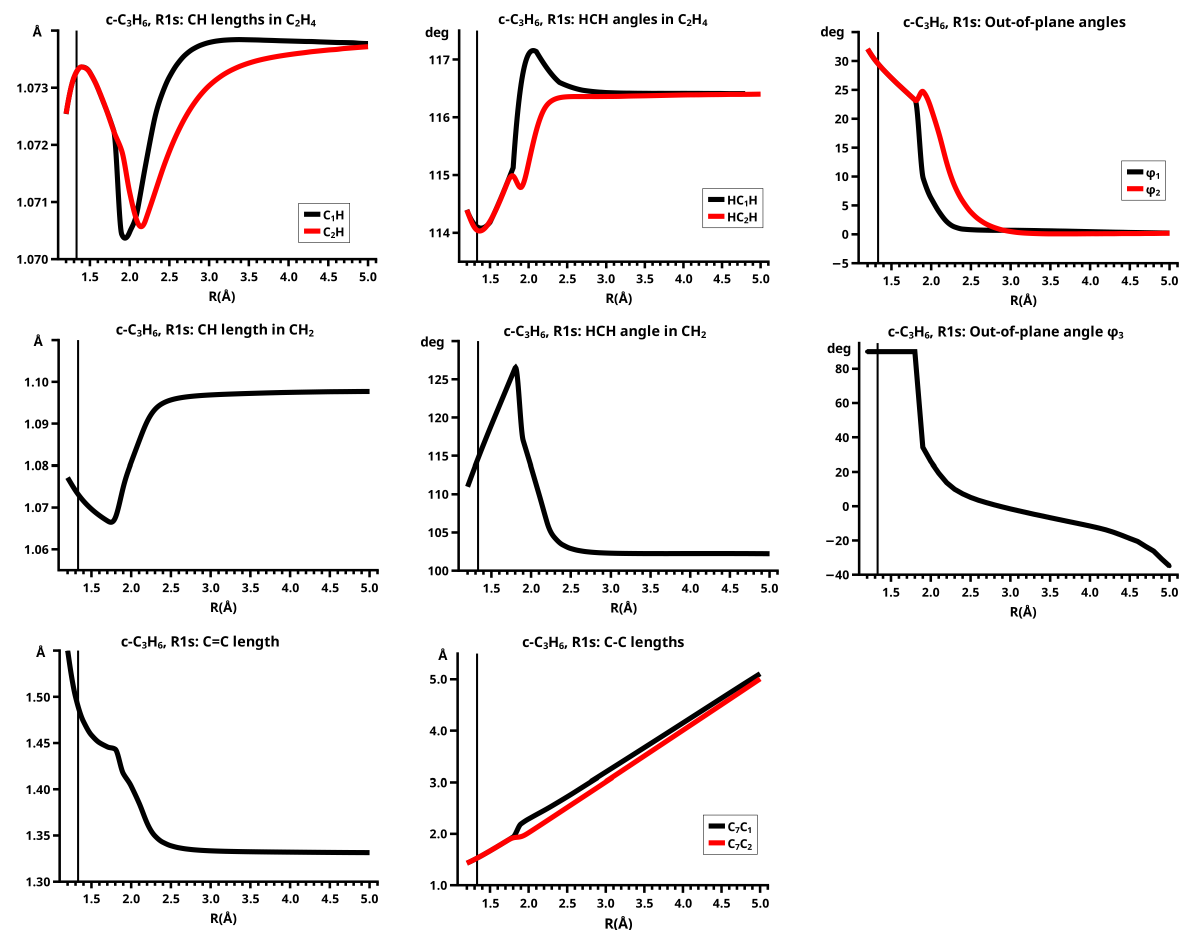}
\caption{Geometry parameters of the methylene and the ethene fragment in reactions \textbf{R1s}.  }\label{fig:cProp_Cs_geo}
\end{figure}

There is no structural and energetic difference between the elimination and recombination reactions, both reactions follow the same trough and the total energy decreases monotonously during the recombination reaction. In the dissociated system, the methylene molecule lies roughly parallel to ethene, indeed $\varphi_3$ is negative, that means the H atoms of the methylene fragment are closer to the ethene molecule than the carbon atom. This ``parallel'' structure remains during the approach in the recombination reaction up to $R=1.8$\,\AA, where the methylene fragment rotates to $\varphi_3 = 90$ degrees, and the full $C_{2v}$ symmetry is reached. Between $R=2.5$\,\AA{} and $R=1.8$\,\AA{} most geometry parameters change significantly. The CH bond length in methylene decreases from 1.098\,\AA{} to 1.066\,\AA{} and the HCH bond angle increases from 102 degrees to about 126 degrees; at $R=1.8$\,\AA{} the geometry of methylene resembles the geometry in the $^3B_1$ state. See Figures \ref{fig:cProp_Cs_geo}.  During the approach of the fragments atom Z is closer to atom Y than to atom X, accordingly, the lengths of the two C-C single bonds are different for $R > 1.8$\,\AA, they are equal only when the molecular structure has adopted $C_{2v}$ symmetry. Because of the asymmetric approach of methylene and ethene the geometry parameters of ethene at the two carbon atoms are significantly different. Note, in $C_s$ symmetry, the  2-$^1A_1$ state of methylene is not involved in both reactions, as the curves of the geometry parameters show.

\begin{figure}[ht]
\includegraphics[width=0.8\textheight]{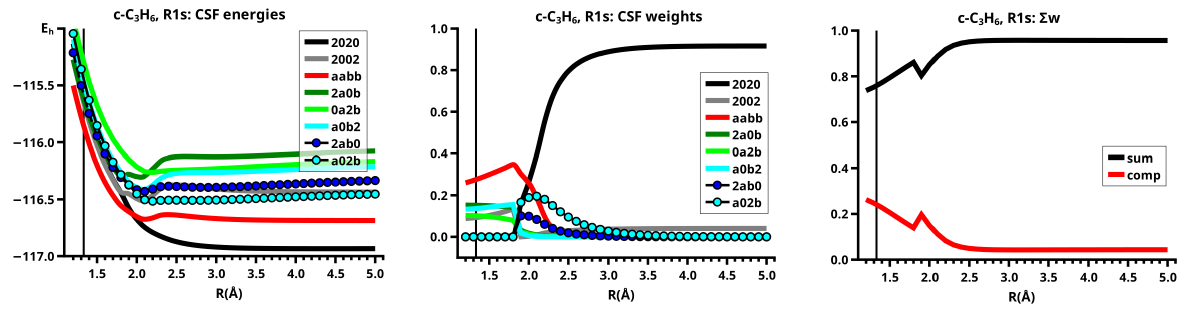}
\caption{CSF energies, weights, and sum of weights of the largest contributions to the wave functions in reactions \textbf{R1s}.}\label{fig:cProp_Cs_Ew}
\end{figure}

When the system's symmetry is reduced to $C_s$ all active MOs have $a'$ symmetry and therefore all 20 CSFs are totally symmetric, but only eight CSFs are large. See Figure \ref{fig:cProp_Cs_Ew}. In the dissociated system, only CSF NB and the DX CSF $|2002|$, describing angular correlation in methylene, are important, they have the same weights as in the reaction in $C_{2v}$.   The neutral TT does not contribute significantly before $R=2.2$\,\AA, the weight increases slowly and reaches its maximum at $R=1.8$\,\AA, when the molecular system has again $C_{2v}$ symmetry. Then the weights of the ionic CSFs become, as expected, larger than the weight of the neutral TT CSF. In contrast to the reaction in $C_{2v}$, the  C CSF $|a02b|$, with a doubly occupied s-type lone pair FMO and the two singly occupied FMOs $\pi$ and the p-type FMO, contributes very early, its weight is significant already at about $R=2.5$\,\AA. At large distances, the overlap of the two FMOs contributes to the stabilization of the dissociated system with methylene lying ``parallel'' to the ethene; at short $R$ values, the CSF describes a charge shift from the doubly occupied $\pi$ FMO to the empty p lone pair FMO, and this is only possible, when the $p$ FMO points towards the $\pi$ FMO, so that they can overlap when $R$ decreases. Indeed, the empty p lone pair FMO does not point somewhere to the $\pi$ FMO, but only to one of the p AOs spanning the $\pi$ MO. The second C CSF $|2ab0|$, which describes a charge shift from the doubly occupied s lone pair of methylene to the empty $\pi^*$ in ethene, is much less important. Both C CSFs are not totally symmetric in $C_{2v}$ and therefore cannot contribute to the wave  function as soon as the molecular system has adopted $C_{2v}$ symmetry. The charge shift during covalent bonding is thus described by different CSFs: in $C_s$ it is mainly the C CSFs, in $C_{2v}$ it is only the CX1 and CX2 CSFs.
The CSF energies in Figure \ref{fig:cProp_Cs_Ew} show that the different composition of the wave function influences also the extension of the FMOs. Since there is no sudden replacement of CSF NB by CSF TT, change of the FMO extensions is not that sudden as it is in the $C_{2v}$ reaction, and the drop in the energies is much more moderate.

Figure \ref{fig:cProp_Cs_Ew} shows that in $C_s$ symmetry, the sum of the weights of the large CSFs decreases between $R=2.5$\,\AA{} and $R=1.9$\,\AA, then after the system has adopted $C_{2v}$ symmetry, the contribution of the large CSFs is larger again, and the slowly decreases, whereas the polarizing small CSFs become more important.

\newpage
\subsection{The reactions \ce{c-Si3H6 -> Si2H4 + SiH2}}
The second  homonuclear 3-ring of this study is cyclotrisilane.

\begin{figure}
\includegraphics[width=0.4\textwidth]{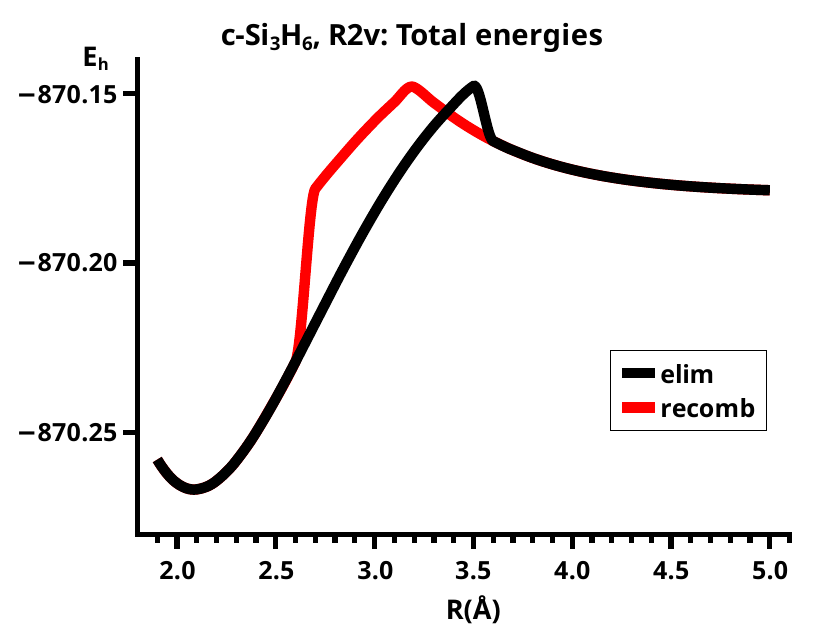}
\includegraphics[width=0.4\textwidth]{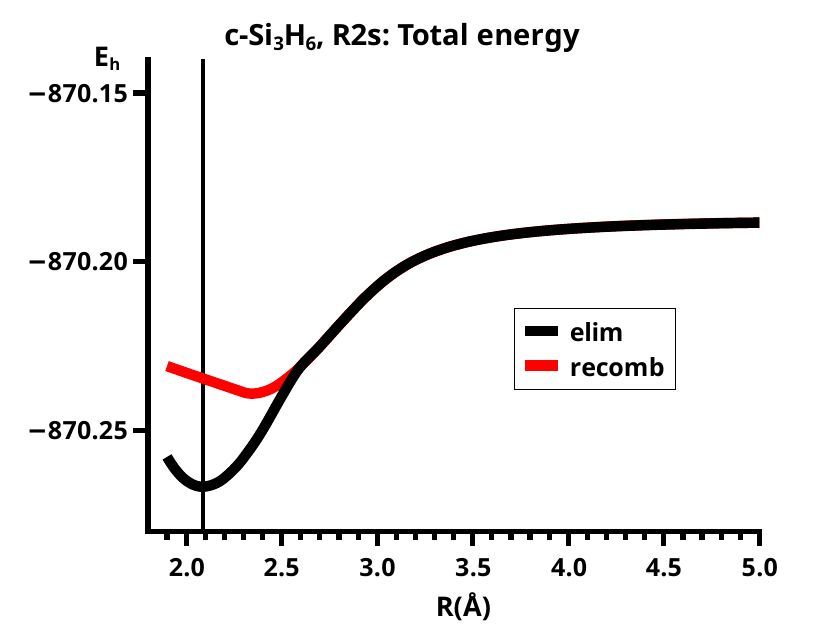}
\caption{Total energies for reactions \textbf{R2v} (left) and \textbf{R2s} (right).}\label{fig:cTri_allE}
\end{figure}

All three atoms X, Y, and Z in the sketch in Figure \ref{fig:molgeom} are  silicon atoms.
We investigated  the elimination of \ce{SiH2} from the 3-ring as well as the recombination of \ce{SiH2} and disilene in both $C_{2v}$ (\textbf{R2v}) and $C_s$ symmetry (\textbf{R2s}). For all reactions, the approximate reaction coordinate $R$ is in the range of $R_m=1.9$\,\AA{}  to $R_d=5.0$\,\AA; at the equilibrium geometry of cyclotrisilane the $R$ value is 2.088\,\AA. We checked also whether, in $C_{2v}$ symmetry,  a CAS(4,4) wave function is sufficient for the description of the reactions studied. The results obtained with a CAS(6,6) wave function can be found in the Supporting Information, they agree very well with those obtained with CAS(4,4) wave functions; the only difference worth mentioning is that the CAS(6,6) wave function yields planar disilene at $R=5.0$\,\AA, whereas with CAS(4,4) disilene is slightly cis-bent. But this has no influence on the bonding process. The trans-bent equilibrium structure of disilene has $C_{2h}$ symmetry; with CAS(4,4) the out-of-plane angles is 43 degrees, with CAS(6,6) the angle is 37 degrees. This, again, shows that for these reactions a CAS(4,4) wave function is sufficient. And it strengthened our opinion that this is the case for all other reactions.

In $C_{2v}$  symmetry, elimination and recombination reaction follow different MEPs; in $C_s$ symmetry the recombination reaction does not lead to the 3-ring but to the open chain trisilylene diradical. It turns out that this recombination reaction is different from all others investigated in this study. The differences between the reactions of cyclopropane and cyclotrisilane, in $C_s$ symmetry, are caused by the low-spin ground state of silylene, and by the flexible, non-planar structure of disilene. Disilene has a trans-bent ground state structure, which it can adopt only when the reactions occur in $C_s$ symmetry, in $C_{2v}$ symmetry the ground state structure is cis-bent. In the dissociated system, both trans-bent and cis-bent disilene have pyramidal structures at both silicon atoms; in the cis-bent structure all four hydrogen atoms point off the silylene fragment, as they do in the 3-ring; in the trans-bent structure,  the hydrogen atoms at one silicon atom point off the silylene fragment and the hydrogen atom at the other silicon atom point towards it. We speak of correct pyramidality, if the pyramidality is the same as in the 3-ring, otherwise of wrong pyramidality. In case of correct pyramidality, the out-of-plane angle $\varphi$ is positive, in case of wrong pyramidality it is negative. In $C_s$ symmetry, two different structures of the dissociated system are possible: in the first, the silicon atom of silylene points to the silicon atom in disilene with the correct pyramidality and the hydrogen atoms point to the silicon atom with the wrong pyramidality, this structure is designated with S1; in the second case, the silylene fragment lies the other way around, this structure is designated with S2.  This classification is used for all systems with non-planar ethene analogs.

\begin{figure}[ht]
\includegraphics[width=0.4\textwidth]{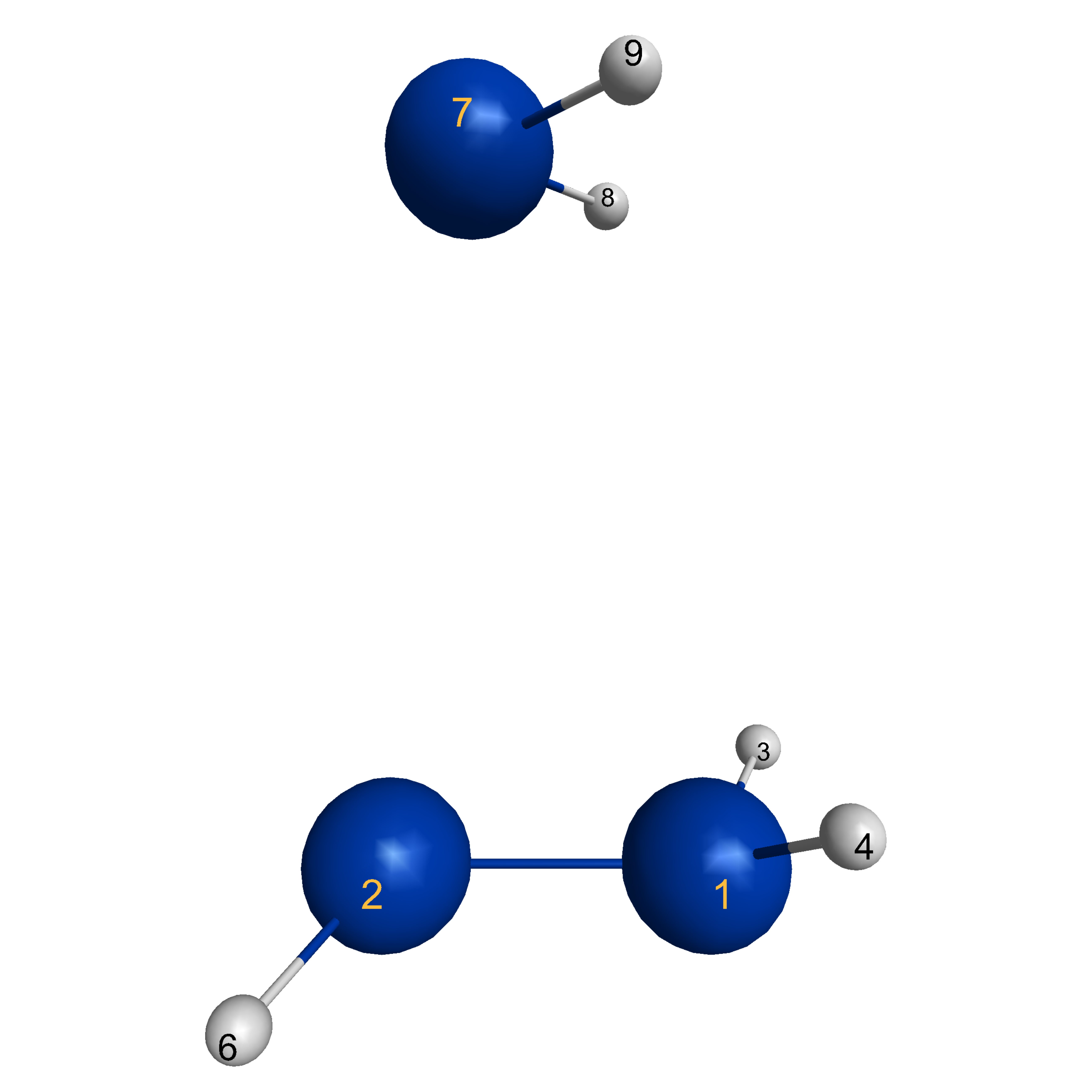}\hfill
\includegraphics[width=0.4\textwidth]{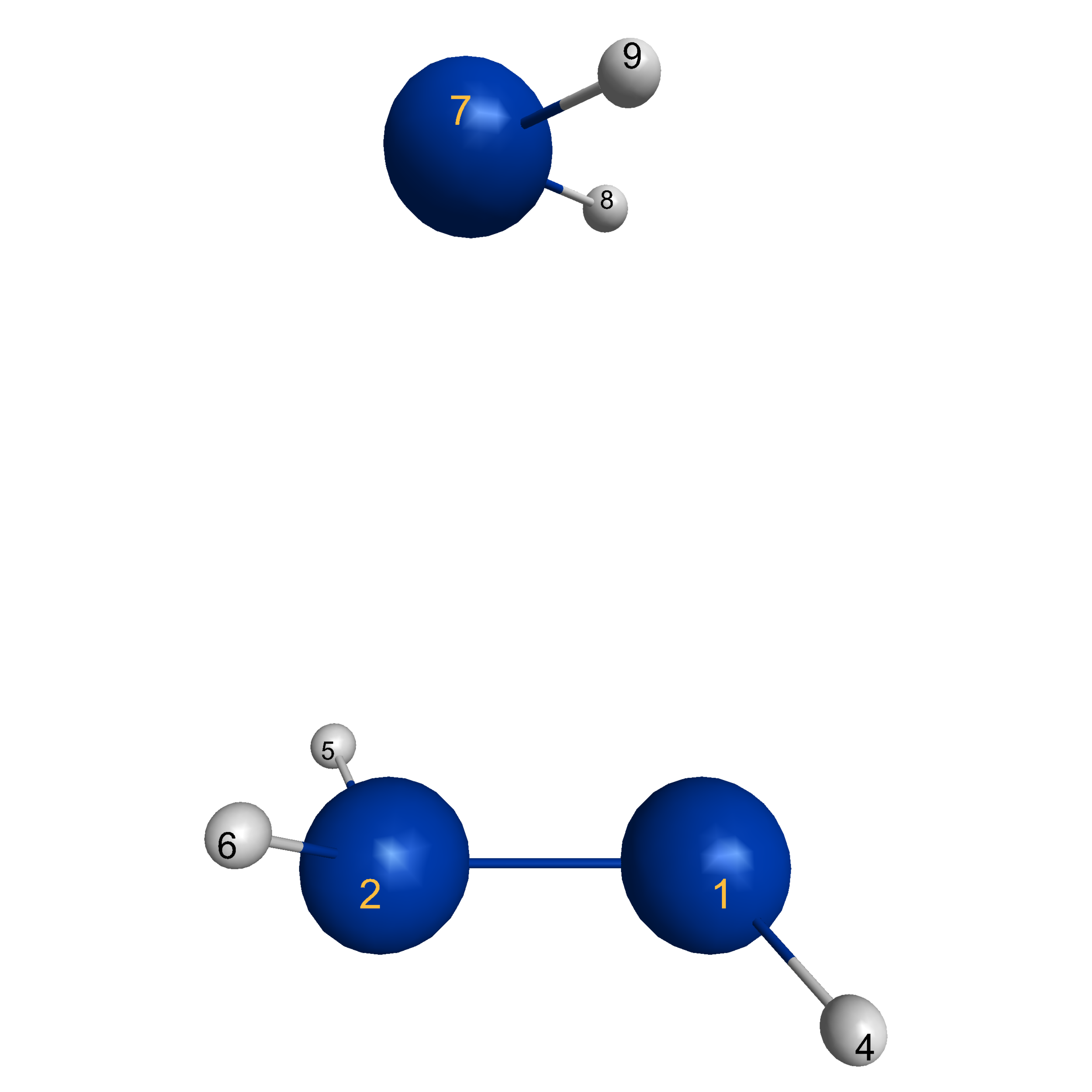}
\caption{Left: Structure S1. Right: Structure S2.}\label{fig:S1_S2}
\end{figure}

\newpage
\subsubsection{\textbf{R2v}: The elimination and recombination reactions in  $C_{2v}$}
For $R\ge 3.6$\,\AA, the description of the dissociated system is identical in both the elimination and the recombination reaction, the same is true for the description of the bonded system for $R\le 2.6$\,\AA.

\begin{figure}[ht]
\includegraphics[width=0.8\textheight]{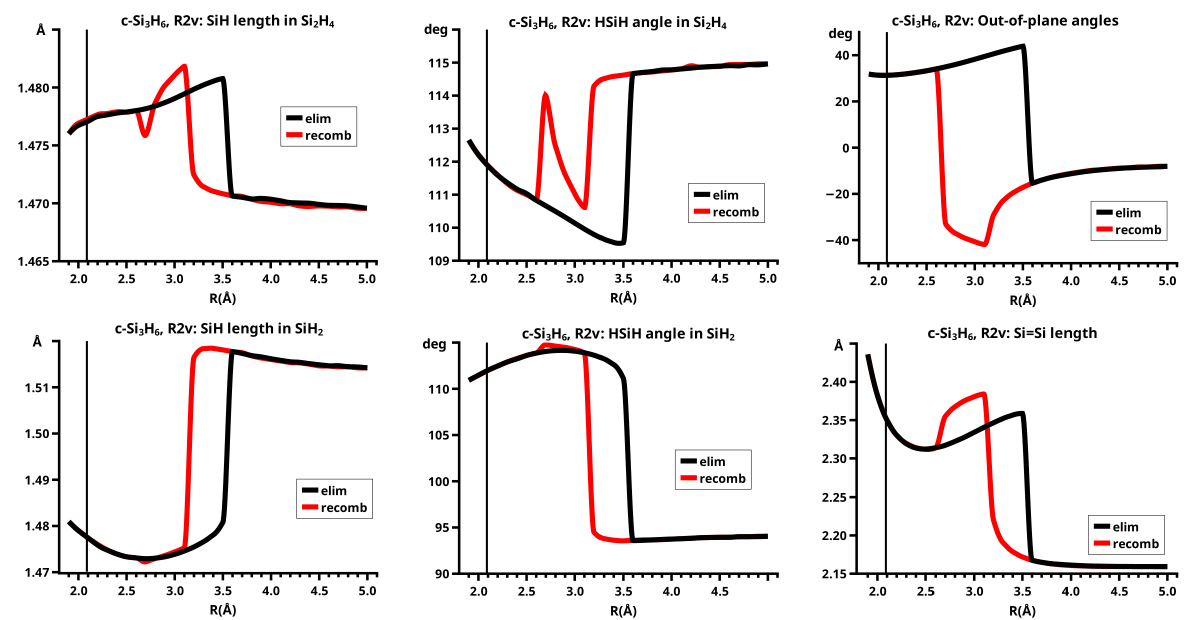}
\caption{Geometry parameters of the silylene and disilene fragments in reactions \textbf{R2v}.}\label{fig:cTri_C2v_sil}
\end{figure}

In the elimination reaction, the PEC increases monotonically from the equilibrium structure up to the crossing point $R_0=3.5$\,\AA, where the silylene fragment has still the typical triplet geometry, an SiH length of about 1.48\,\AA, and an HSiH angle of about 113 degrees. The SiSi length of $R=2.35$\,\AA{} in the disilene fragment is typical for an Si-Si single bond, in cis-bent disilene both silicon atoms have correct pyramidality with an out-of-plane angle of about 40 degrees. Between $R_0=3.5$\,\AA{} and  $R=3.6$\,\AA, the system jumps from the state with the high-spin characteristics to the state with low-spin characteristics: silylene has long SiH bonds (about 1.52\,\AA) and a small HSiH bond angle (about 95 degrees); the disilene fragment has a short Si-Si bond length of about 2.17\,\AA, the SiH bond length is about 1.471\,\AA, and the HSiH angle is about 115 degrees, the fragment is still cis-bent with an out-of-plane angle of only -20\, deg, but now the pyramidality is wrong. This is the result of an inversion of the pyramid, which is also called ``umbrella inversion''.  When the separation of the fragments increases the  geometry of silylene remains nearly constant,  the geometry of the disilene fragment changes slowly, at $R=5.0$\,\AA, disilene is only slightly cis-bent with an out-of-plane angle of about -10 degrees, the HSiH angle is 115 degrees, which is equal to the HSiH angle in planar disilene. See Figure \ref{fig:cTri_C2v_sil}.

\newpage

\begin{figure}
\includegraphics[width=0.8\textheight]{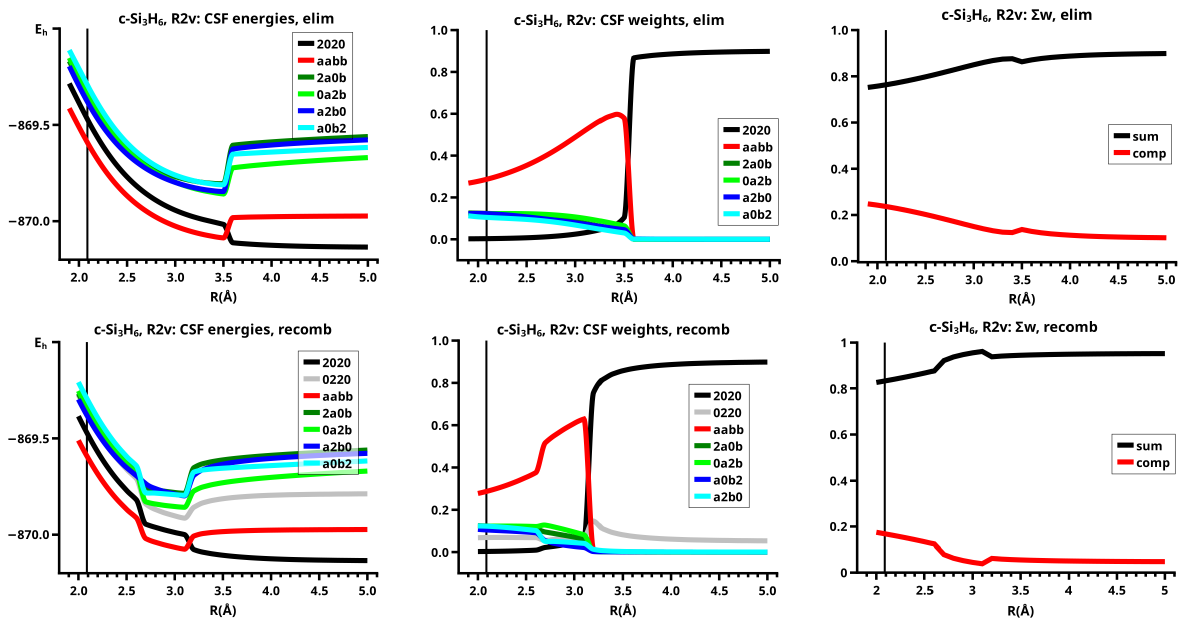}
\caption{CSF energies, weights, and sum of weights of the largest contributions to the wave functions in reactions \textbf{R2v}.}\label{fig:cTri_C2v_Ew}
\end{figure}
\newpage

The recombination reaction is remarkably different from that of cyclopropane. Starting from the structure obtained in the elimination reaction, the silylene fragment remains nearly unchanged up to the crossing point $R_0=3.2$\,\AA, the same holds for the SiH length and the HSiH angle in disilene, only the Si=Si length and the out-of-plane angle change. At the crossing point,  the Si=Si length has increased by 0.05\,\AA{}  and the pyramidality has increased, the out-of-plane angle has a value of -29 degrees, and the energy has reached its highest value. Then, again, the system jumps from the MEP describing the low-spin part of the reaction to the MEP corresponding to the high-spin part: The geometry of silylene has changed to that of triplet silylene; the Si-Si bond length in disilene has increased by 0.16\,\AA{} to  2.38\,\AA, which is longer than that of Si-Si single bonds in silanes, the pyramidality is still wrong and more pronounced than before the jump, the out-of-plane angle is -42 degrees, the SiH length and the HSiH angle in disilene are 1.48\,\AA,{} and 111 degrees, respectively. The energy at $R=3.1$\,\AA{} is lower than that at $R_0$ but it is about 23 millihartree higher than the energy of the elimination reaction for the same $R$ value. The change from low-spin to high-spin characteristics explains the moderate decrease in energy, the high energy value, compared with the elimination reaction, is caused by the wrong pyramidality. To reach the equilibrium structure of cyclotrisilane, an ``umbrella inversion'' is necessary; this happens between $R=2.7$\,\AA{}  and $R=2.6$\,\AA{} and causes the jump discontinuity in the CSF energy curves. For $R$ values smaller than $R=2.6$\,\AA, the geometry of the system and its energy are identical with those in the elimination reaction.

Pyramidal inversion in XY$_3$ molecules, like ammonia, can be described by a single MEP, the out-of-plane angle changes continuously from negative to positive values. Reduction of the pyramidality means a decrease of the sp$^3$ hybridization of atom X, and this causes an increase of the YXY angles and a decrease of the XY length; in the planar structure atom Y is sp$^2$ hybridized, the YXY bond angle has its maximum and the XY bond length has its minimum value; then the pyramidality increases again, the hybridization of atom X changes from sp$^2$ to sp$^3$, and the geometry parameters change accordingly. Such a continuous change is, however,  not observed for the inversion in the disilene fragment. Instead one finds rather sudden geometry changes of the out-of-plane angle, the HSiH angle, and the SiH length, indicating jumps between different MEPs. At both singular points of the PEC one observes discontinuous changes of the disilene fragment. At the crossing point ($R_0=3.2$\,\AA), the wave function changes its characteristics from low-spin to high-spin with corresponding changes of the system structure without a reversion of the pyramidality of disilene; at the second singular point ($R_0=2.7$\,\AA) the pyramidality is reversed without a change of the high-spin characteristics of the wave function.

\begin{figure}[ht]
\includegraphics[width=0.8\textwidth]{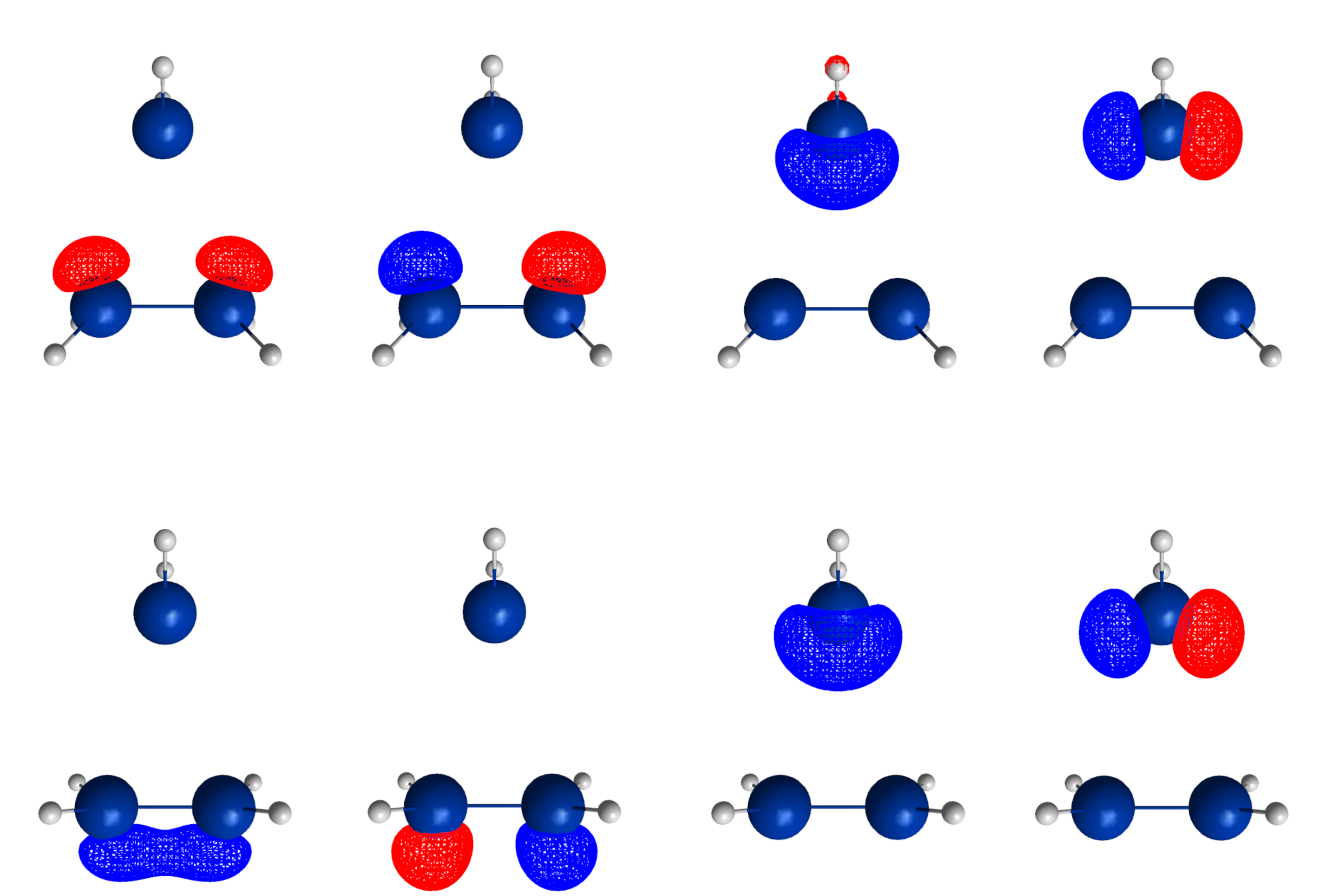}
\caption{The active FMOs $\pi$, $\pi^*$, $s$, and $p$ in the elimination reaction in $C_{2v}$.
Top: The high-spin FMOs at $R=3.4$\,\AA.  Bottom: The low-spin FMOs at $R=3.6$\,\AA.}\label{fig:elim_MO}
\end{figure}

\begin{figure}[ht]
\includegraphics[width=0.8\textwidth]{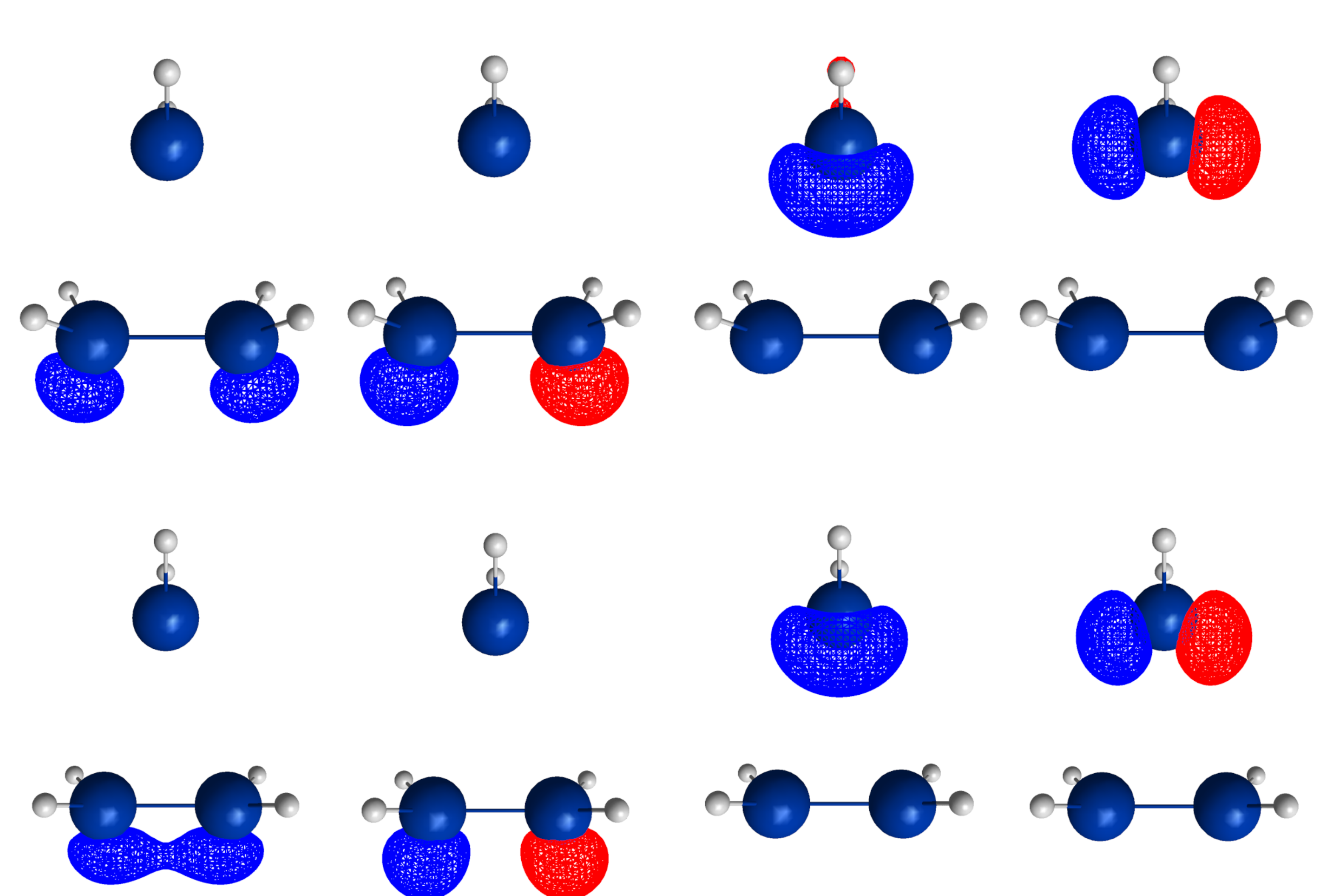}
\caption{The active FMOs $\pi$, $\pi^*$, $s$, and $p$ in the recombination reaction in $C_{2v}$.
Top: The high-spin FMOs at $R=3.1$\,\AA.  Bottom: The low-spin FMOs at $R=3.3$\,\AA.}\label{fig:recom_MO}
\end{figure}

The elimination reaction is described by six large OVB CSFs, in the bonded system the neutral CSF TT has  largest weight, whereas the weight of NB at $R=3.5$\,\AA{} is only 0.1. The four ionic CSFs CX1 and CX2, which describe the charge polarization during the covalent bonding, have nearly identical energies and weights, the sum of their weights is about twice as large as that of TT. In the dissociated system, NB has a weight of about 0.9, the CSF $|2002|$ describing the angular correlation in silylene  has too small a value to be shown in Figure \ref{fig:cTri_C2v_Ew}. The change of the wave function from  high-spin to  low-spin characteristics at the crossing point, is accompanied by a strong change of the shape of the active FMOs, see Figure \ref{fig:elim_MO}. For $R< 3.5$\,\AA, the lobes of the $\pi$ orbitals point towards the silylene fragment; at $R=3.6$\,\AA, where the disilene fragment has  wrong pyramidality, the lobes of the $\pi$ orbitals point in the opposite direction.
The recombination reaction is dominated by the same CSFs as for the elimination reaction. When the system approaches $R=3.3$\,\AA, the disilene fragment is strongly but wrongly pyramidalized, with the lobes of the $\pi$ orbitals pointing off the silylen fragment, see Figure \ref{fig:recom_MO}. When the wave function changes from low-spin to high-spin characteristics, there is a spin flip, but the geometry with the wrong pyramidality is retained, and the wave function, dominated by TT, is therefore constructed with FMOs where the lobes of the $\pi$ orbitals point off the silylene fragment. The shape of the $\pi$ FMOs and the local high-spin configuration reduces the Coulomb repulsion strongly, the CSF energies are therefore very low, only the energy of CSF NB increases. The shape of the $\pi$ FMOs prohibits also the stabilization of the system by forming new bonds, and therefore the total energy is higher than in the elimination reaction.
After the umbrella inversion, new covalent bonds can be formed, this decreases the total energy, but at the same time, the Coulomb repulsion increases and this causes higher CSF energies. This explains the upward jumps in the CSF energy curves at the second singular point.

Note the difference between the reactions of \ce{c-Si3H6} and \ce{c-C3H6} caused by the different ground state structures of the ethene analogs and the different ground state multiplicities of the carbene analogs.

\subsubsection{\textbf{R2s}: The elimination and recombination reactions in  $C_s$}
Atom X in the sketch in Figure \ref{fig:molgeom} is Si$_1$ in the graphs, Y is Si$_2$, and Z is Si$_7$.

\newpage
\begin{figure}
\includegraphics[width=0.8\textheight]{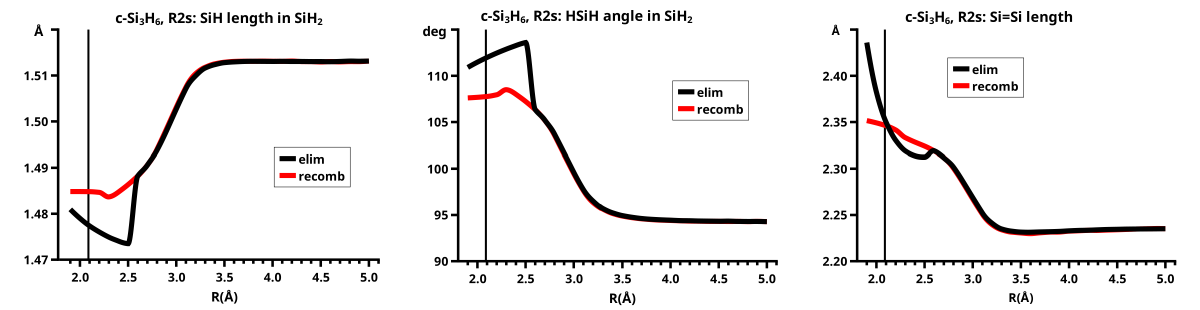}
\caption{Geometry parameters of the silylene and disilene fragments in reactions \textbf{R2s}.}\label{fig:cTri_Cs_sil}
\end{figure}

\begin{figure}
\includegraphics[width=0.8\textheight]{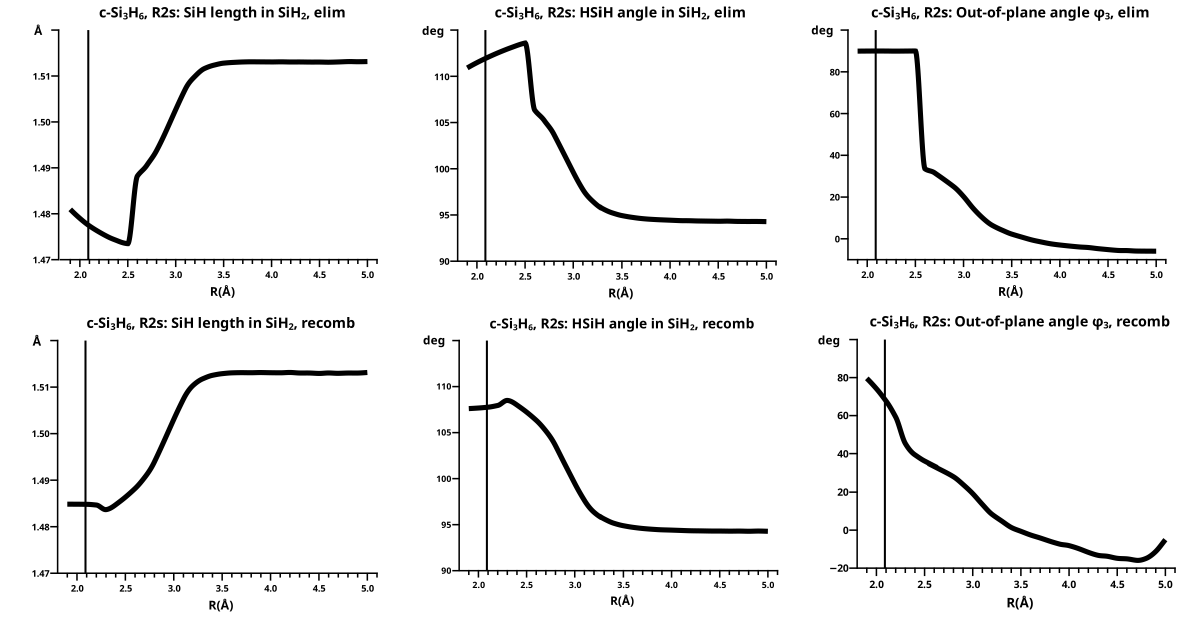}
\caption{Geometry parameters of the disilene fragment in reactions \textbf{R2s}.}\label{fig:cTri_Cs_disi1}
\end{figure}

\begin{figure}
\includegraphics[width=0.8\textheight]{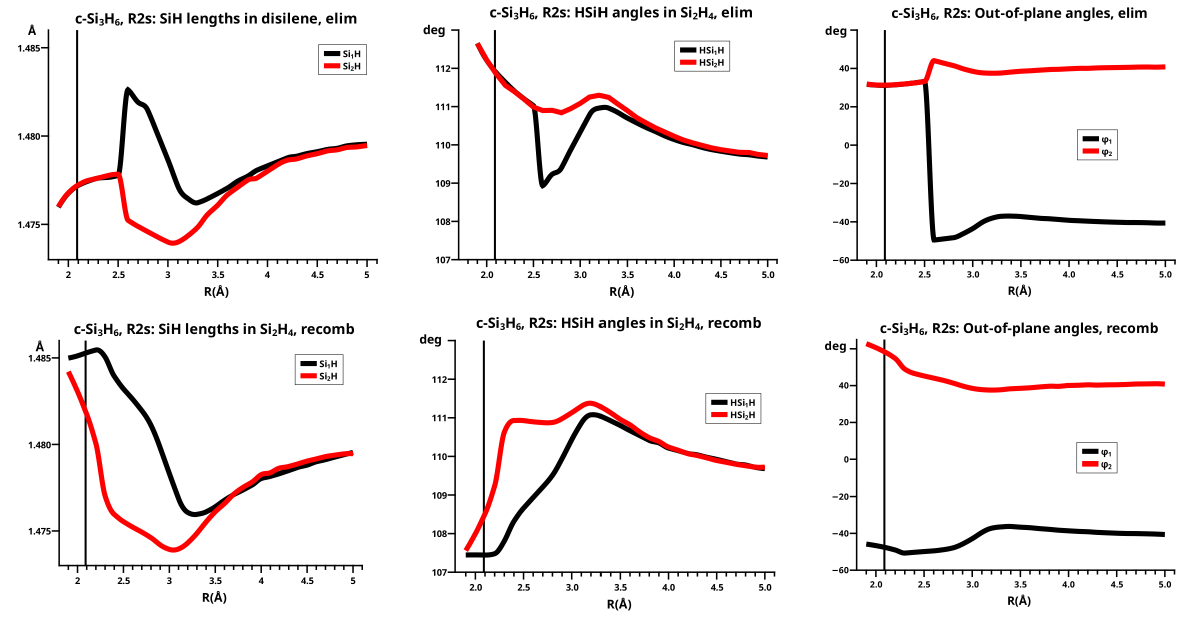}
\caption{Geometry parameters of the disilene fragment in reactions \textbf{R2s}.}\label{fig:cTri_Cs_disi2}
\end{figure}

\newpage
The fully relaxed dissociated cyclotrisilane system consists of trans-bent disilene and silylene. Because disilene has a trans-bent ground state structure, the energy of the dissociated system in $C_s$ lies below the dissociated system in $C_{2v}$ where the disilene fragment has a less favorable cis-bent structure. See Figures \ref{fig:cTri_allE}.

The elimination reaction follows in the beginning the same MEP as that for the reaction in $C_{2v}$, but between $R=2.6$\,\AA{} and $R=2.7$\,\AA{} the system's symmetry changes from $C_{2v}$ to $C_s$. At $R=2.7$\,\AA, the disilene fragment has already changed from cis-bent to trans-bent, the pyramidality of atom X has changed from correct to wrong. The silylene fragment, which stands initially normal to the disilene fragment, has folded down, angle $\varphi_3$ has changed from 90 degrees to about 30 degrees. As the reaction proceeds, the disilene fragment adopts the ground state geometry  with a short Si=Si bond length, and the silylene geometry develops into  the silylene 1-$^1A_1$ structure, lying parallel to the disilene fragment; the dissociated system has S1 structure.

\begin{figure}
\includegraphics[width=0.8\textwidth]{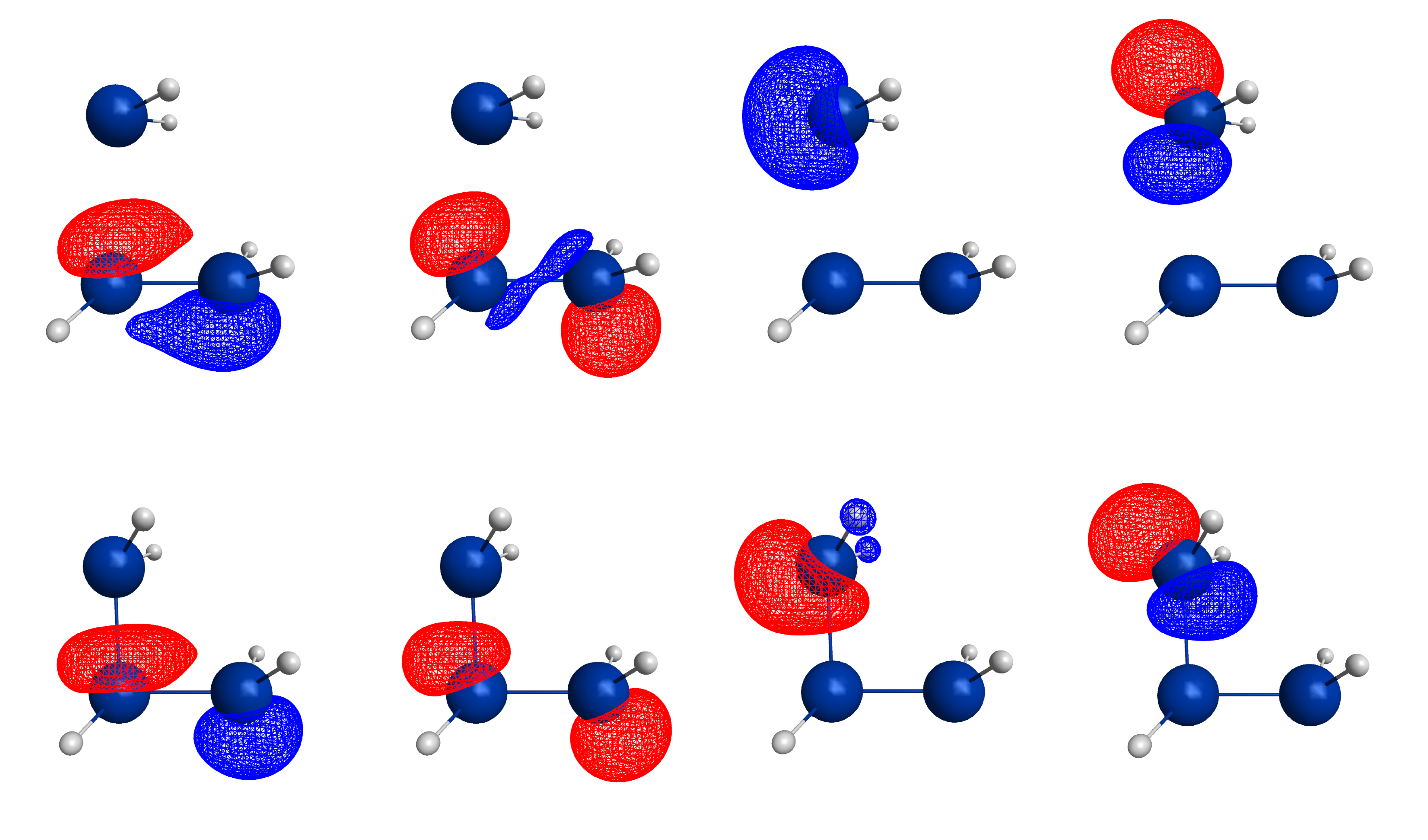}
\caption{The active FMOs $\pi$, $\pi^*$, $s$, and $p$ in the recombination reaction in $C_s$.
Top: The FMOs at $R=3.2$\,\AA.  Bottom: The FMOs at $R=2.4$\,\AA.}\label{fig:320-240-FMO}
\end{figure}

The recombination reaction is an exception to all other recombination reactions in $C_s$ as it does not lead to \ce{c-Si3H6} but to the open chain trisilylene \ce{(SiH2)3}.
Starting from the S1 structure, the PECs of the elimination and the recombination reactions are identical between $R=5.0$\,\AA{} and $R=3.2$\,\AA, as are the curves of the geometry parameters; but for $R\le 3.1$\,\AA{} all physical quantities calculated for the two reactions are different and they change in a non-uniform way, except for the distance between atoms Z and Y, which is the silicon atom in the disilene fragment closest to the silicon atom in silylene. Between these two silicon atoms, the new covalent single bond is formed; in the equilibrium structure, it will have the same bond length as the former Si=Si double bond, which becomes also an Si-Si single bond. Note that the two Si-Si single bonds, X-Y and Y-Z, are not equivalently described, because the CAS(4,4) wave function allows only for the new bond between atoms Y and Z a correlated description but not for the bond between X and Y, which is the former $\sigma$ of the double bond in disilene. Accordingly, also the two terminal SiH$_2$ moieties in trisilylene are not equivalently described, only a CAS(6,6) wave function for the three Si-Si bonds in the 3-ring could do this. In the relaxed trisilylene, there is no covalent bond between the two terminal SiH$_2$ moieties with atoms Z and X, the distance between these two atoms is about 80 percent larger than the lengths of the two Si-Si single bonds. The following description of the bonding process leading to the open chain molecule is only rudimentary. In trisilylene, the terminal SiH$_2$ moieties have the pyramidal structure of silyl groups, both carry a single electron, but the pyramidality of the terminal SiH$_2$ moieties is such that the lobes of the singly occupied FMOs do not point to each other. Therefore, the creation of an Si-Si bond is impossible. The formation of the new Si-Si single bond needs an optimum overlap of the FMOs between Z and Y. For this, the silylene must shift towards atom Y; as long as silylene lies parallel to disilene, the empty p FMO can only overlap with the doubly occupied $\pi$ FMO; the spatial extension of the doubly occupied s FMO and the empty $\pi^*$ FMO does not allow the necessary overlap. With decreasing $R$ the silylene fragment folds up and then the lone pair electrons of silylene can contribute to bonding. See Figure \ref{fig:320-240-FMO}. The recombination follows a single MEP, there are no discontinuous changes in the geometry parameters.

When the recombination reaction starts from structure S2, there is an umbrella inversion at $R=3.0$\,\AA, but after that the reaction proceeds as in the reaction starting from S1.

Figures \ref{fig:cTri_Cs_sil}, \ref{fig:cTri_Cs_disi1}, and \ref{fig:cTri_Cs_disi2} show the different internal coordinates in the elimination reaction and in the recombination reaction.

\newpage
\begin{figure}
\includegraphics[width=0.8\textheight]{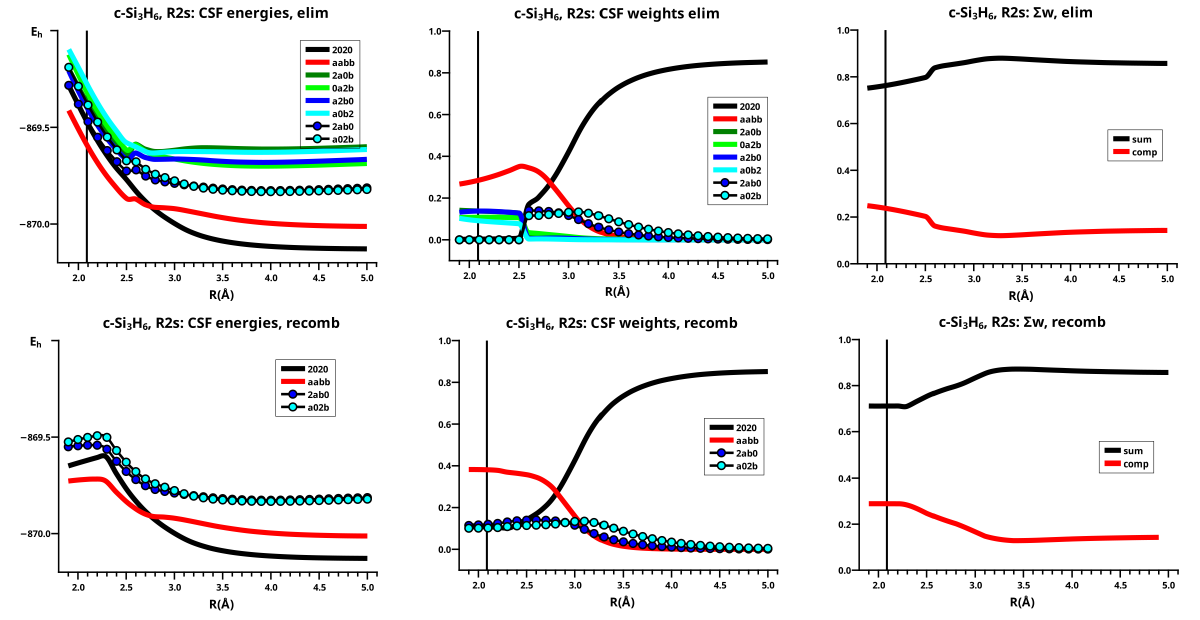}
\caption{CSF energies, weights, and sum of weights of the large contributions to the wave function in reactions \textbf{R2s}.}\label{fig:cTri_Cs_trisi}
\end{figure}
\newpage

In the initial phase of the elimination reaction, when the system has $C_{2v}$ symmetry, the neutral CSF TT has largest weight, and CSF NB has zero weight; polarization is described by the two  CSFs CX1 $|0a2b|$ and $|2a0b|$, and the two CX2 CSFs $|a0b2|$ and $|a2b0|$. When the symmetry is reduced to $C_s$, the two C CSFs $|a02b|$ and $|2ab0|$ replace the CX1 and CX2 CSFs, and NB gains weight. In the dissociated system, the NB CSF dominates, its weight is 0.85, the sum of the weight of all other 19 CSFs is only 0.15.

Between $R=5.0$\,\AA{} and $R=3.0$\,\AA, the description of the dissociated system is identical in both the recombination and the elimination reactions: going from $R=5.0$\,\AA{} to $R=3.0$\,\AA, the weight of CSF NB decreases and the C CSF $|a02b|$ becomes slowly important; the weight of the second C CSF $|2ab0|$ increases more slowly, both CSFs become equally important around $R=3.0$\,\AA. In the wave function of the bonded system,  CSF TT gives the largest contribution, at the equilibrium structure, the weight of TT is 0.38, and CSF NB and the two C CSFs have there equal weight of about 0.1.

A comment on the singly ionic CSFs $|a02b|$ and $|2ab0|$, which are very important in all reactions in $C_s$ symmetry, not just the currently considered recombination reaction.
CSF $|a02b|$  describes a cation/anion fragment pair, which is the result of the transfer of an electron from the doubly occupied $\pi$ FMO of disilene to the empty p FMO of silylene. At large distances, this CSF contributes to the stabilization of the parallel structure of the fragments; with decreasing distance between the fragments, it describes a polarization of the $\pi$ electrons of disilene, which prepares the fragments for bonding. The second  C CSF $|2ab0|$ describes the anion/cation fragment pair that is the result of  the transfer of an electron from the doubly occupied s FMO to the empty $\pi^*$ FMO; it is not obvious that this CSF helps to stabilize the dissociated system at large distances, but at medium distances both CSFs polarize mutually the charges in both fragments. The distance dependent mix of the ionic CSFs and the two neutral CSFs NB and TT describe  distance dependent charge and spin distributions, which cause the oscillation of the geometry parameters, especially for the disilene moiety at silicon atom 1, which has the weakest interaction with the silylene fragment in the early stage of the recombination reaction.

\newpage
\subsection{\ce{c-CSi2H6 -> Si2H4 + CH2}}
In \ce{c-CSi2H6} the atoms X and Y from the sketch in Figure \ref{fig:molgeom} are equal to Si$_1$ and Si$_2$, respectively, atom Z is the carbon atom C.

The creation of disilacyclopropane \ce{c-CSi2H6} by addition of methylene to disilene and the elimination of methylene from the 3-ring is studied in  high $C_{2v}$ (\textbf{R3v}) and low $C_s$ symmetry (\textbf{R3s}).  The approximate reaction coordinate $R$ is in the range of $R_m=1.4$\,\AA{} to $R_d=5.0$\,\AA.  At the equilibrium of \ce{c-CSi2H6} the approximate reaction coordinate is $R=1.568$\,\AA.

\begin{figure}[ht]
\includegraphics[width=0.4\textwidth]{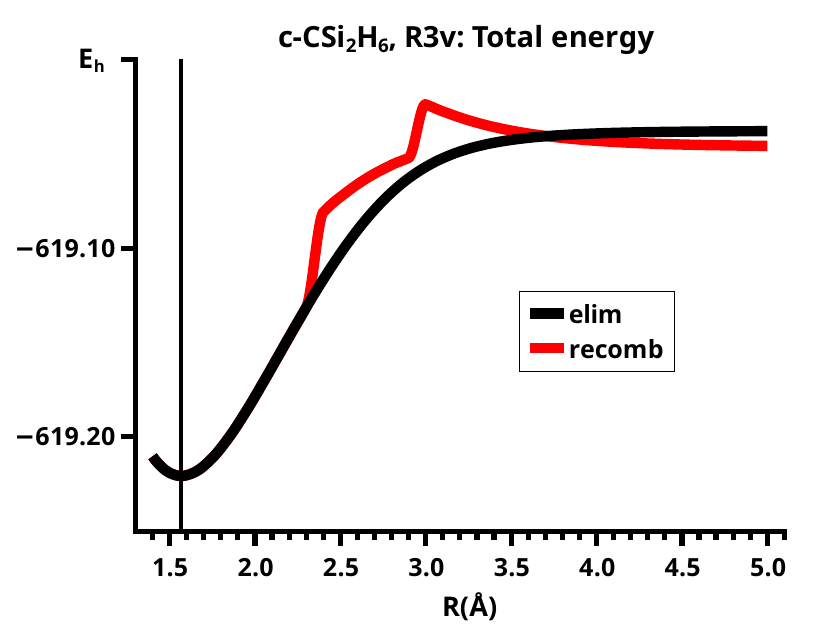}
\includegraphics[width=0.4\textwidth]{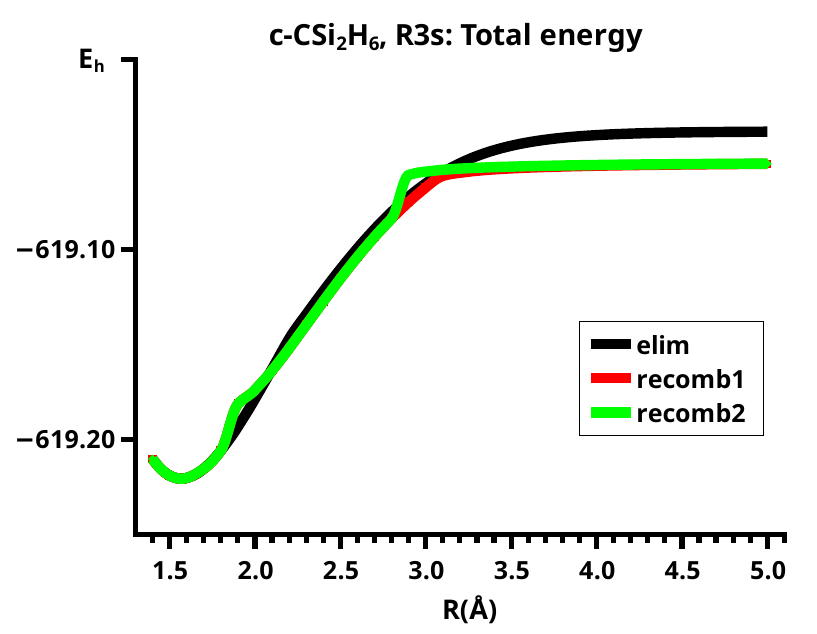}
\caption{Total energies for the reactions \textbf{R3v} and \textbf{R3s}.}\label{fig:CSi2_E}
\end{figure}

The PECs (Figure  \ref{fig:CSi2_E}) show, first, that in $C_{2v}$ symmetry, elimination and recombination reactions of the \ce{c-CSi2H6} system  follow different MEPs; second, in contrast to the elimination reactions for cyclopropane and cyclotrisilane, the \ce{c-CSi2H6} system does not jump during the elimination reaction to the lower reaction channel, therefore the dissociated system consists of methylene and disilene both in their lowest triplet states. The reaction has the properties of a diabatic reaction. The PEC for the recombination reaction resembles that for the cyclotrisilane system. The reactions in $C_s$ symmetry are different from those of cyclotrisilane. First, the elimination reaction also leads to triplet fragments, as the shapes of the PECs and of the geometry parameters suggest; second, the recombination reactions starting from both structures S1 and S2 lead to disilacyclopropane, so three different reactions in $C_s$ were investigated.

\subsubsection{\textbf{R3v}: The elimination and recombination reactions of \ce{c-CSi2H6} in  $C_{2v}$}
\begin{figure}[ht]
\includegraphics[width=0.8\textheight]{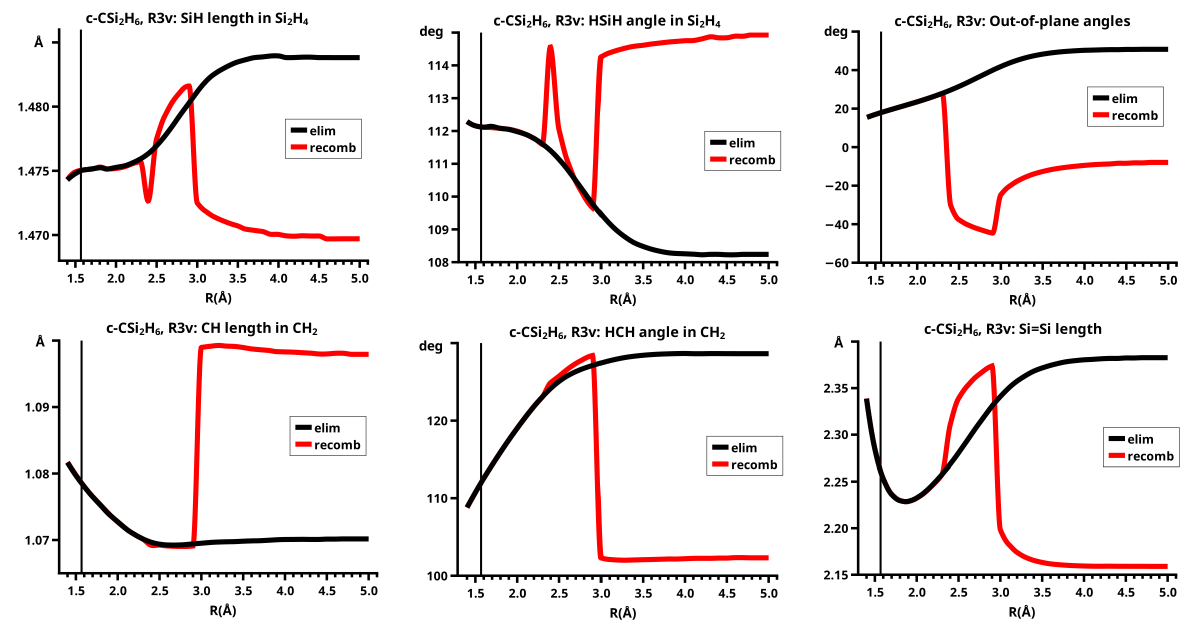}
\caption{Geometry parameters of  \ce{c-CSi2H6}  in reactions \textbf{R3v}.}\label{fig:CSi2C2v_geo}
\end{figure}

The methylene elimination follows a single MEP,  as the PEC (Figure \ref{fig:CSi2_E}) and all curves of the geometry parameters show, see figure \ref{fig:CSi2C2v_geo}, there is no jump to the lower lying entrance valley of the recombination reaction. In the dissociated system, both fragments are in their respective triplet states.
During the recombination reaction between $R=5.0$\,\AA{} and  $R=3.0$\,\AA,  both fragments are in low-spin states, and both silicon atoms of disilene have wrong pyramidality, disilene is cis-bent; the methylene fragment is in its 1-$^1A_1$ state. The total energy increases and reaches its highest value at $R=3.0$\,\AA{}, then the fragments change to high-spin states; the geometry parameters of methylene are the best indicators for the change of multiplicity. In disilene, the Si-Si distance increases from 2.20\,\AA{} to 2.37\,\AA, the Si-H length increases and the pyramidality at the silicon atoms increases as well, but it is still wrong. With decreasing $R$, the energy decreases also, but is still higher than in the elimination reaction. In this phase of the reaction, only the disilene parameters change but not the methylene parameters. At $R=2.3$\,\AA, umbrella inversion allows the formation of the new bonds and the total energy is lowered to that of the elimination reaction.
The change from low-spin to high-spin at the crossing point at $R=3.0$\,\AA{} means for the disilene fragment a reduction of the Coulomb repulsion, and,  thus, a lowering of the total energy for $R< 3.0$\,\AA. This is similar to the reaction in the \ce{c-Si3H6} system. In methylene the change means the replacement of an excited low-spin state by the  high-spin ground state, and causes a lowering of the total energy. In this aspect, the reaction in the \ce{c-CSi2H6} system is different from the reaction in the \ce{c-Si3H6} system.  However, because of the wrong pyramidality, the $\pi$ lobes point off the methylene fragment, so that no new bonds can be formed. This is only possible after the umbrella inversion.

The weight of CSF TT supports the description of the elimination reaction: in the dissociated system TT has weight 1.0, no other CSF contributes to the wave function; in the bonded system the CX1  CSF $|0a2b|$ and the CX2 CSF $|a0b2|$ describe the charge shift in the covalent C-Si bonds. Around the equilibrium geometry, the doubly ionic CSF $|0022|$ enforces the charge shift towards the carbon atom, this is in agreement with chemical explanations based on the different electronegativities of silicon and carbon. Altogether, the wave function describing the elimination reaction is a linear combination of only four large CSFs, but, as the sum of the weights shows, the CSFs with weights smaller than 0.1 contribute together about 30 percent at the equilibrium geometry, most of them are ionic CSF describing the polarization of the electron densities in the fragments. That means, during the elimination reaction the system's wave function does not change its characteristics, the elimination reaction has the character of a diabatic reaction.

\begin{figure}[ht]
\includegraphics[width=0.8\textheight]{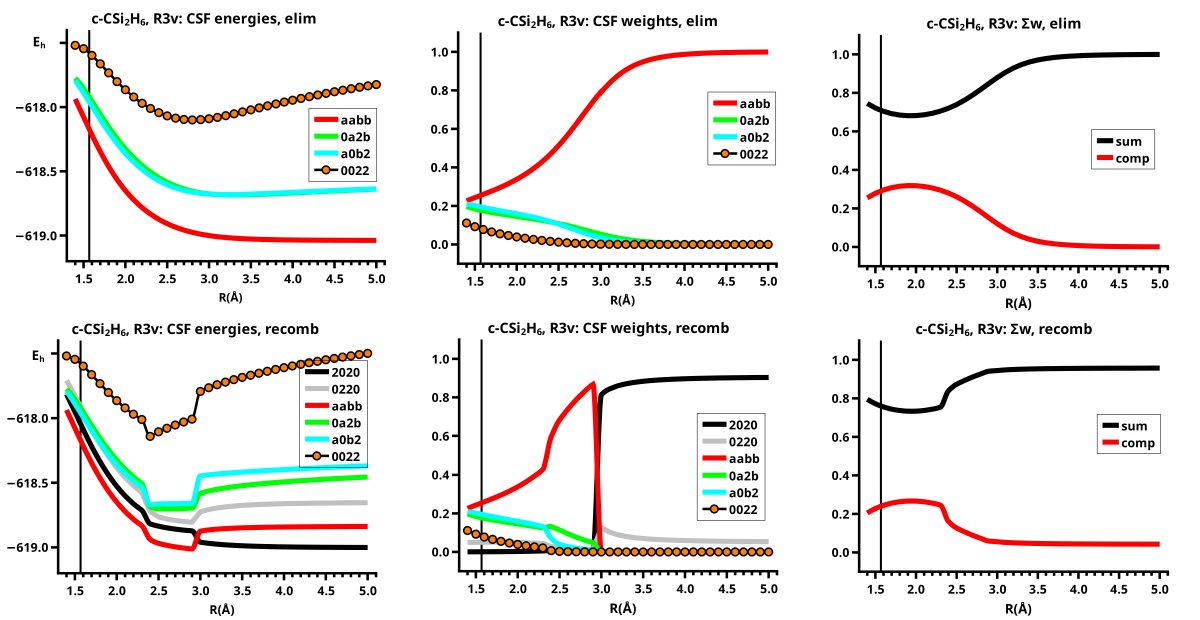}
\caption{CSF energies, weights, and sum of weights of the large contributions to the wave functions  in reactions \textbf{R3v}.}\label{fig:cTri_C2v_sum}
\end{figure}

The dissociated system in the recombination reaction is dominated by the CSF NB and the CSF $|0220|$, which describes the left-right correlation of the $\pi$ bond, the $|2002|$ CSF for the angular correlation of the lone pair electrons in methylene has too small weight to be shown. Between $R=5.0$\,\AA{} and $R=3.0$\,\AA, the sum of their weights is larger than 0.94. When the wave function acquires high-spin characteristics at about $R=3.0$\,\AA, TT  is the dominating CSF; the necessary charge shifts during covalent bonding are described by CSFs CX1  $|0a2b|$, CX2 $|a0b2|$, and  DC $|0022|$.
Before the umbrella inversion, only $|0a2b|$ contributes to the wave function, after the umbrella inversion, also $|a0b2|$ contributes and then  both  CSFs are equally important, as the weights show. See Figure \ref{fig:cTri_C2v_sum}.

\subsubsection{\textbf{R3s}: The elimination and recombination reactions of \ce{c-CSi2H6} in $C_s$}
Also the elimination reaction in $C_s$ symmetry follows a single MEP, in the dissociated system both fragments are in triplet states. Up to $R=2.10$\,\AA{}, the MEP is the same as for the reaction in $C_{2v}$, but then the methylene fragment folds down to $\phi_3\approx 50$ degrees, and the methylene fragment shifts along the Si=Si direction so that the Si-C distances have different lengths, Z-X is longer than Z-Y. The Si=Si distance increases continuously, the out-of-plane angles increase slightly from 20 to 50 degrees, and the disilene fragment remains cis-bent; the pyramidality of the SiH$_2$ moiety at atom 1 becomes more pronounced than that of the second  SiH$_2$ moiety at atom 2. Note, that the energy of the dissociated system is equal for the reactions in $C_{2v}$ and in $C_s$ symmetry, see Figure \ref{fig:CSi2_E}.

\begin{figure}[ht]
\includegraphics[width=0.8\textheight]{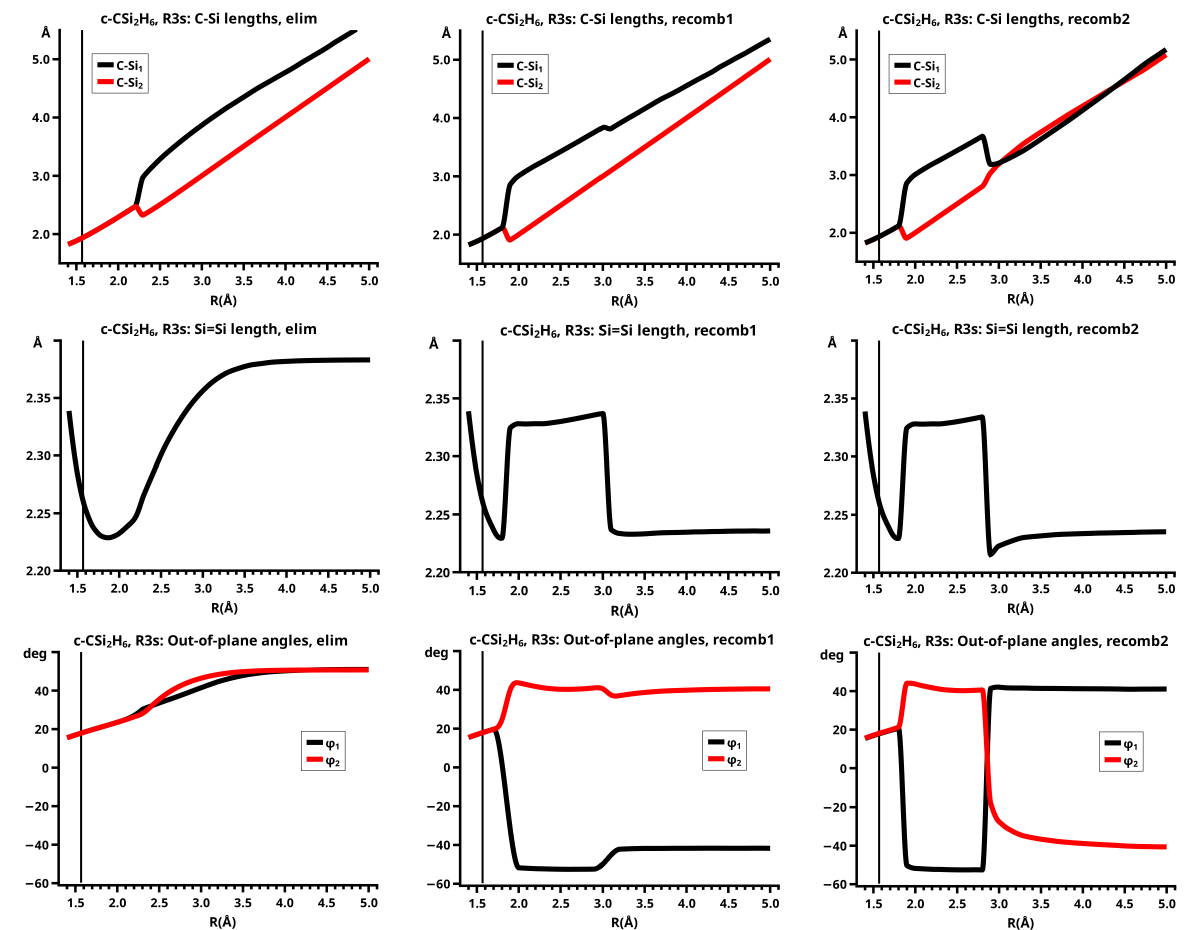}
\caption{Geometry parameters of the disilene fragment  in reactions \textbf{R3s}.}\label{fig:CSi2Cs_disilene}
\end{figure}

\begin{figure}[ht]
\includegraphics[width=0.8\textheight]{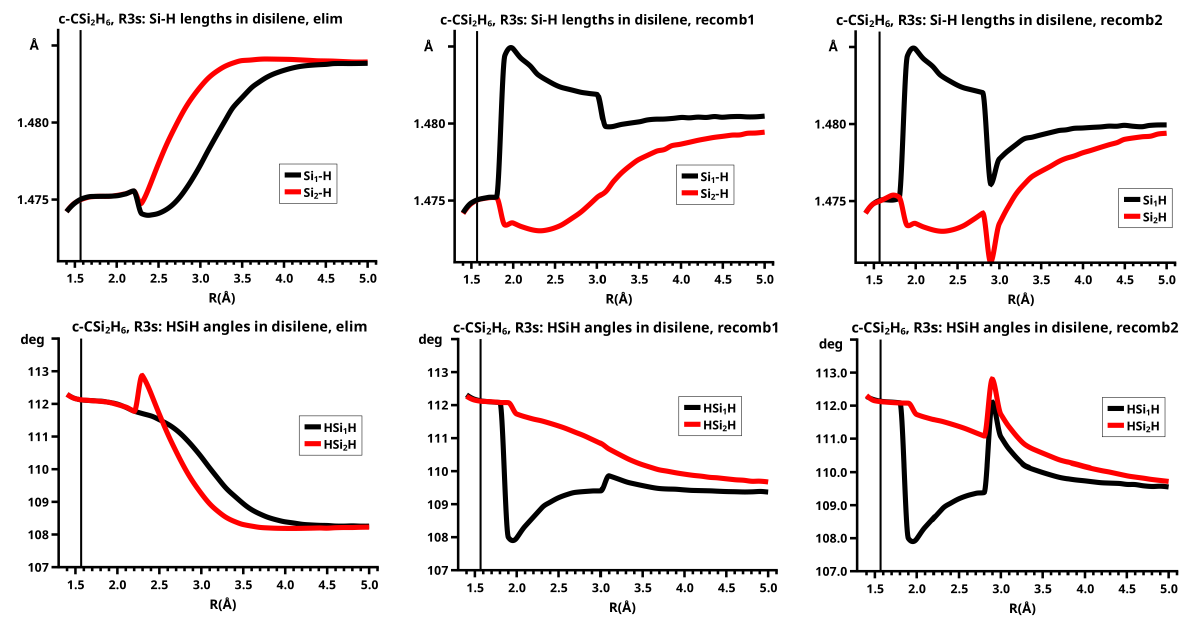}
\caption{Geometry parameters of the disilene fragment  in reactions \textbf{R3s}.}\label{fig:CSi2Cs_disilene2}
\end{figure}

\begin{figure}[ht]
\includegraphics[width=0.8\textheight]{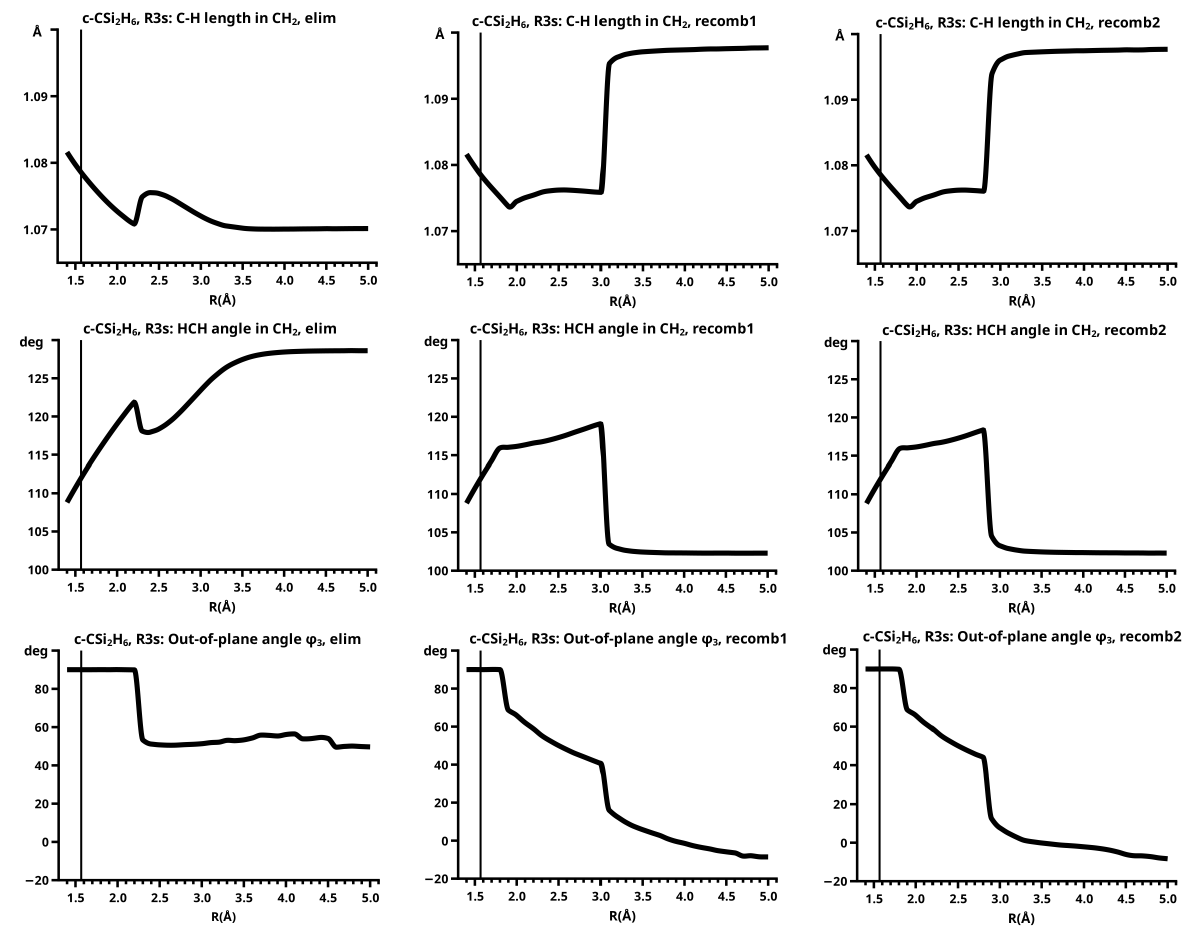}
\caption{Geometry parameters of the methylene fragment  in reactions \textbf{R3s}.}\label{fig:CSi2Cs_methylene}
\end{figure}

The recombination reactions, labelled recomb1 and recomb2, respectively, can start from structure S1 or S2, and both recombination reactions yield the 3-ring. In reaction recomb1, the system's energy is nearly constant (Figure \ref{fig:CSi2_E}) between $R=5.0$\,\AA{} and $R=3.1$\,\AA, the geometry parameters of both fragments are either nearly constant or change only moderately. See Figures \ref{fig:CSi2Cs_disilene}, \ref{fig:CSi2Cs_disilene2}, \ref{fig:CSi2Cs_methylene}. In this phase of the reaction, methylene is in the 1-$^1A_1$ state and disilene is in its ground state with trans-bent geometry. At about $R=3.0$\,\AA{} the wave function has changed from low-spin to high-spin characteristics,  methylene has changed from singlet to triplet geometry, the short Si=Si double bond is now a long Si-Si single bond, and the geometries of the two SiH$_2$ moieties in disilene have changed very differently. In the \ce{SiH2} moiety at silicon atom 2, to which the first new Si-C bond is formed, both the SiH bond lengths and the HSiH bond angle change rather smoothly from the values in disilen to the values in the 3-ring, the SiH lengths decrease, and the HSiH angle increases. Neither geometry parameter changes suddenly at $R=3.0$\,\AA. This is different from the other \ce{SiH2}  moiety, where both parameters show jumps at $R=3.0$\,\AA. In the further course, the SiH bond lengths increase and the HSiH angles decrease in both moieties, their geometries develop towards the pyramidal structure of a silyl group. At $R=2.10$\,\AA{}, the symmetry changes from $C_s$ to $C_{2v}$ and the two SiH$_2$ moieties become equivalent, this happens with a jump in the total energy.

In structure S2, the carbon atom points to the silicon atom with the wrong pyramidality, the distance of the carbon atom to both silicon atoms is equal. In reaction recomb2, the change from low-spin to high-spin occurs at  $R=2.9$\,\AA, at this point the pyramidality of both silicon atoms changes, as the out-of-plane angles show, and the methylene fragment shifts relative to the disilene fragment, so that the two C-Si distances are now different, as it was when the reaction starts at structure S1. From now on, reaction recomb2 proceeds as reaction recomb1.

The wave function that describes the elimination reaction is composed of five  CSFs, the dominant neutral TT CSF and four ionic CSFs. In the $C_{2v}$ part of the reaction, that is  up to $R=2.10$\,\AA, the same four CSFs TT, CX1, CX2, DC dominate the wave function, as in the $C_{2v}$ reaction; when the system changes to $C_s$ symmetry, the C CSF $|a02b|$ is the most important ionic CSF,  CX2 and DC disappear, and CX1 has weights smaller than 0.1. For $R>4.0$\,\AA, the wave function is described solely by CSF TT, at $R=5.0$\,\AA{} its weight is 1.00.

\begin{figure}[ht]
\includegraphics[width=0.8\textheight]{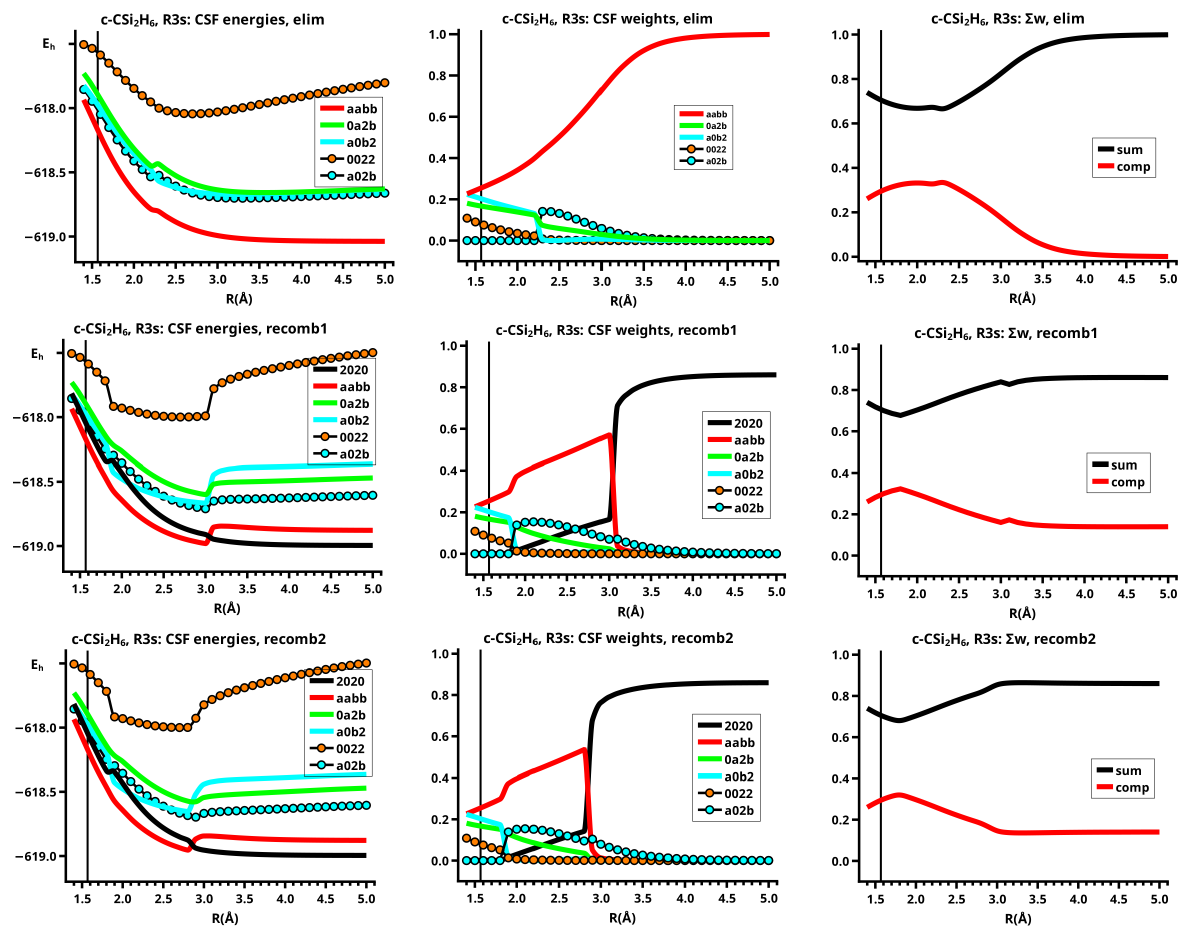}
\caption{CSF energies, weights, and sum of weights of the largest contributions to the wave function in reactions \textbf{R3s}.}\label{fig:CSi2Cs_Ew}
\end{figure}

In the recombination reactions, the wave function is composed of six CSFs; CSF NB dominates the dissociated system, at $R=5.0$\,\AA, its weight is 0.86, the remaining 14 percent stem from ``small'' CSFs, but not from any other ``large'' CSF. At $R=3.1$\,\AA, NB has still  a weight OF 0.72, the weight of the second large CSF, the ionic CSF C $|a02b|$, is only 0.07. At $R=3.0$\,\AA, the weight of NB has dropped to 0.17, the largest CSF is TT with a weight of 0.57. The weights of the ionic CSFs C, CX1, and CX2 increase as expected; at $R=1.8$\,\AA, the system has changed to $C_{2v}$ symmetry, and then the reaction proceeds as in the $C_{2v}$ elimination reaction. See Figure \ref{fig:CSi2Cs_Ew}.

\newpage
\subsection{\ce{c-SiC2H6 -> C2H4 + SiH2}}
The creation of silacyclopropane \ce{c-SiC2H6} by addition of silylene to ethene and the elimination of silylene from the 3-ring is studied in  high $C_{2v}$ (\textbf{R4v}) and low $C_s$ symmetry  (\textbf{R4s}).  The approximate reaction coordinate $R$ is in the range of $R_m=1.4$\,\AA{} to $R_d=5.0$\,\AA.  At the equilibrium of \ce{c-SiC2H6} the value of the approximate reaction coordinate is $R=1.705$\,\AA.
In \ce{c-SiC2H6} the atoms X and Y from the sketch in Figure \ref{fig:molgeom} are equal to C,  atom Z is the silicon atom.

\begin{figure}
\includegraphics[width=0.4\textwidth]{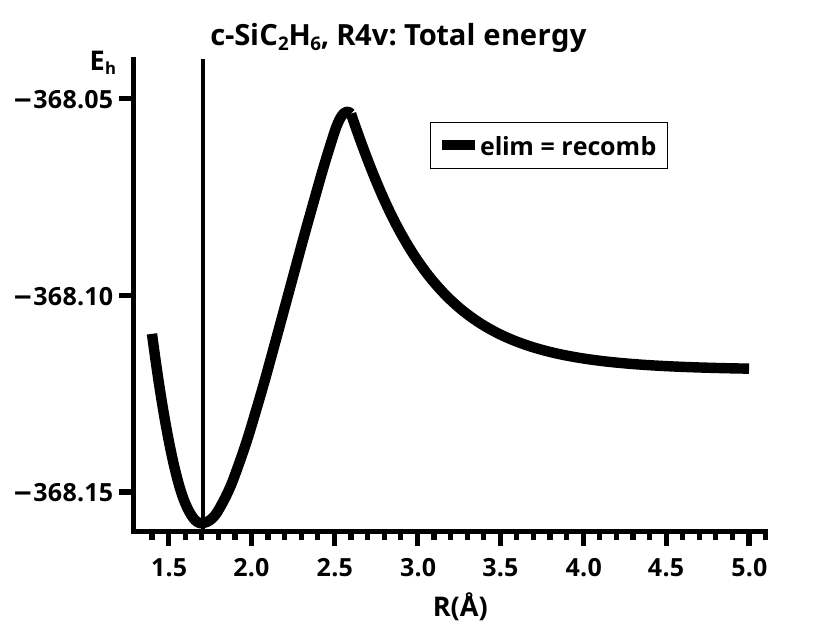}
\includegraphics[width=0.4\textwidth]{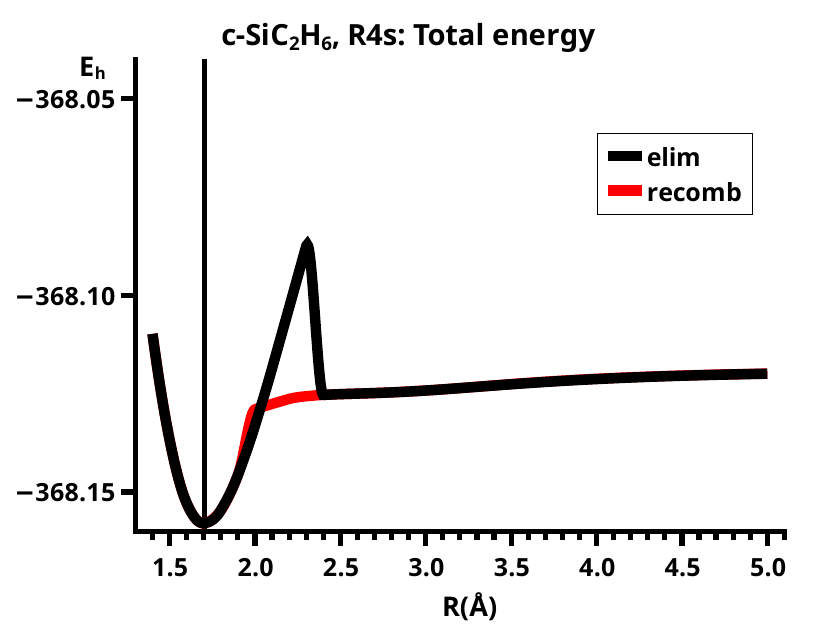}
\caption{Total energies for the reactions \textbf{R4v} and \textbf{R4s}.}\label{fig:SiC2_E}
\end{figure}

Both the elimination and the recombination reaction give the same PEC, which has a prominent cusp at $R=2.6$\,\AA{} (Figure \ref{fig:SiC2_E}). The two branches describe the reactions in the bonded and the dissociated system, respectively, at the crossing point the system jumps immediately between the two troughs, the system never follows the MEPs beyond the crossing point. Therefore there is only a single curve for each geometry parameter, and there is only a single set of CSF energy and weight curves.

\subsubsection{\textbf{R4v}: The elimination and recombination reactions of \ce{c-SiC2H6} in $C_{2v}$}
\begin{figure}[ht]
\includegraphics[width=0.8\textheight]{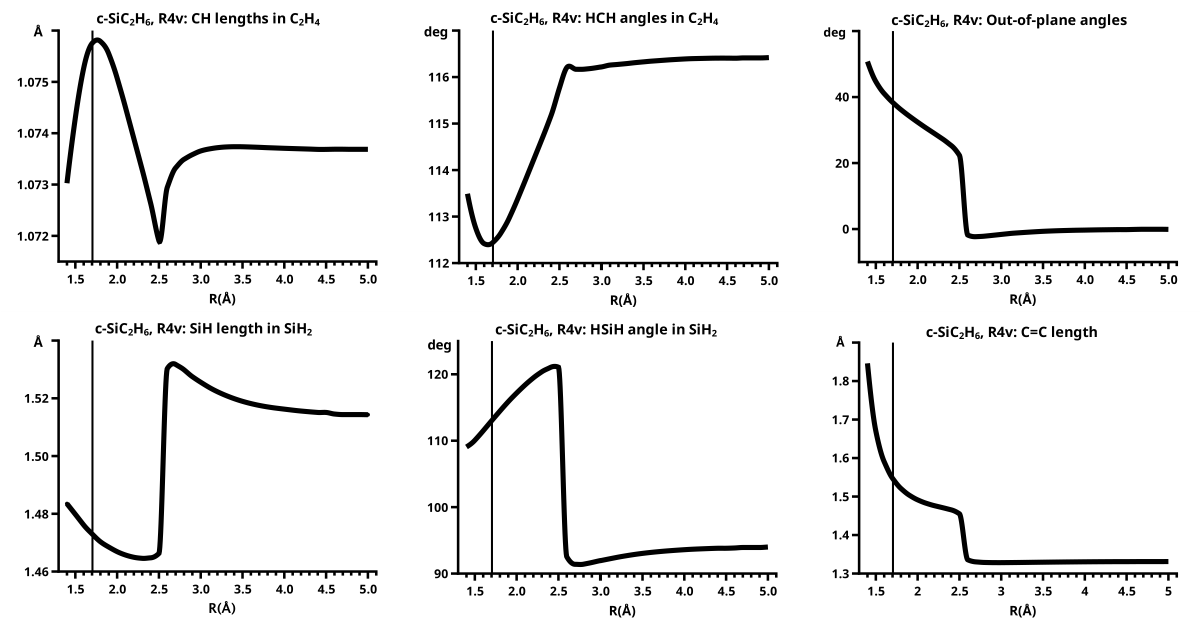}
\caption{Geometry parameters of both fragments.}\label{fig:SiC2C2v_geo}
\end{figure}

\begin{figure}[ht]
\includegraphics[width=0.8\textheight]{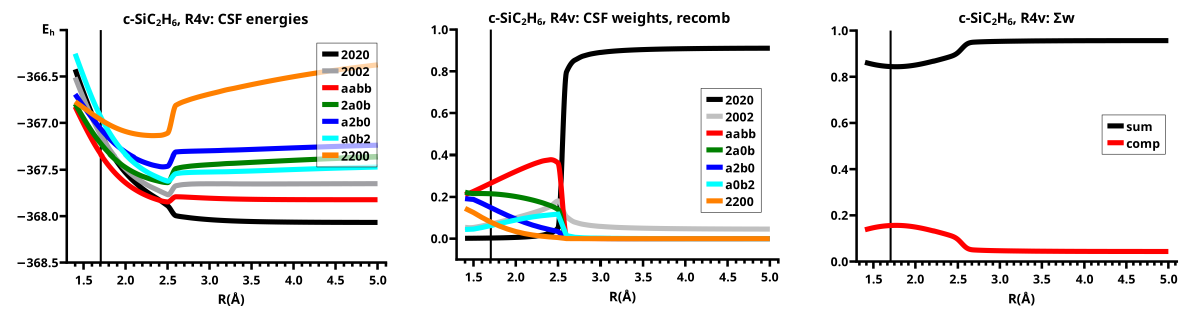}
\caption{CSF energies, weights, and sum of weights of the large contributions to the wave functions. }\label{fig:SiC2C2v_Ew}
\end{figure}

The PEC (Figure \ref{fig:SiC2_E}) and all curves of the geometry parameters (Figure \ref{fig:SiC2C2v_geo} show that the silylene elimination and the recombination follow the same two MEPs; the crossing point for both reactions is at $R=2.6$\,\AA. When the recombination starts at the dissociated geometry obtained in the elimination reaction, the reaction yields effectively the same geometries and energies (Figure \ref{fig:SiC2C2v_Ew}. In the bonded system both fragments are in high-spin states, in the dissociated system  both fragments are in their respective singlet ground states, the ethene fragment is planar.

The dissociated system is described by the neutral CSFs NB and DX $|2002|$, which describes the angular correlation in the silylene fragment. The contribution of the small CSFs is only 7 percent. In the bonded system the neutral CSF TT has largest weight, the CX1 CSF $|2a0b|$, the CX2 CSFs $|a2b0|$ and the C CSF $|a0b2|$ describe the charge shifts in the covalent C-Si bonds. The DC CSF $|2200|$ has weight larger than 0.1 only for  $R$  smaller than the equilibrium value, it is thus irrelevant. The wave function is thus a linear combination of seven large CSFs, two dominate the dissociated system, four the bonded system, and one is irrelevant.

\subsubsection{\textbf{R4s}: The elimination and recombination reactions of \ce{c-SiC2H6} in  $C_s$}
The elimination reaction follows the $C_{2v}$ MEP up to $R=2.3$\,\AA, where the total energy is about 0.038 millihartrees above the energy in the entrance channel of the recombination reaction. At this geometry, the system symmetry changes from $C_{2v}$ to $C_s$ and the wave function changes from high-spin to low-spin characteristics. The geometry parameters change very differently. Least changes are observed for the C-Si bond lengths, there is  a slight shift of the silylene fragment parallel to the ethene fragment, so that atom Z is closer to carbon atom 2 than to carbon atom 1. The geometry parameters of the \ce{CH2} moiety at atom 2 change pronouncedly smoother than those of the other \ce{CH2} moiety. The most abrupt changes shows the \ce{SiH2} moiety.  Because of the planarity of ethene there is only one starting geometry for the recombination reaction, and, thus, a single recombination reaction. Up to $R=2.0$\,\AA, the total energy  is nearly constant during the approach of the fragments; and the wave function retains the low-spin characteristics. The geometry parameters are rather constant up to $R=3.0$\,\AA, then all parameters start changing, but, up to $R=2.3$\,\AA, the parameter curves are identical with those for the elimination reaction. Between $R=2.3$\,\AA{} and $R=2.0$\,\AA, the changes increase, the largest differences are found for the parameters describing the different CH$_2$ moieties of the ethene fragment. At $R=2.0$\,\AA, the total energy is higher than that of the elimination reaction, but at $R=1.90$\,\AA, the symmetry  changes to $C_{2v}$ and the recombination follows the $C_{2v}$  MEP.

\begin{figure}[ht]
\includegraphics[width=0.8\textheight]{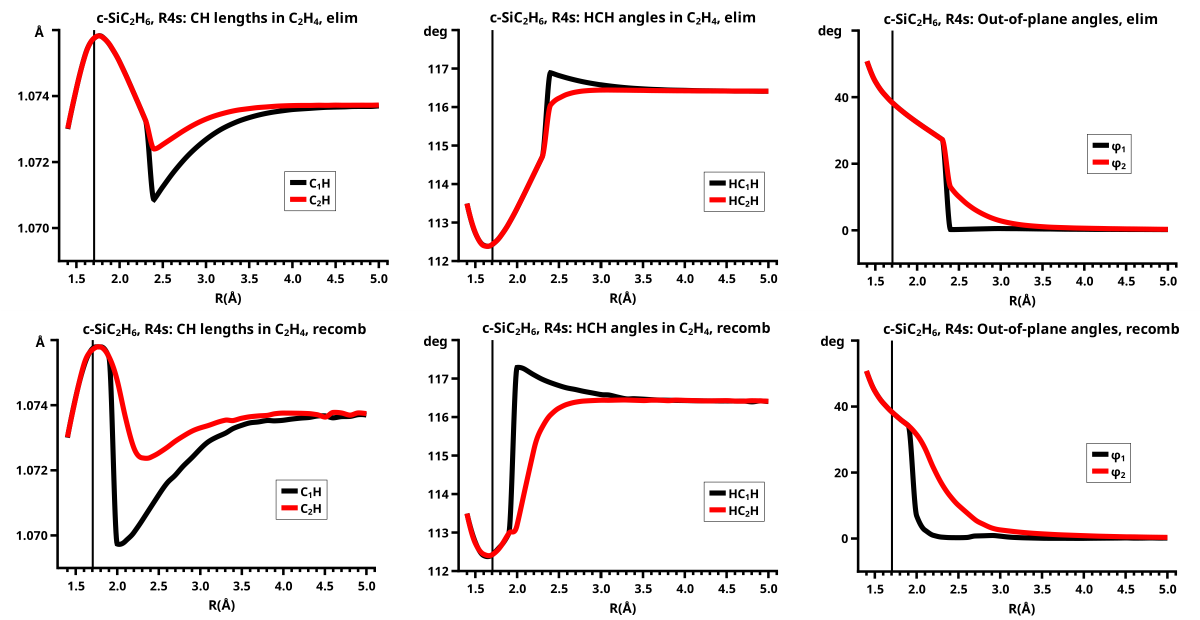}
\caption{Geometry parameters of both fragments.}\label{fig:SiC2Cs_geo1}
\end{figure}

\begin{figure}[ht]
\includegraphics[width=0.8\textheight]{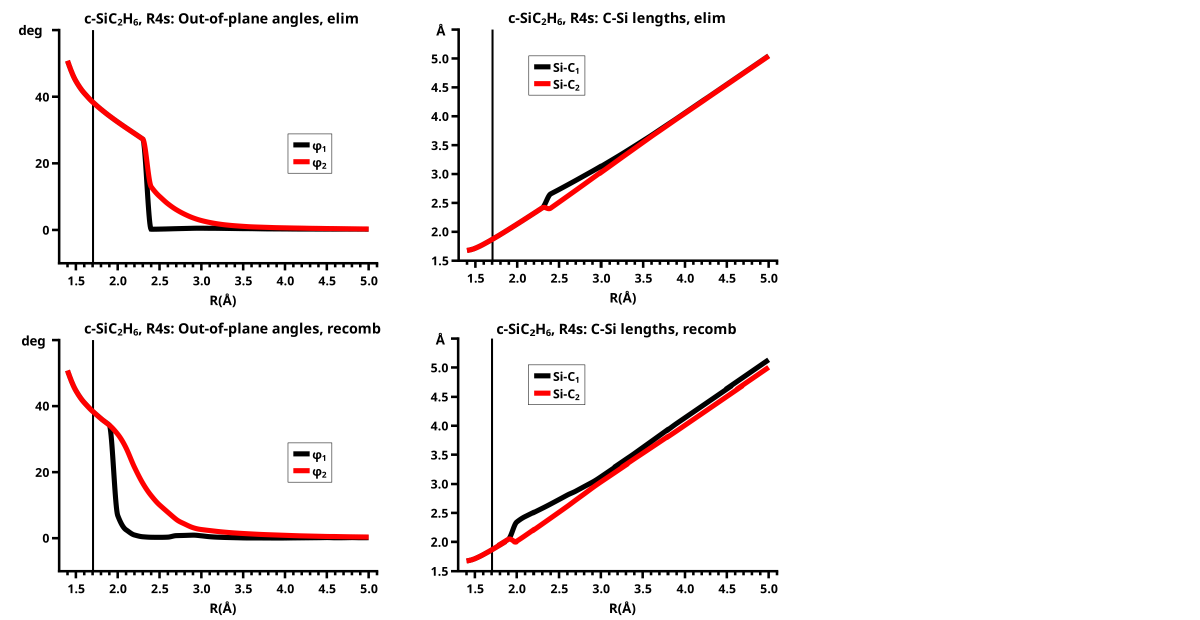}
\caption{Geometry parameters of both fragments.}\label{fig:SiC2Cs_geo2}
\end{figure}

\begin{figure}[ht]
\includegraphics[width=0.8\textheight]{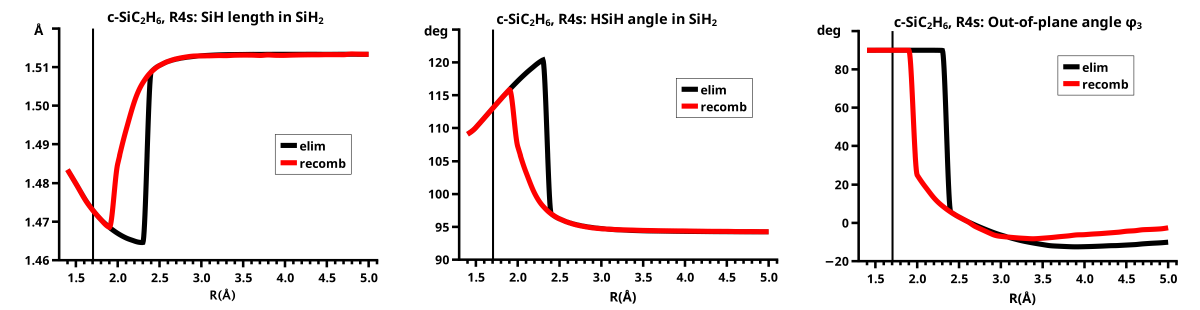}
\caption{Geometry parameters of both fragments.}\label{fig:SiC2Cs_geo3}
\end{figure}

\begin{figure}[ht]
\includegraphics[width=0.8\textheight]{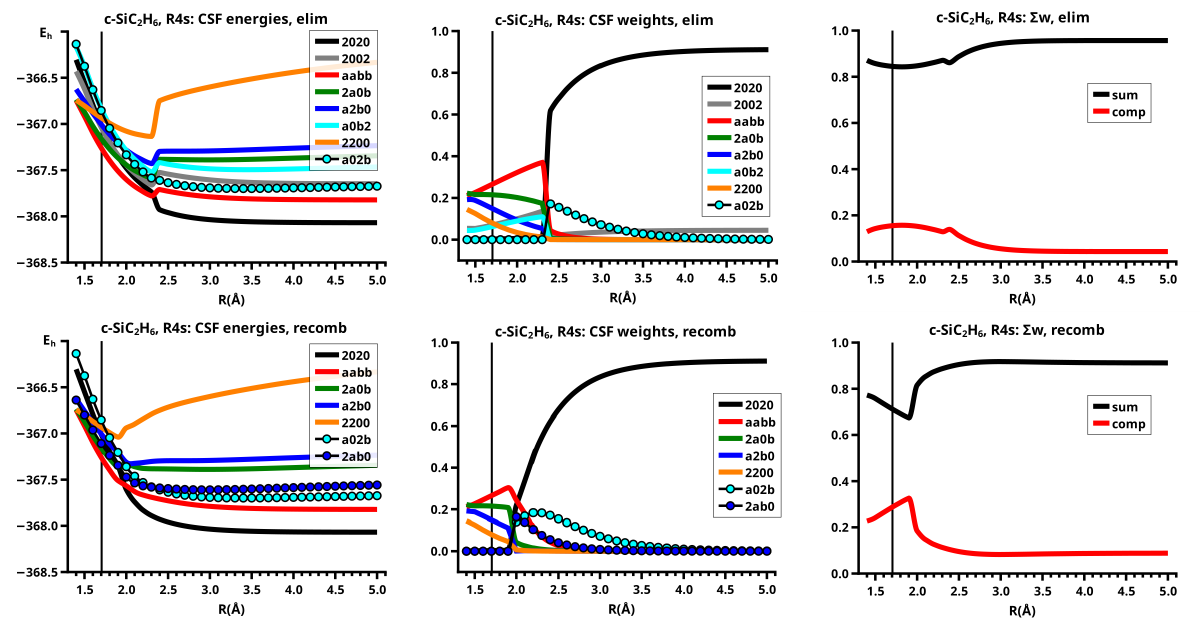}
\caption{CSF energies and weights and the sum of the weights of the largest contributions to the wave function and the complementary contribution of the small contributions.}\label{fig:SiC2Cs_Ew}
\end{figure}

As long as the system has $C_{2v}$ symmetry in the elimination reaction, the wave function is a linear combination of the neutral CSF TT, the CX1 CSF $|2a0b|$,  the two CX2 CSFs $|a2b0|$ and $|a0b2|$, and the DC CSF $|2002|$. The CX1 and CX2 CSFs describe the charge shifts in the covalent Si-C single bonds, CSF CSF $|2002|$ describes  in the dissociated system the angular correlation in the silylene $^1A_1$ ground state; in the bonded system it helps to describe the deformation of the electron distribution in the silylene fragment. The doubly ionic CSF $|2200|$ has a weight larger than 0.1 only at small $R$ values, indeed this CSF is rather unimportant for the description of the reaction.  When the symmetry changes to $C_s$, the wave function changes the characteristics from high-spin to low spin, CSF TT is replaced by CSF NB and the C CSF $|a02b|$; with increasing $R$ the weight of NB increases and that of $|a02b|$ decreases; at $R=5.0$\,\AA{} only NB and  the correlating CSF $|2002|$ survive, all other CSFs contribute not more than 4 percent.
In the recombination reaction, up to $R=2.5$\,\AA, the wave function is dominated by CSF TT contributes;  at $R=3.5$ the weight of the C CSF $|a02b|$ increases, at $R=2.5$\,\AA{} also that of the C CSF $|2ab0|$. At $R=2.0$\,\AA, the system symmetry becomes $C{2v}$, the C CSFs no longer contribute to the wave functions, the wave function has the same the composition as it has in the elimination reaction.

\newpage
\subsection{\ce{c-CSi2H6 -> CSiH4 + SiH2}}
The elimination of silylene from disilirane and the reverse addition of silylene to silaethene, reaction  \textbf{R5s}, proceeds only in $C_s$ symmetry; the approximate reaction coordinate $R$ is in the range of $R_m=1.3$\,\AA{} to $R_d=5.0$\,\AA, at the equilibrium of \ce{c-CSi2H6} the value of the approximate reaction coordinate is $R=1.834$\,\AA.
In the molecular system,  the atoms X and Y from the sketch in Figure \ref{fig:molgeom} are carbon and silicon, respectively, atom Z is silicon.

\begin{figure}[ht]
\includegraphics[width=0.8\textwidth]{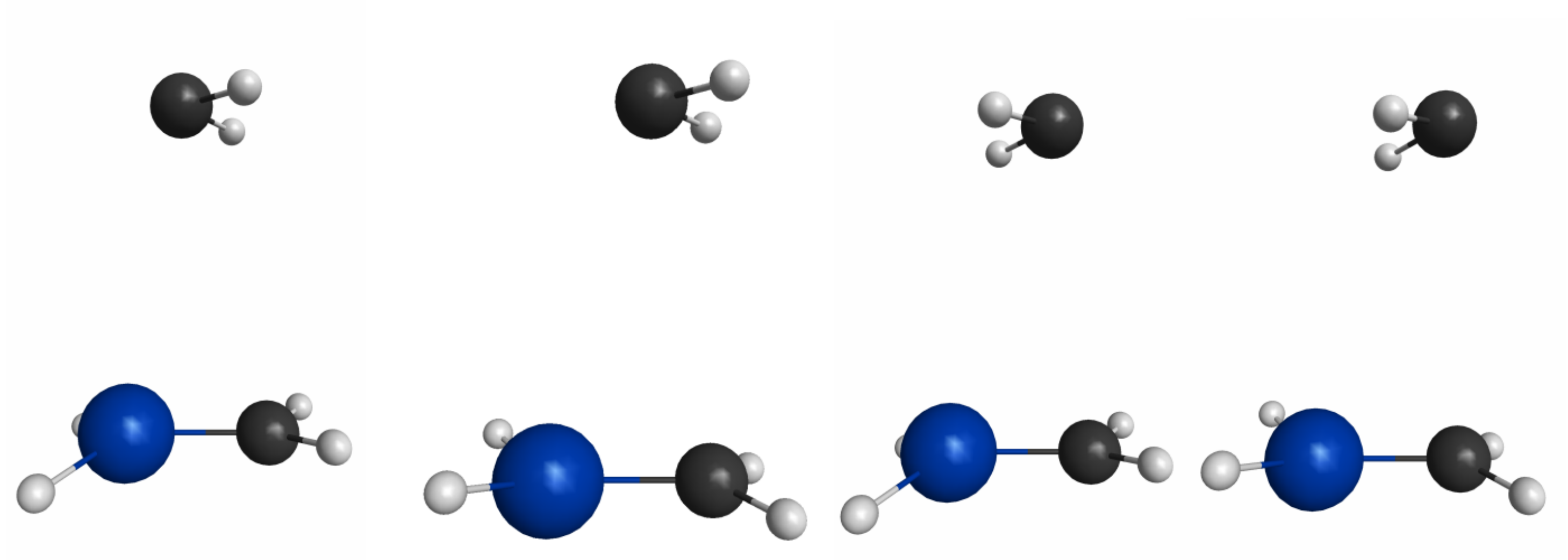}
\caption{From left to right: The four possible structures S1, S2, S3, and S4. In \ce{CSi2H6}, atom Z is silicon; in \ce{SiC2H6}, atom Z is carbon. }\label{fig:reacSiC2Cs_OI123}
\end{figure}

This is the first of two reactions with silaethene as ethene analog; it has a heteroatomic double bond and a non-zero dipole moment; and it is not planar but trans-bent, the pyramidality at the silicon atom is much larger than at the carbon atom. To each of the two trans-bent structures of silaethene there are two possible orientations of the carbene analog, giving four possible structures as starting structures of recombination reactions; this is different from the reactions studied above.
In structure S1, the silicon atom of silylene is close to the silicon atom in silaethene having correct pyramidality; in structure S2, the pyramidality is wrong. When the orientation of silylene in structure S1 is reverted, one gets structure S3; reverting the orientation of silylene in structure S2 yields structure S4. In all four possible structures, the silylene fragment lies ``parallel'' to the double bond. See Figure \ref{fig:reacSiC2Cs_OI123}.

\begin{figure}[ht]
\includegraphics[width=0.4\textwidth]{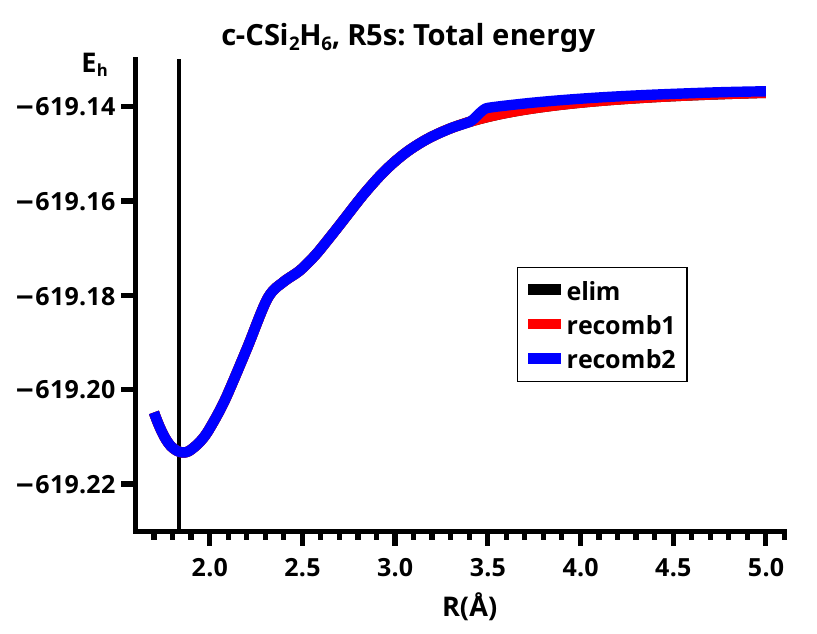}
\includegraphics[width=0.4\textwidth]{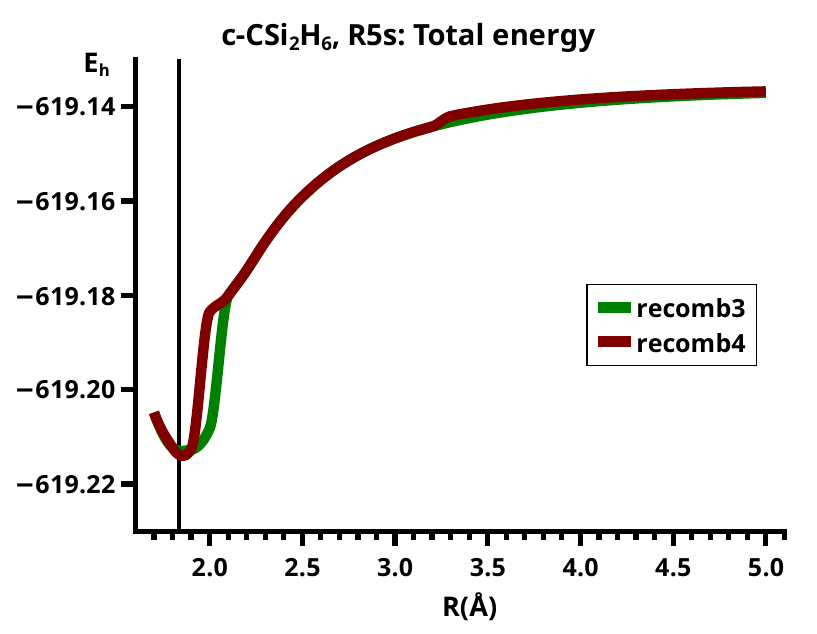}
\caption{Total energies for the elimination reaction and the four recombination reactions.}\label{fig:rCSi2_E}
\end{figure}

The elimination reaction follows a single MEP, the total energy increases monotonously, the dissociated system has structure S1. In this structure, the dipole moments of silylene and silaethene are  antiparallel, so one can assume that this structure is also electrostatically stabilized, when the fragments approach each other.
The PECs of the elimination reaction and the recombination reaction recomb1, starting from structure S1, are essentially identical, the maximum difference between the energies is 30 microhartrees; in Figure \ref{fig:rCSi2_E}, left, the PECs cannot be distinguished. Also the corresponding geometry parameter curves are identical. Therefore, only the recombination reactions are discussed.

At $R=5.0$\,\AA, atom Z has roughly the same distance from atoms X and Y; at $R=3.0$\,\AA, the Z-Y distance, i.e., the Si-Si distance, is by 0.33\,\AA{} shorter than the Z-X distance (Si-C); the interaction between the two silicon atoms is apparently larger than between the silylene silicon atom and the carbon atom.  Around $R=2.4$\,\AA, the silaethene fragment becomes cis-bent; the silylene fragment changes from singlet to triplet geometry, and the ``parallel'' relative position is changed to a ``vertical'' one. The Z-Y distance is shorter than the Z-X distance, so the new Si-Si bond is created before the C-Si bond. In the silaethene fragment, the SiH and CH lengths increase and the HSiH and HCH angles decrease, the ``planar'' fragment adopts the non-planar structure in the 3-ring. In the 3-ring, the Si-Si distance is longer than the Si-C distance.

\begin{figure}[ht]
\includegraphics[width=0.8\textheight]{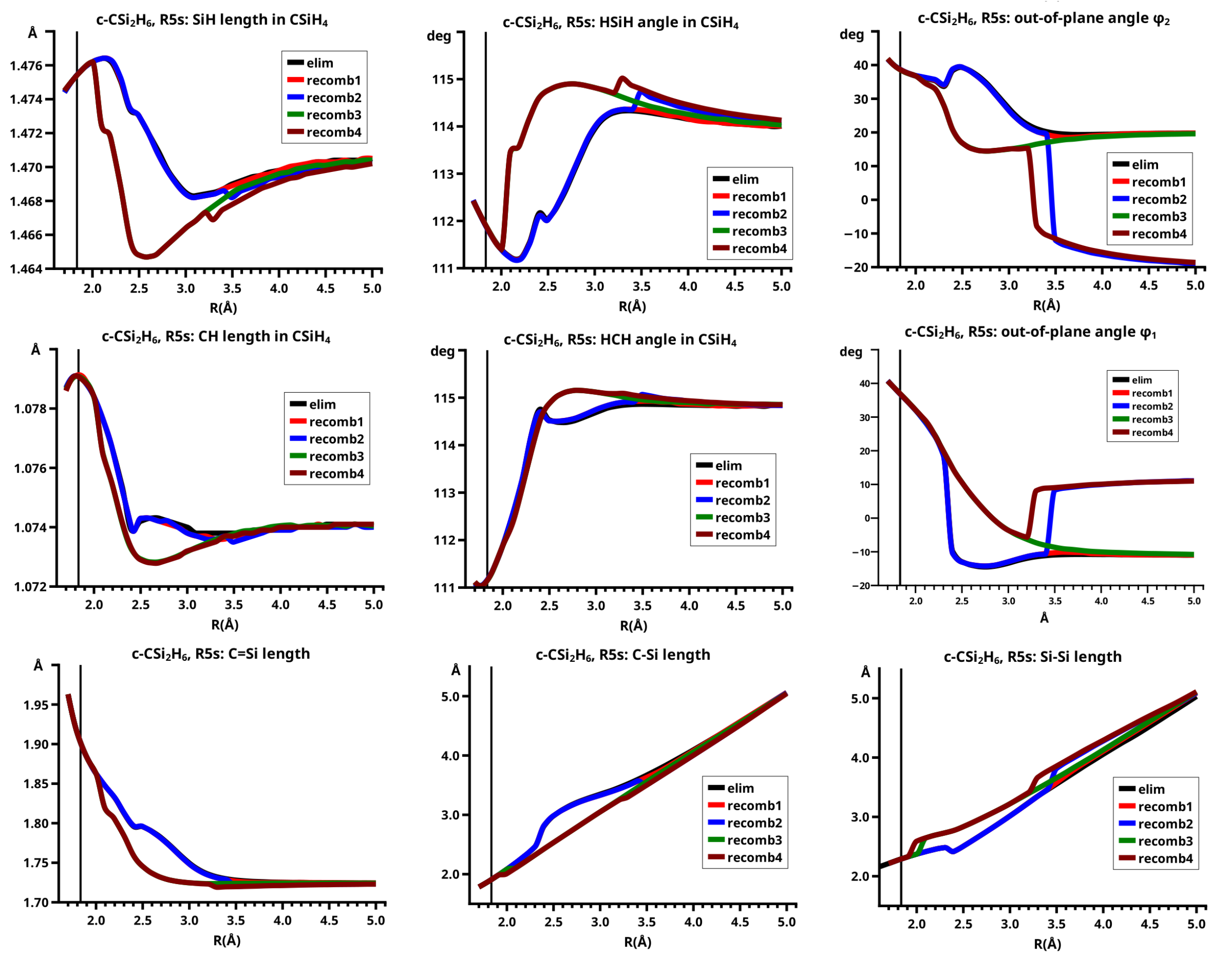}
\caption{Geometry parameters of the silaethene fragment.}\label{fig:rCSi2-geom1}
\end{figure}

\begin{figure}[ht]
\includegraphics[width=0.8\textheight]{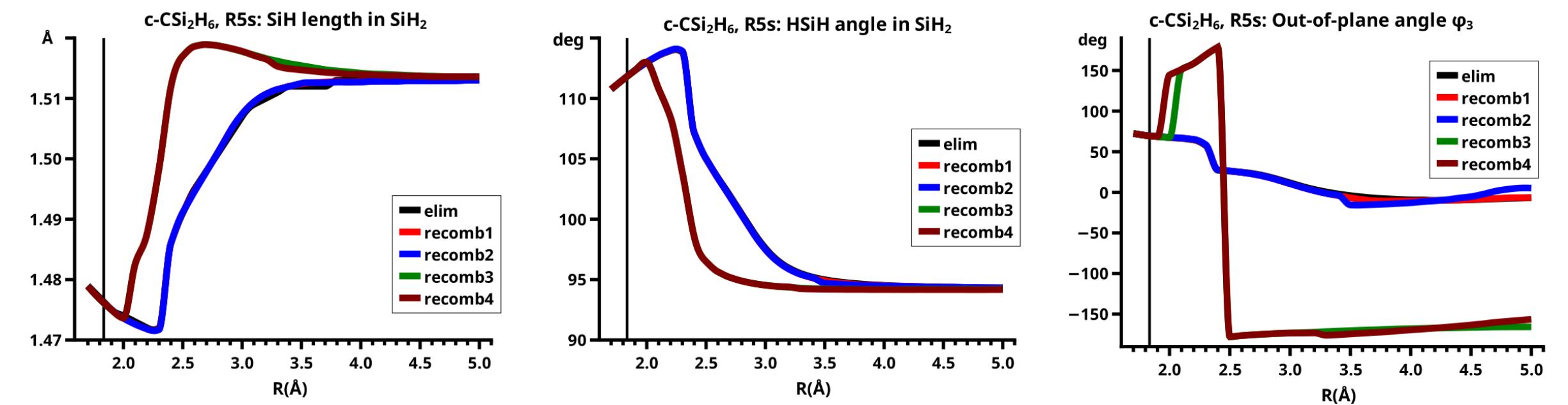}
\caption{Geometry parameters of the silylene fragment.}\label{fig:rCSi2-geom2}
\end{figure}

In reaction recomb2, starting from structure S2, the PEC lies slightly above that of reaction recomb1 up to $R=3.5$\,\AA,  the energy differences are not larger than 1.4 millihartrees. All geometry parameters, except for the out-of-plane angles are equal to those of recomb1. At $R=3.5$\,\AA, the pyramids in silaethene are inverted, the energy drops only by 3 millihartrees; for $R< 3.5$\,\AA, reactions recomb1 and recomb2 are equal, as are all geometry parameters. See Figures \ref{fig:rCSi2-geom1} and \ref{fig:rCSi2-geom2}.

Whereas in reactions recomb1 and recomb2 ``parallel'' means that  $\varphi_3$  varies in a small interval $-x \le \varphi_3 \le x$ degrees around $\varphi_3 = 0$ degrees, in reactions recomb3 and recomb4 $\varphi_3$ has values $\varphi_3 \ge -180+x$ or $\varphi_3 \le 180-x$.
This is the origin of the jump in the $\varphi_3$ curve for recomb3 and recomb4 in Figure  \ref{fig:rCSi2-geom2}. In recomb3 and recomb4, silylene remains ``parallel'' up to $R=2.4$\,\AA, at this point the characteristics of the wave function changes from low-spin to high-spin.
The silicon atom in silylene is close to the carbon atom in silaethene, which has wrong pyramidality. Because of the parallel orientation of the dipole moments of the fragments one might assume that this structure is electrostatically destabilized, but the energy difference of 100 microhartrees between structures S1 and S3 shows that the dipol-dipole interaction is of minor importance for the stabilization at large distances. More important is the different interaction of the silicon atom in silylene with the atoms in silaethene. In structure S3, the Z-X (Si-C) distance is about 0.06\,\AA{} shorter than the Z-Y (Si-Si) distance, and the difference becomes even larger with decreasing $R$ values. During the approach of the two fragments, angle $\varphi_1$ at the carbon atom changes from negative to positive values; angle $\varphi_2$ is already positive as it must be in the 3-ring. Comparison of the total energies shows that the interaction between the two silicon atoms in recomb1 is much larger than that between silicon and carbon in recomb3; between $R=2.8$,\AA{} and $R=2.1$,\AA, the energy of recomb3 is approximately 10 millihartrees higher that the energy of recomb1. At about $R=2.4$\,\AA, the wave function changes its characteristics from low-spin to high-spin, the CH$_2$ and the SiH$_2$ moieties in silaethene change to the typical values in the 3-ring, the silylene fragment changes the orientation and $\varphi_3$ jumps to a value of about 45 degrees, only at $R=2.0$\,\AA, when the 3-ring is formed, the silylene fragment changes to an upright position.
Structure S4, the starting structure of reaction recomb4, can be regarded as the result of an umbrella inversion at both atoms in the double bond (double umbrella inversion) in structure S3. The only difference between recomb3 and recomb4 are the different out-of-plane angles of silaethene up to $R=3.4$\,\AA, the total energy of reaction recomb4 lies slightly above that of reaction recomb3, the largest difference is 2.3 millihartrees. At $R=3.4$\,\AA, the molecular system changes by a double umbrella inversion to that of reaction recomb3. The difference between recomb3 and recomb4 is minimal, as is the difference between recom1 and recomb2, the large difference between recomb1 and recomb3, and between recomb2 and recomb4, respectively, is caused by the much larger interaction of the silicon atom in silylene with the silicon atom than with the carbon atom in silaethene.

\newpage

\newpage

\begin{figure}[ht]
\includegraphics[width=0.8\textheight]{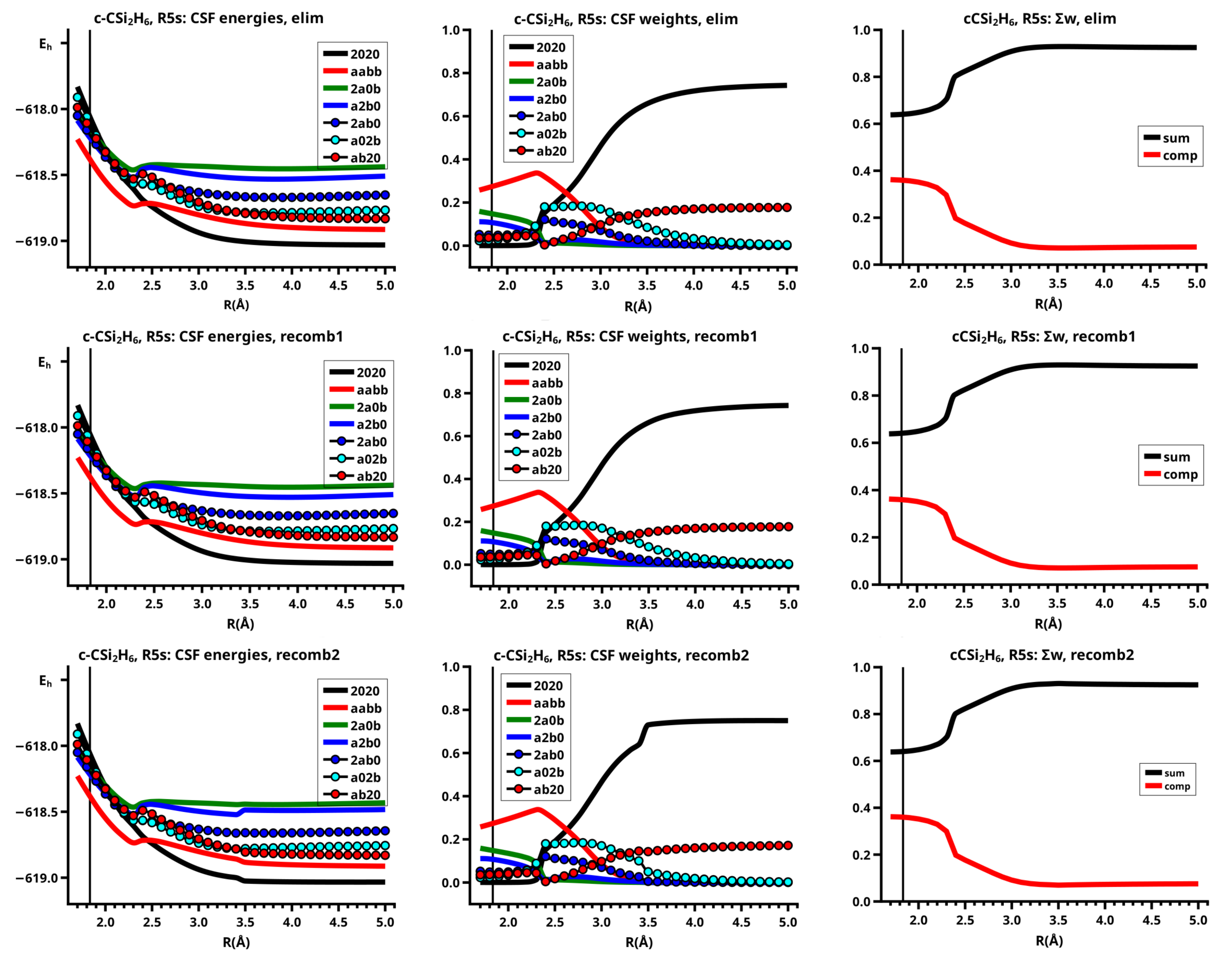}
\caption{CSF energies and weights for all five reactions.}\label{fig:rCSi2-Ew1}
\end{figure}

\begin{figure}[ht]
\includegraphics[width=0.8\textheight]{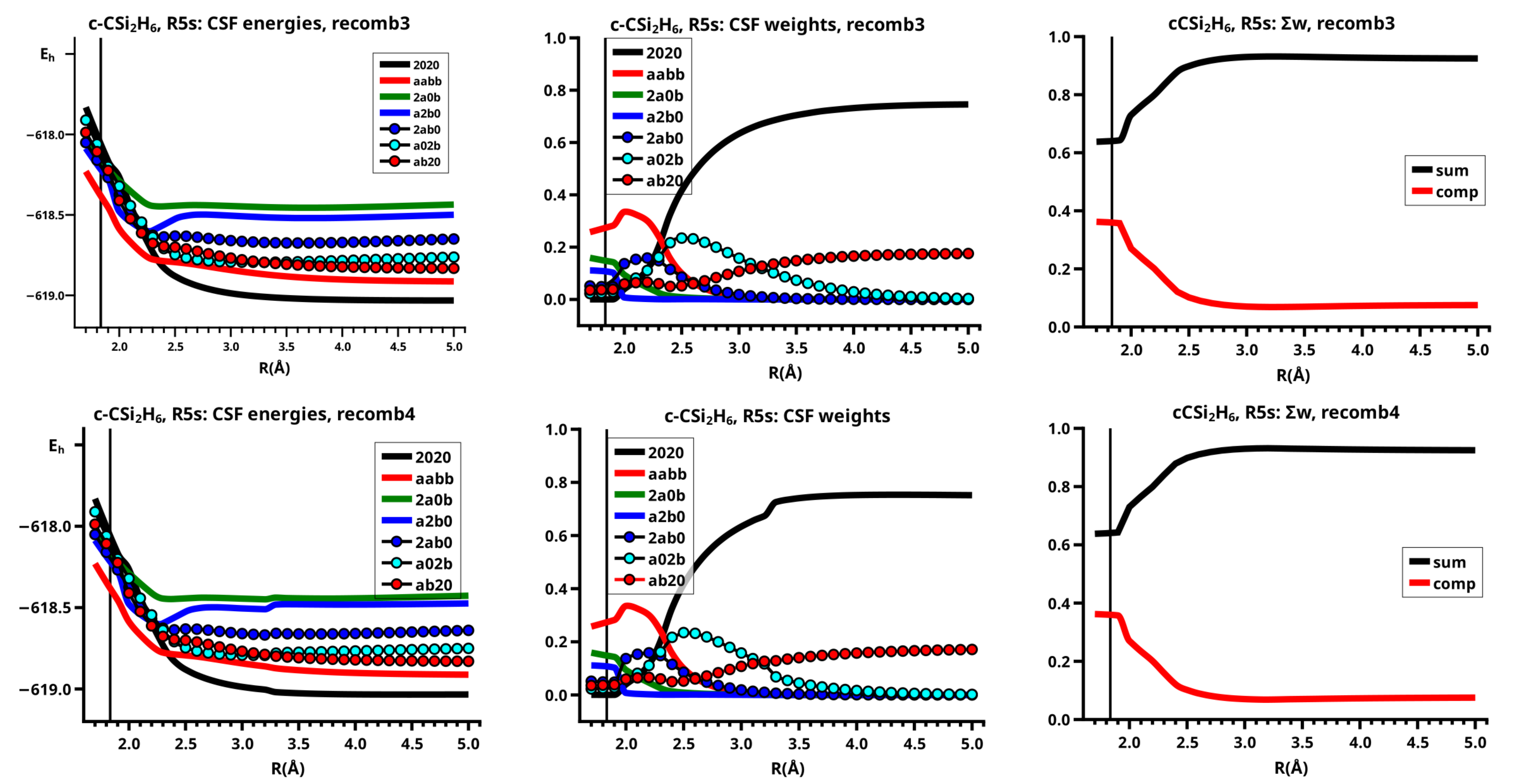}
\caption{CSF energies and weights for all five reactions.}\label{fig:rCSi2-Ew2}
\end{figure}

The wave functions describing all five reactions are linear combinations of seven CSFs. In the dissociated system, the wave functions are spanned by only the neutral CSFs NB and X $|ab20|$, which describes the polarized $\pi$ density in silaethene. The sum of their weights is larger than 90 percent.  When the fragments approach, the weights of both CSFs decrease, the weight of CSF TT increases as do the weights of the C CSFs $|a02b|$ and $|2ab0|$, the first one describes a charge shift from the bonding $\pi$ FMO to the empty p lone pair FMO in silylene, the second one a shift from the doubly occupied s lone pair FMO to the antibonding $\pi$ FMO. If one considers that, in the dissociated system, carbon AOs dominate the $\pi$ FMO and silicon AOs the $\pi^*$ FMO, CSF $|a02b|$ describes mainly a charge shift from the $\pi$ electron density, located at the carbon atom, to the empty lone pair FMO in the silylene fragment;  CSF $|2ab0|$ describes  a charge shift from the doubly occupied s lone pair FMO in the silylene fragment to the empty silicon AOs in the $\pi^*$ FMO. So the first CSF is relevant for the description of the new C-Si bond, and the second CSF for the description of the new Si-Si bond.  In reactions recomb1 and recomb2, both CSFs contribute between $R=3.5$\,\AA{} and $R=2.4$\,\AA; at $R=2.5$\,\AA, the weight of $|a02b|$ is about 60 percent larger than that of $|2ab0|$; then the contribution of both CSFs is suddenly only half of that before. As the sums of the weights show, several small CSFs must replace the two ionic CSFs.
In reactions recomb3 and recomb4, only C CSF $|a02b|$ contributes up to $R=2.7$\,\AA, CSF $|2ab0|$ starts to contribute when the weight of $|a02b|$ has already reached its maximum. This is when the pyramidality of the CH$_2$ moiety in silaethene is inverted, and the wave function changes its characteristics from low-spin to high-spin. See Figures \ref{fig:rCSi2-Ew1} and \ref{fig:rCSi2-Ew2}.

All five reactions follow obviously a single MEP, there are no significant jumps in the energy and geometry curves; also the CSF energies vary smoothly, even there where the characteristics of the wave functions change.

\newpage
\subsection{\ce{c-SiC2H6 -> CSiH4 +CH2}}
Reaction \textbf{R6s}, the elimination of methylene from silirane and the  addition of methylene to silaethene,  proceeds only in $C_s$ symmetry; the approximate reaction coordinate $R$ is in the range of $R_m=1.3$\,\AA{} to $R_d=5.0$\,\AA, at the equilibrium of \ce{c-SiC2H6} the value of the approximate reaction coordinate is $R=1.452$\,\AA.
In \ce{c-CSi2H6} the atoms X and Y from the sketch in Figure \ref{fig:molgeom} are carbon and silicon, atom Z is carbon.

The four  starting structures for the recombination reaction (see Figure \ref{fig:reacSiC2Cs_OI123}) are the same as for reaction \textbf{R5s}.

Structure S1 is the result of the elimination reaction and the starting structure for reaction recomb1; both reactions are essentially identical, the maximal difference between the PECs is 27 microhartrees.
Structure S1 consists of methylene in the 1-$^1A_1$ state and trans-bent silaethene in the singlet ground state. The \ce{SiH2} moiety has already correct pyramidality, angle $\varphi_2$ has 10 degrees. The \ce{CH2} moiety has wrong pyramidality with $\varphi_1=-11$ degrees, the pyramid is very flat. The methylene fragment is ``parallel'' to silaethene with the carbon atom close to the silicon atom in silaethene.

Between $R=5.0$\,\AA{} and $R=3.2$\,\AA, all geometry parameters of the fragments are essentially constant; only the Z-X (C-C) and Z-Y (C-Si) bond lengths change linearly, C-Si up to $R=2.0$\,\AA, C-C only up to $R=3.2$\,\AA, C-C is always slightly longer than C-Si.
At $R=5.0$\,\AA, the methylene fragment has an HCH angle of 102 degrees and a CH length 0f 1.098\,\AA; at $R=3.2$\,\AA{} the wave function changes its characteristics from low-spin to high-spin, and the interaction between the fragments increases.  Between $R=2.9$\,\AA{} and $R=2.7$\,\AA{} the HCH angle in methylene widens rapidly to 115 deg, which is neither close to the angle in the $^1A_1$ state, nor in the $^1B_1$  state, or in the $^3B_1$  state; then it increases further and reaches a maximum value at $R=1.8$\,\AA. The CH length decreases from 1.098\,\AA{} to 1.08\,\AA{} and reaches a minimum of 1.072\,\AA{} again at $R=1.8$\,\AA. The value of neither geometry parameter can be compared to those of methylene in either of the pure states.
The out-of-plane angle $\varphi_3$ increases from about 0 degrees at $R=5.0$\,\AA{} to 40 degrees at $R=2.8$\,\AA{} and then nearly linearly to 75 degrees in the 3-ring. The geometry parameters of silaethene change very differently. The out-of-plane angle of the CH$_2$ moiety, $\varphi_1$, is essentially constant between $R=5.0$\,\AA{} and $R=2.0$\,\AA, then it changes rapidly to 25 degrees in the 3-ring. The CH bond length and the HCH bond angle are also constant until the flat pyramid inverts  at $R=1.8$\,\AA, the continuous inversion of the pyramid via a flat intermediate structure is accompanied by a change from sp$^3$ hybridization to sp$^2$ and then again to sp$^3$ hybridization.  The out-of-plane angle $\varphi_2$ of the \ce{SiH2} moiety increases from 20 degrees to 40 degrees, and then decreases to 25 degrees in the 3-ring; both the SiH and the CH bond lengths have a minimum at $R=3.2$\,\AA, and both the HSiH and the HCH angles have a maximum at the same position. Again, these changes correspond with the planarization of both pyramids, and, thus, with a change of hybridization from sp$^3$ towards sp$^2$. Between $R=3.0$\,\AA{} and $R=2.7$\,\AA{}, the length of the Si=C bond increases from 1.73\,\AA{] to 1.81\,\AA, this values is kept up to $R=2.1$\,\AA, then it decreases and has a minimal value of 1.79\,\AA{} at $R=1.8$\,\AA, and then increases again. Although several geometry parameters change considerably between $R=3.2$\,\AA{} and $R=1.6$\,\AA, the total energy decreases nearly linearly.

The change of the initial structure from S1 to S2 results in very small difference between reactions recomb1 and recomb2: up to $R=3.2$\,\AA, all geometry parameters change slowly but significantly stronger than in recomb1, most prominent are the different signs of angles $\varphi_1$ and $\varphi_2$. Nevertheless, the differences in the total energy are not larger than 1.4 millihartrees before the double umbrella inversion at $R=3.2$\,\AA, after the umbrella inversion both reactions have identical energies.

The different orientation of the methylene fragment in structure S3 influences the total energy of reactions recomb3 and recomb1 only between $R=3.2$\,\AA{} and $R=2.6$\,\AA, see Figures \ref{fig:rSiC2_E}. The only geometry parameter that is essentially different from the corresponding recomb1 parameter is $\varphi_3$. All other parameters have similar shapes; the position, where the largest changes of the parameter values occur, are where the wave function its characteristics changes, this is in recomb3 at about $R=2.6$\,\AA, and in recomb1 at about $R=2.9$\,\AA. For recomb1 and recomb3, the curves of the HCH angle in silaethene are very similar; in contrast to the recomb1 curve for the CH length, the recomb3 curve has a minimum at $R=2.7$\,\AA. The parameter curves with the largest differences between recomb1 and recomb3 are the HSiH curves. As expected are the Z-X lengths smaller and the Z-Y lengths larger than in case of recomb1.

The major differences between reactions recomb3 and recomb4 are again due to the different pyramidality of silaethene in structure S4. Nevertheless, the PECs for recomb3 and recomb4 are identical, and they are nearly constant between $R=5.0$\,\AA{} and  $R=2.5$\,\AA, where the energy drops to the recomb1 and recomb2 values  and then decreases linearly  between $R=2.8$\,\AA{} and $R=1.6$\,\AA. All significant changes of geometry parameters occur at about $R=2.5$\,\AA, where the characteristics of the wave function changes. See Figures \ref{fig:rSiC2-geom} and \ref{fig:rSiC2-geom1}.

\begin{figure}[ht]
\includegraphics[width=0.4\textwidth]{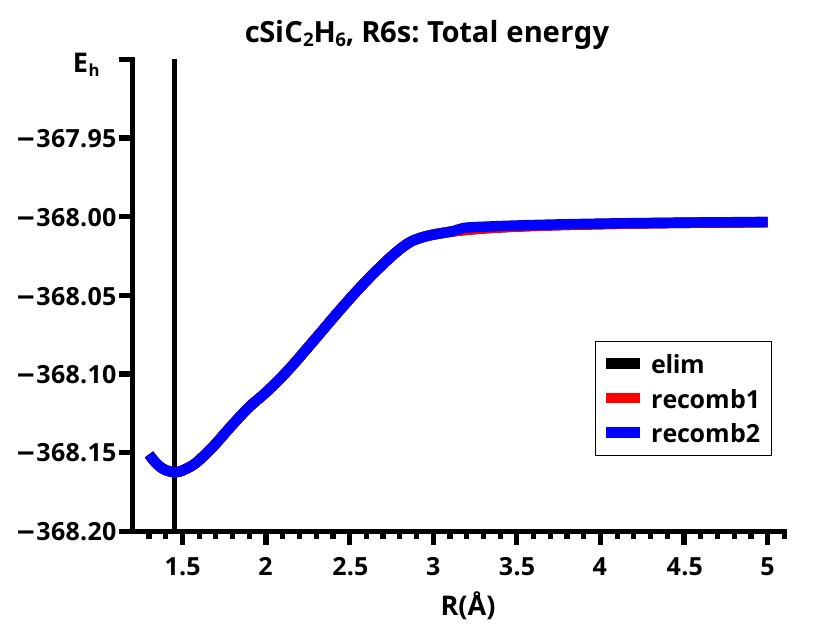}
\includegraphics[width=0.4\textwidth]{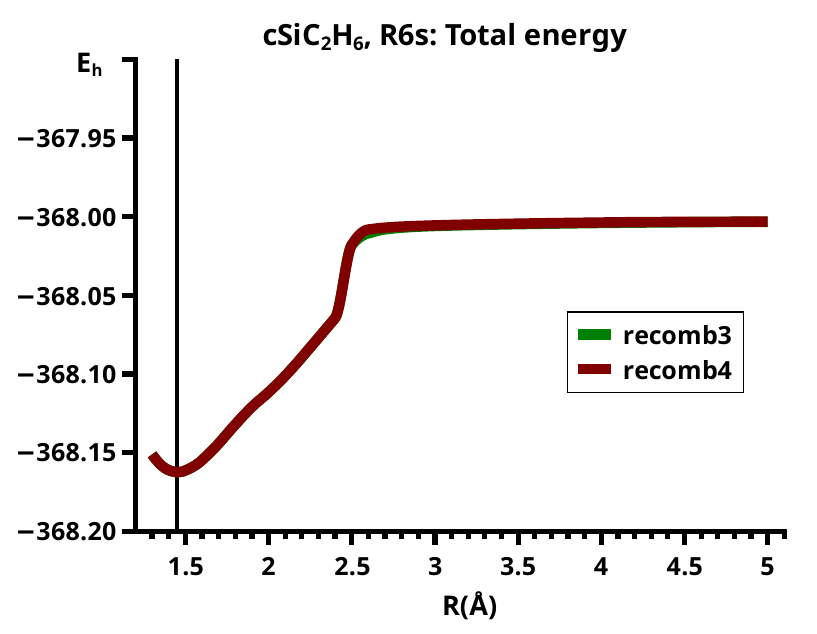}
\caption{Total energies for the elimination reaction and the four recombination reactions.}\label{fig:rSiC2_E}
\end{figure}

\newpage
\begin{figure}[ht]
\includegraphics[width=0.8\textheight]{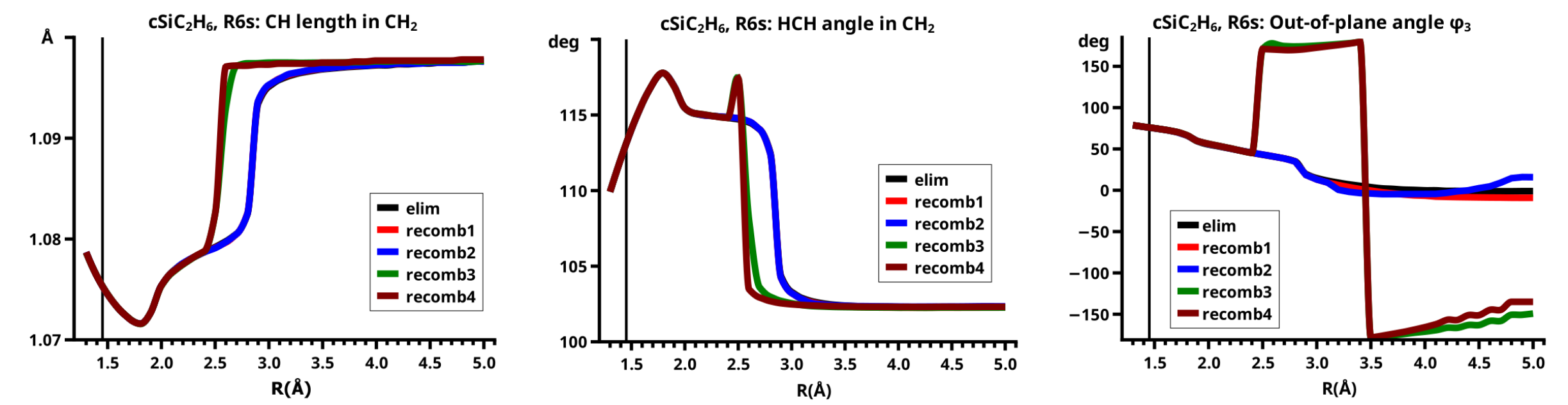}
\caption{Geometry parameters of the molecular system.}\label{fig:rSiC2-geom}
\end{figure}

\begin{figure}[ht]
\includegraphics[width=0.8\textheight]{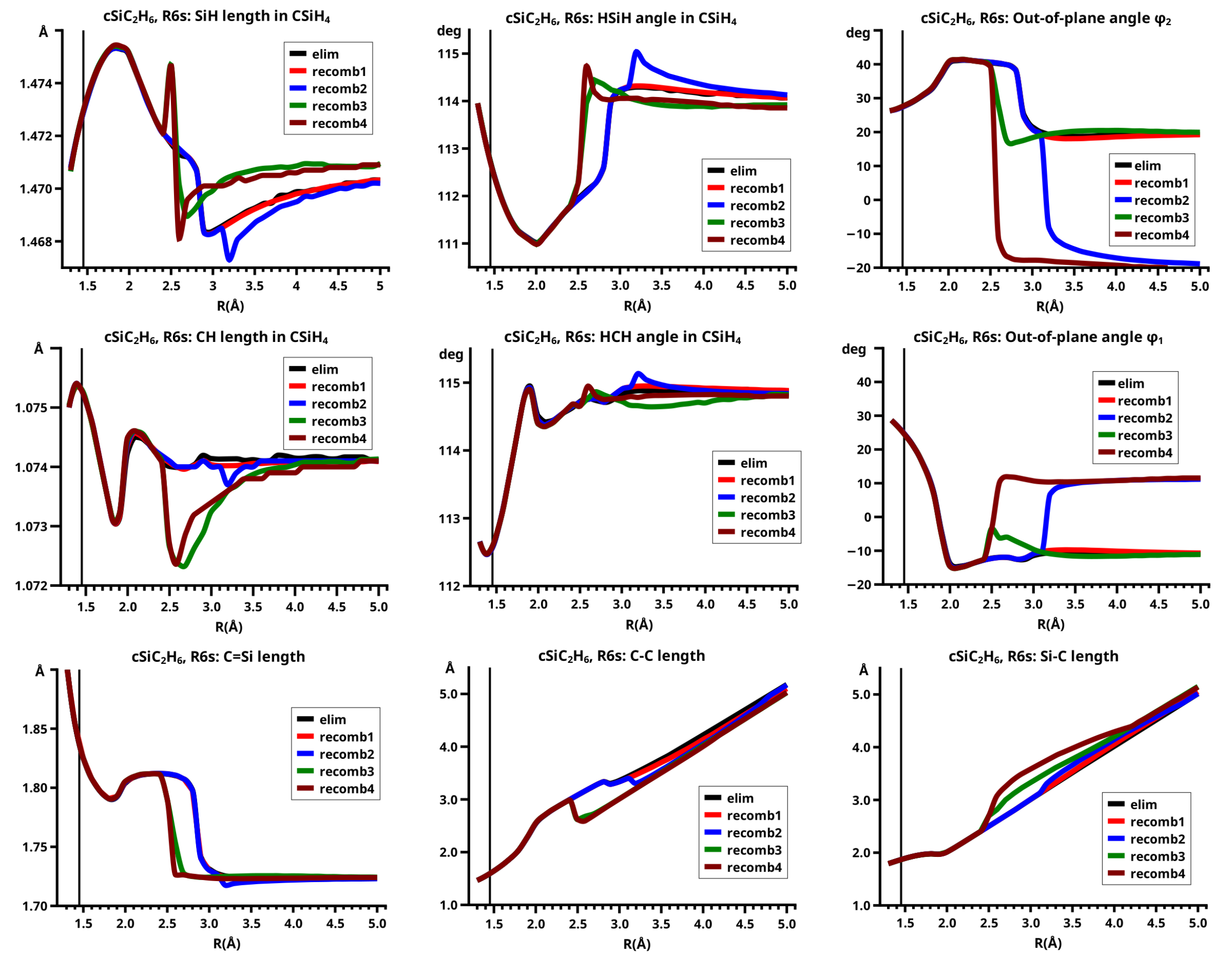}
\caption{Geometry parameters of the molecular system.}\label{fig:rSiC2-geom1}
\end{figure}
\newpage

The wave functions describing all five reactions are linear combinations of the three neutral CSFs NB, TT, and the X CSF $|ab20|$, and of the C CSF $|a02b|$, the CX1 CSF $|0a2b|$, and the CX2 CSF $|a0b2|$. The dissociated system is dominated by CSF NB and the local excitation
$|a0b2|$, the sum of their weights is 93 percent for all reactions. Between $R=3.5$\,\AA{} and $R=3.0$\,\AA{} the C CSF starts to contribute, at the same time the weight of NB declines. The weight of NB drops suddenly to about 0.2 and the weight of CSF TT increases to about 0.4 at those $R$ values, where the fragment start to interact, in recomb1 and recomb2, it is between $R=2.9$\,\AA{} and $R=2.7$\,\AA,
in recomb3 and recomb4, it is between $R=2.6$\,\AA{} and $R=2.4$\,\AA. Up to $R=1.8$\,\AA, CSFs NB, TT, and C are the only dominating large CSFs; in this interval, the weight of NB goes to zero, the weight of TT approaches 0.3, and the weight of CSF C is below 0.05. Now CSFs CX1 and CX2 start to contribute. The graphs with the sum of all large CSFs and their complement show, that a large part of the fragment interaction is represented by small CSFs; as soon as the interaction starts, the sum of the weights of the large CSFs decreases linearly to about 0.6 at the equilibrium of the 3-ring. So, even when only three large CSFs dominate the wave function of the interacting fragments, the large number of small CSFs makes it impossible to attribute one of them to incomprehensible geometric features. An example is the methylene structure with a CH bond length of 1.072\,\AA{} and a HCH bond angle of 115 deg. See Figures \ref{fig:rSiC2-Ewall} and  \ref{fig:rSiC2-Ewall1}.

\newpage
\begin{figure}[ht]
\includegraphics[width=0.8\textheight]{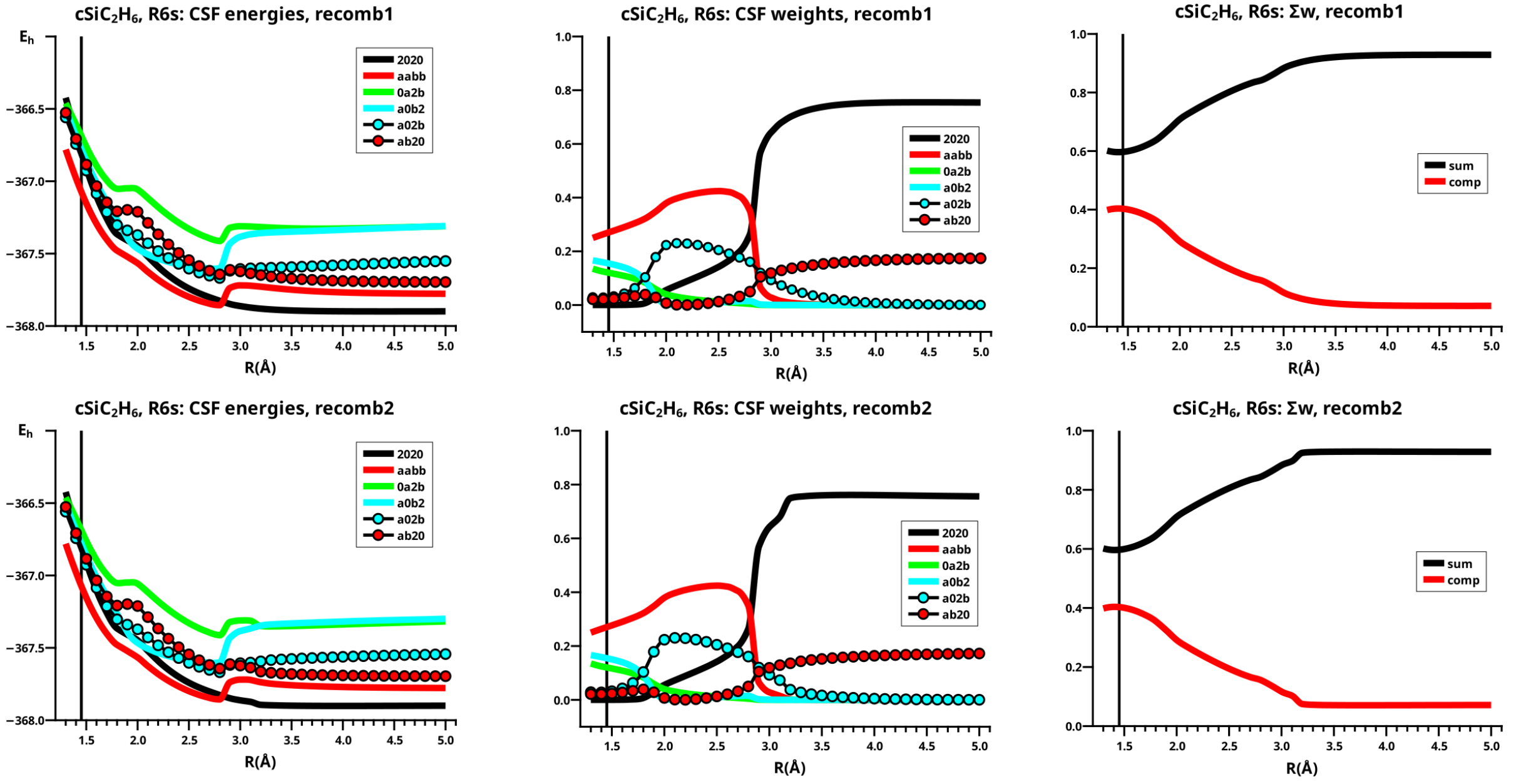}
\caption{CSF energies and weights for all four recombination reactions.}\label{fig:rSiC2-Ewall}
\end{figure}

\begin{figure}[ht]
\includegraphics[width=0.8\textheight]{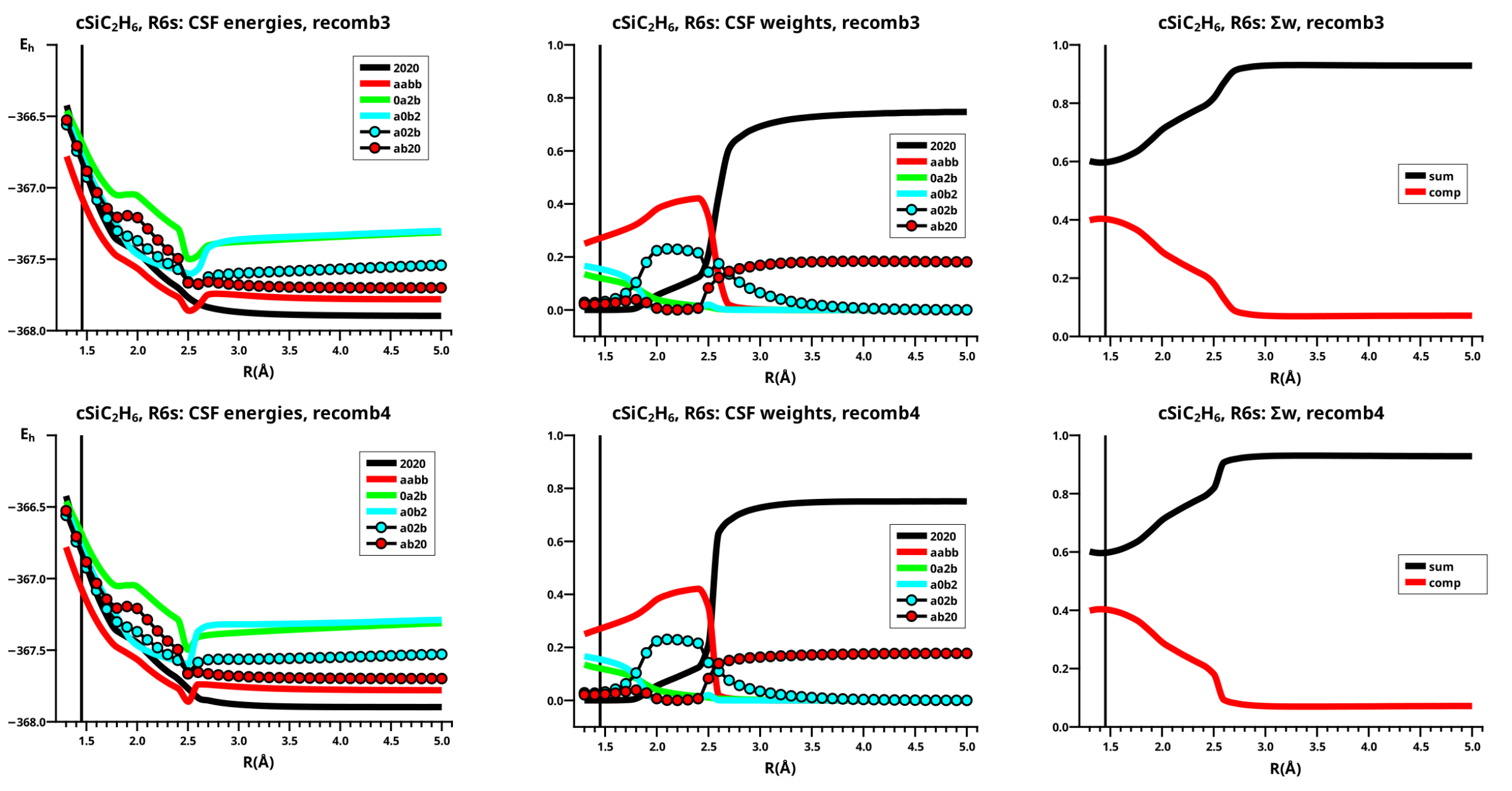}
\caption{CSF energies and weights for all four recombination reactions.}\label{fig:rSiC2-Ewall1}
\end{figure}

\section{Discussion}
\begin{figure}[ht]
\includegraphics[width=0.4\textheight]{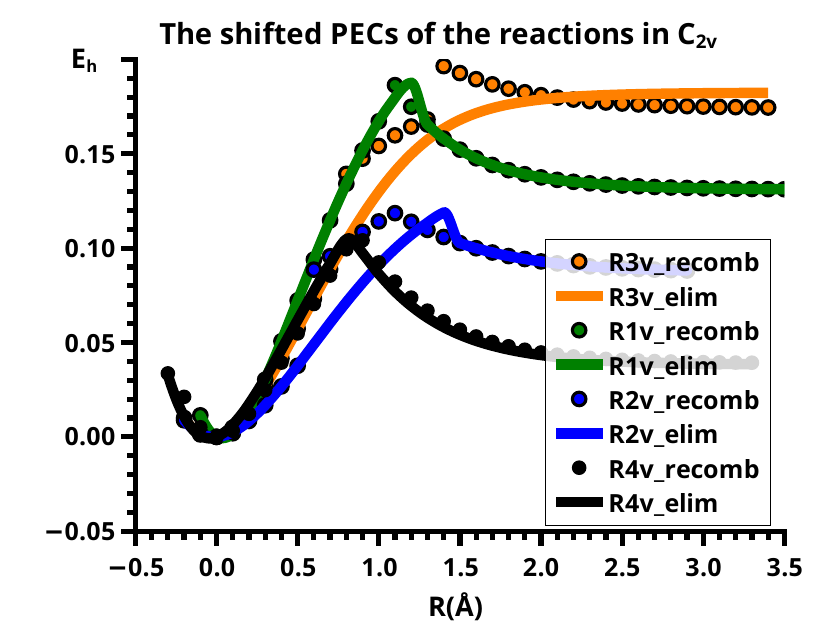}
\includegraphics[width=0.4\textheight]{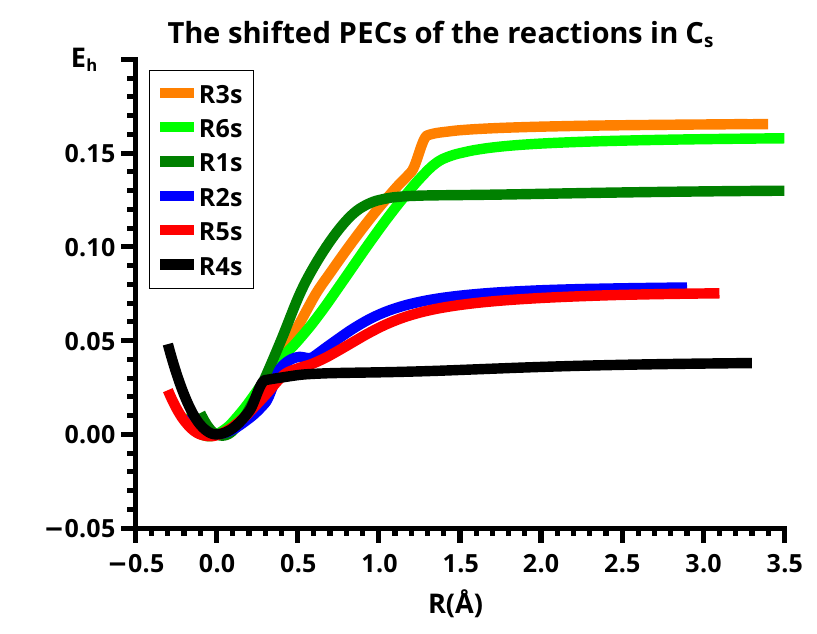}
\caption{Comparison of the PECs of all investigated reactions in $C_{2v}$ symmetry (left), and $C_s$ symmetry (right), all PECs shifted so that the minima coincide and the energy at the minimum is set to zero.}\label{fig:shiftedPECs}
\end{figure}

Although MCSCF reaction energies do not contain the large contributions of dynamic correlation, comparison of the shifted PECs in Figure \ref{fig:shiftedPECs} gives nevertheless a reasonable estimate of the stability of the 3-rings. In $C_{2v}$, \ce{CSi2H6} is most stable with respect to elimination of a carbene analog, followed by \ce{C3H6}, \ce{Si3H6}, and \ce{SiC2H6}, which  is least stable. But one must not forget that the energy at the crossing point is not an adiabatic reaction barrier, it is the energy of the two isoenergetic structures at the bottom of the troughs of the diabatic elimination and recombination reactions. As the PECs show the resistance against elimination of methylene is much higher than against the elimination of silylene. The conventional thermodynamical measure of the stability of the 3-rings is the reaction energy of the elimination reaction with all reactants in their respective ground states. In our calculations, the molecular system is always in the singlet ground state, therefore, some fragments can be in excited states, and the energies must be corrected. See the Supporting Information for the details. According to the corrected $\Delta E_{\rm reac}$ values,  \ce{cC2SiH6} (412.7\,kJ/mol) is most stable against elimination of \ce{CH2} in $C_{2v}$ symmetry, followed by \ce{cC3H6} (300.9\,kJ/mol); the elimination of \ce{SiH2} costs less energy, \ce{cSi3H6} (231.4\,kJ/mol), and  \ce{cSiC2H6} (103.2\,kJ/mol). That \ce{cCSi2H6} is more stable than  \ce{cC3H6} is surprising, that  \ce{cSi3H6}  is less stable than \ce{cC3H6} was expected.

The difference between the crossing point energy and the energy of the stable 3-ring is  more important when the elimination reaction is forced to follow the $C_{2v}$ MEP, for example because of large substituents of the 3-rings that prohibit a deformation of the system towards $C_s$ geometries. Because the elimination of \ce{CH2} from \ce{cC2SiH6} is a diabatic reaction, there ist no crossing point and therefore no estimate of a possible reaction barrier. The largest ``barrier'' is thus found for the methylene elimination from \ce{cC3H6} (492.4\,kJ/mol); for the silylene elimination we find 311.2\,kJ/mo) for  \ce{cSi3H6}, and 273.4\,kJ/mol for \ce{cSiC2H6}. Also for the recombination reaction we get estimates of the barrier, and now the order is different. The largest barrier is found for the addition of methylene to ethene,  \ce{cC3H6} (188.3\,kJ/mol), the for the addition of silylene to ethene, \ce{cSiC2H6} (170.2\,kJ/mol), then follows the addition of methylene to disilene, \ce{cCSi2H6} (101.9\,kJ/mol), and  finally the addition of silylene to disilene, \ce{cSi3H6} (79.8\,kJ/mol).

The PECs for the different system symmetries show that resistance against elimination of a methylene analog is highest in $C_{2v}$ symmetry. Since most stable 3-rings have large ligands, folding down of the carbene analogs is thus hampered and the elimination proceeds along a MEP that is close to the MEP in $C_{2v}$. Also the elimination of carbene analogs from \ce{cCSi2H6} and \ce{cSiC2H6} in reaction \textbf{R5s}  and \textbf{R6s} will proceed along a MEP that is close to the MEPs in $C_{2v}$ when the 3-rings have large substituents, and there will be reaction barriers considerably above the reaction energies.
These results are relevant for all reactions, where steric factors force the elimination of the carbene analog to proceed in high $C_{2v}$ symmetry. Without such steric hindrances the reactions can proceed in $C_s$ symmetry and the high barriers disappear.

With respect to the stability of \ce{cSi3H6} one can say that there are no intrinsic reasons for a thermodynamic instability; that cyclotrisilanes could not be made for a long time had only kinetic reasons.

To find out why a molecule is stable or unstable, it is not sufficient just to look just at the energetics of a reaction, a reliable description of a  reaction must consider both the physical interactions in a molecular system and all consequences of the fermionic character of electrons. Strange enough, although it is well known that the latter is of utmost importance for the description of many-electron systems, the role of charge and therefore the Coulomb interaction dominates too many discussions and explanations of chemical reactions and molecular structure. One can only speculate about the reasons for this, one possible reason could be that MO based wave functions use eigenfunctions of an energy operator, the Fock operator; or it is simply the fact that arguments regarding energy or charge of electrons are more comprehensible than those regarding the spin. It is, nevertheless, sheer ignorance of what Lennard-Jones said seventy years ago about the importance of the Pauli exclusion principle for molecular structure and of the reactions leading to it: ''\dots a property which holds for all electronic systems, whether they are atoms, molecules or solids: \emph{Electrons of like spins tend to avoid each other}. This effect is most powerful, much more powerful than that of electrostatic forces. It does more to determine the shapes and properties of molecules than any other single factor. It is the exclusion principle which plays the dominant role in chemistry.''\cite{Lennard1954} And L\'evy-Leblond and Balibar said that the exclusion principle, ``plays the role of a fictitious, although highly effective, mutual repulsion being exerted within the system, irrespective of any other actual forces or interactions [...] that might be present.''\cite{Levy1990} The fictitious short-range repulsion, also called ``Pauli repulsion'', was addressed by Daudel \etal, who simply stated: ``This shows that the `Pauli repulsion' between electrons possessing the same spin is very significant. In fact, this \emph{repulsion is the main origin of bond angles}.\cite{Daudel1984} Because the ``Pauli repulsion'' applies to identical spins and because it is short-range, it concerns mainly electrons with a  high-spin arrangement localized on an atom. Any spin flip in an electron distribution that increases the number of identical spins has immediate consequences not only for the angular structure of the atom, also bond lengths are affected. In a composite system, any change of the spin arrangement in one subsystem affects the spin arrangement in another, spatially close subsystem. Monitoring the geometry changes during a chemical reaction allows often to conjecture which spin arrangement was changed in which subsystem. The full information about the change of the molecular structure plus possible information about local spin and charge arrangements is stored in the MEP, that is the set of the $n$ internal coordinates $q_i(R)$ as a function of the approximate reaction coordinate. If, for whatever reason, some internal coordinates are not optimized, in technical parlance, if they are frozen, one obtains an approximate MEP, which can be simply wrong. If a change of the spin arrangement causes a strong change of some internal coordinates, and if these coordinates are frozen, the reaction is erroneously  described, instead of a correct diabatic PEC one gets, for example, a pseudo-adiabatic one.
But even if all geometry parameters are fully optimized, and the information delivered by them is not properly interpreted, one can miss important informations, such as the  existence of different MEPs and jumps but also the absence of jumps between the corresponding reaction valleys. Reactions in different valleys are described by wave functions with different essential characteristics, representing different local charge and spin distributions, which are hardly seen when wave functions based on delocalized MOs are used, but which can easily be seen when localized MOs, for example FMOs, are used.

In $C_{2v}$ symmetry, elimination reaction and recombination reaction occur in (at least) two different reaction valleys with corresponding MEPs; one valley describes the stretching of two single bonds in the 3-ring, the second describes the compression of the ethene and the carbene analogs. The wave function describing these reactions have different essential characteristics. A reaction following a third MEP
is the recombination reaction to \ce{cCSi2H6} where after the crossing point, when the wave function describes already fragments with unpaired electrons, the total energy shows another jump. This jump is caused by the double umbrella inversion, which does not change the spin distribution but the shape of the $\pi$ FMOs. Only this change of the orbital shape makes the formation of the new covalent bonds possible. Wave functions made with such different FMOs have again different essential characteristics.

The metaphoric description of the change of the MEP during a reaction as jump between the troughs of the reaction valleys has its origin in the neglect of nuclear motions in the Born-Oppenheimer approximation; indeed the elementary processes should be described by an electron-phonon model, which can explain the coupling of the electronic states in the two different regions in the configuration space, and how the system can overcome the barrier between the trough. Note that even an electron-phonon model would be rather approximate, because it demands that only vibrations are allowed that retain the $C_{2v}$ symmetry.

All reactions in this study that occur in $C_{2v}$ are orbital symmetry forbidden according to the Woodward-Hoffmann rules. In the dissociated system, each reactant has a doubly occupied $a_1$ HOMO and an empty $b_1$ LUMO, for each symmetry there are one positive and one negative linear combinations with  bonding and antibonding character, respectively. In the dissociated system, the positive and the negative linear combination of the $a_1$ HOMOs are doubly occupied,  the  two linear combinations of the $b_1$ LUMOs are unoccupied. In the 3-ring,  the two $\sigma$ bonds between the fragments are described by the doubly occupied bonding $a_1$ and the bonding $b_1$ MOs, the antibonding $a_1$ and $b_1$ MOs are unoccupied. This implies that the  occupied bonding  $a_1$ orbitals in the 3-ring and the dissociated system are correlated, as are the unoccupied antibonding $b_1$ orbitals; but the occupied bonding $b_1$ orbital in the 3-ring correlates with the unoccupied bonding $b_1$ linear combination in the dissociated system, and the unoccupied antibonding $a_1$ orbital in the 3-ring correlates with the occupied antibonding $a_1$ linear combination in the dissociated system. This shows that the orbital symmetry of the occupied and the unoccupied orbitals is not conserved, and, therefore, the reaction is forbidden. If, however, the system adopts the lower $C_s$ point group symmetry, all occupied and all unoccupied orbitals belong to the same $a'$ IRREP, the orbital symmetry is conserved, and the reactions is orbital symmetry allowed. With the help of the Woodward-Hoffmann rules it is possible to understand, in which point group symmetry of the system  a certain reaction is allowed and in which it is forbidden, but it does not give a physical explanation of the local processes involved in the making or breaking of covalent bonds. This can only be done when the so called static correlation is accounted for using proper multi-configurational wave functions from methods like CI (configuration interaction), MCSCF, or CC (coupled cluster), which are based on delocalized MOs. Such MOs lack, in general, the ability to describe local effects and processes. VB (valence bond) wave functions, which  are constructed with atom centered orbitals, and are able to do this. See the excellent book by LucjanPiela for a discussion of the benefits of different methods for the interpretation of reactions using multi-configurational wave functions.\cite{Piela2014}
An early study of the recombination reactions with \ce{cC3H6} was done by Reuter \etal some 30 years ago\cite{Reuter1991}; the increase of the HCH angle to the triplet value was found as was the increase of the C-C length in ethene, but, using MOs that are delocalized over the whole molecular system, an interpretation in terms of change of the fragment multiplicity was not possible. The same was true in the study of the insertion of silylene into the hydrogen molecule by one of the present authors.\cite{Sax1985} The assumption that a key step in the reaction is the simultaneous change of the multiplicities in both fragments was confirmed with an OVB analysis of the CASSCF wave function much later.\cite{Sax2015,Sax2017}

The question that also an OVB analysis cannot answer as yet is: What causes the reactants to change their electron distribution so that they have unpaired electrons, or in other words, what causes the simultaneous spin flip in the reactants by which the geometries change from typical low-spin to typical high-spin geometries? Any attempt to explain this by referring to energies fails, be it orbital energies of delocalized orbitals, as used in Woodward-Hoffmann-type studies, or the lowering the total energy as driving force for the stabilization of  the system. The total energy is the sum of many different contributions, caused by various interactions or just properties of a many fermion system; it is the most important scalar physical quantity that measures the result of the interactions, but it is not the cause of a system change. After all, in many reactions there is first an energy increase, before, after some changes of the structure, the system develops into a thermodynamically stabilized system. Metaphorically speaking, the system had to know in advance, that after surmounting barriers it is on the way leading to the desired stable molecular structure. Such types of argumentation have the odor of quantum teleology. So, what triggers a system, when the reactants have reached a certain distance, to change simultaneously the spin distribution that equips  the reactants with the unpaired electrons needed for covalent bonding? We know, that the PEP plays the dominant role for the molecular structure, with quantum theoretical methods one can monitor the structural changes during a reaction; the OVB results show that, at certain distances between the reactants, the reactants undergo simultaneously changes in their spin distribution, and this is connected with the fermionic character of the electrons. Currently, no mathematical description exists for the physical causes of such changes. In our opinion, it must reflect the significance of the fermionic character of electrons in general, and the role of the PEP in particular, which are  indisputable for any correct description of processes in molecular systems. Or, to quote again L\'evy-Leblond and Balibar: ``The fermionic nature of electrons plays an absolutely essential role and the Pauli exclusion principle reigns supreme, \'without which things would never be what they are.\' ''\cite{Levy1990}

\section{Method}
For all calculations we used CAS(4,4) wave functions, the comparative calculations with CAS(6,6) wave functions were only done for two systems; all non-valence MOs were kept frozen. The geometry optimizations were done with delocalized MOs, for the subsequent OVB analysis localized FMOs were used.  All calculations were done with a local version of GAMESS.\cite{Gamess}
The 6-311G(2d) basis set was used throughout. For the calculation of the PECs, the approximate reaction coordinate $R$, was incremented in steps of 0.1\,\AA; all other geometry parameters were optimized. In $C_s$ and $C_{2v}$ symmetry, 11 and 6 degrees of freedom must be  optimized, respectively. The elimination reactions started with MOs optimized at the equilibrium geometry of the 3-rings, as active MOs the two bonding and the corresponding antibonding MOs were chosen that describe the two $\sigma$ bonds that are broken during the elimination reaction; the active MOs for the recombination reactions were the $\pi$ and $\pi^*$ MOs of the ethene analogs and the s-type and p-type lone pair AOs of the carbene analogs calculated for the molecular system with the fragments separated by  $R=5.0$\,\AA. For the calculation of the PECs following strategy was used: take the system structure optimized for a fixed $R$ value; increment the $R$ value; use the new geometry as starting point for the next constrained optimization. Similarly, the MOs optimized for the structure at $R$ were  used as initial MOs for the energy calculation at the incremented geometry. With this strategy,  the electron structure along the approximate reaction coordinate was conserved.

For each optimized structure, the optimized CASSCF MOs were localized on the fragments, using an orthogonal Procrustes transformation.\cite{Sax2012} In brief: Assuming that the electron structure of the reactants changes slowly during a reaction, the fragment MOs of the non-interacting reactants at the optimized structure at $R$ must be contained in the delocalizes MOs of the molecular system. So take the geometry of the fragments in the optimized geometry of the molecular system and calculate for each fragment simple fragment MOs by a cheap SCF method (RHF or UHF), fragment MOs of different fragments are not orthogonal to each other. The overlap matrix $S_{ni}$ of these fragment MOs contains all information about the non-interacting but deformed fragments in the interacting molecular system. The optimized CASSCF MOs contain all information on the electron structure of the interacting system. By an orthogonal transformation of the MOs one can localize the MOs on the fragments, the localized fragment MOs (FMO) form an orthogonal set. The overlap between the non-orthogonal and the orthogonal FMOs  be matrix $S$. The orthogonal transformation is obtained by demanding that the norm of the matrix difference $S_{ni} - S$ is a small as possible. This procedure is the Procrustes transformation. As a result, the doubly occupied non-active MOs are transformed into doubly occupied fragment MOs (FMO); active MOs are transformed into active FMOs, which resemble either AOs or hybrid AOs of the carbene analogs, or the $\pi$ and $\pi^*$ orbitals of the ethene analogs. CSFs constructed with these FMOs are dubbed OVB CSFs. Finally, the CI matrix is set up with the OVB CSFs and diagonalized. The diagonal elements of the CI matrix are the CSF energies; the lowest eigenvalue of the CI matrix must be identical with the CAS(4,4) energy obtained with delocalized MOs; the squared CI coefficients are the weights for the OVB CSFs.

\bibliography{Threerings}

\begin{thebibliography}{36}%
\makeatletter
\providecommand \@ifxundefined [1]{%
 \@ifx{#1\undefined}
}%
\providecommand \@ifnum [1]{%
 \ifnum #1\expandafter \@firstoftwo
 \else \expandafter \@secondoftwo
 \fi
}%
\providecommand \@ifx [1]{%
 \ifx #1\expandafter \@firstoftwo
 \else \expandafter \@secondoftwo
 \fi
}%
\providecommand \natexlab [1]{#1}%
\providecommand \enquote  [1]{``#1''}%
\providecommand \bibnamefont  [1]{#1}%
\providecommand \bibfnamefont [1]{#1}%
\providecommand \citenamefont [1]{#1}%
\providecommand \href@noop [0]{\@secondoftwo}%
\providecommand \href [0]{\begingroup \@sanitize@url \@href}%
\providecommand \@href[1]{\@@startlink{#1}\@@href}%
\providecommand \@@href[1]{\endgroup#1\@@endlink}%
\providecommand \@sanitize@url [0]{\catcode `\\12\catcode `\$12\catcode
  `\&12\catcode `\#12\catcode `\^12\catcode `\_12\catcode `\%12\relax}%
\providecommand \@@startlink[1]{}%
\providecommand \@@endlink[0]{}%
\providecommand \url  [0]{\begingroup\@sanitize@url \@url }%
\providecommand \@url [1]{\endgroup\@href {#1}{\urlprefix }}%
\providecommand \urlprefix  [0]{URL }%
\providecommand \Eprint [0]{\href }%
\providecommand \doibase [0]{https://doi.org/}%
\providecommand \selectlanguage [0]{\@gobble}%
\providecommand \bibinfo  [0]{\@secondoftwo}%
\providecommand \bibfield  [0]{\@secondoftwo}%
\providecommand \translation [1]{[#1]}%
\providecommand \BibitemOpen [0]{}%
\providecommand \bibitemStop [0]{}%
\providecommand \bibitemNoStop [0]{.\EOS\space}%
\providecommand \EOS [0]{\spacefactor3000\relax}%
\providecommand \BibitemShut  [1]{\csname bibitem#1\endcsname}%
\let\auto@bib@innerbib\@empty
\bibitem [{\citenamefont {West}, \citenamefont {Fink},\ and\ \citenamefont
  {Michl}(1981)}]{West1981}%
  \BibitemOpen
  \bibfield  {author} {\bibinfo {author} {\bibfnamefont {R.}~\bibnamefont
  {West}}, \bibinfo {author} {\bibfnamefont {M.~J.}\ \bibnamefont {Fink}},\
  and\ \bibinfo {author} {\bibfnamefont {J.}~\bibnamefont {Michl}},\ }\bibfield
   {title} {\enquote {\bibinfo {title} {{Tetramesityldisilene, a Stable
  Compound Containing a Silicon-Silicon Double Bond}},}\ }\href@noop {}
  {\bibfield  {journal} {\bibinfo  {journal} {Science}\ }\textbf {\bibinfo
  {volume} {214}},\ \bibinfo {pages} {1343--1344} (\bibinfo {year}
  {1981})}\BibitemShut {NoStop}%
\bibitem [{\citenamefont {Masamune}\ \emph {et~al.}(1982)\citenamefont
  {Masamune}, \citenamefont {Murakami}, \citenamefont {Blount},\ and\
  \citenamefont {Bally}}]{Masamune1982}%
  \BibitemOpen
  \bibfield  {author} {\bibinfo {author} {\bibfnamefont {S.}~\bibnamefont
  {Masamune}}, \bibinfo {author} {\bibfnamefont {S.}~\bibnamefont {Murakami}},
  \bibinfo {author} {\bibfnamefont {J.~F.}\ \bibnamefont {Blount}},\ and\
  \bibinfo {author} {\bibfnamefont {T.}~\bibnamefont {Bally}},\ }\bibfield
  {title} {\enquote {\bibinfo {title} {{Cyclotrisilane (R$_2$Si)$_3$ and
  Disilene (R$_2$Si=SiR$_2$) Systems: Synthesis and Characterization}},}\
  }\href {https://doi.org/10.1021/ja00368a065} {\bibfield  {journal} {\bibinfo
  {journal} {Journal of the American Chemical Society}\ }\textbf {\bibinfo
  {volume} {104}},\ \bibinfo {pages} {1150--1153} (\bibinfo {year}
  {1982})}\BibitemShut {NoStop}%
\bibitem [{\citenamefont {Masamune}, \citenamefont {Tobita},\ and\
  \citenamefont {Murakami}(1983)}]{Masamune1983}%
  \BibitemOpen
  \bibfield  {author} {\bibinfo {author} {\bibfnamefont {S.}~\bibnamefont
  {Masamune}}, \bibinfo {author} {\bibfnamefont {H.}~\bibnamefont {Tobita}},\
  and\ \bibinfo {author} {\bibfnamefont {S.}~\bibnamefont {Murakami}},\
  }\bibfield  {title} {\enquote {\bibinfo {title} {{Hexaalkylcyclotrisilanes
  (R$_2$Si)$_3$: Hexakis(1-ethylpropy1) and Hexaisopropyl Derivatives}},}\
  }\href {https://doi.org/10.1021/ja00359a046} {\bibfield  {journal} {\bibinfo
  {journal} {Journal of the American Chemical Society}\ }\textbf {\bibinfo
  {volume} {105}},\ \bibinfo {pages} {6524--6525} (\bibinfo {year}
  {1983})}\BibitemShut {NoStop}%
\bibitem [{\citenamefont {Dasent}(1965)}]{Dasent1965}%
  \BibitemOpen
  \bibfield  {author} {\bibinfo {author} {\bibfnamefont {W.}~\bibnamefont
  {Dasent}},\ }\href@noop {} {\emph {\bibinfo {title} {Nonexistent Compounds:
  Compounds of Low Stability.}}}\ (\bibinfo  {publisher} {Marcel Dekker},\
  \bibinfo {address} {New York},\ \bibinfo {year} {1965})\BibitemShut {NoStop}%
\bibitem [{Sys()}]{System}%
  \BibitemOpen
  \href@noop {} {}\bibinfo {note} {In science, a system is defined by a
  scientists by stating of what it is composed and which properties it has. If
  a subset of the system has all properties a system has, it constitutes a
  subsystem of the system. A set of educts in a chemical reaction can be
  considered as a set of non-interacting subsystems of a molecular system; if
  the reaction yields a single product molecule, this molecule is the molecular
  system having several strongly interacting susbsystems called its fragments.
  Alternatively, one could think of the educt molecules as a set of systems and
  the molecular system as a large supersystem composed of them.}\BibitemShut
  {Stop}%
\bibitem [{\citenamefont {Preuss}(1963)}]{Preuss1963}%
  \BibitemOpen
  \bibfield  {author} {\bibinfo {author} {\bibfnamefont {H.-W.}\ \bibnamefont
  {Preuss}},\ }\href@noop {} {\emph {\bibinfo {title} {Quantentheoretische
  Chemie I}}}\ (\bibinfo  {publisher} {Bibliographisches Institut},\ \bibinfo
  {address} {Mannheim, Z\"urich},\ \bibinfo {year} {1963})\BibitemShut
  {NoStop}%
\bibitem [{\citenamefont {Sax}(2015)}]{Sax2015}%
  \BibitemOpen
  \bibfield  {author} {\bibinfo {author} {\bibfnamefont {A.~F.}\ \bibnamefont
  {Sax}},\ }\bibfield  {title} {\enquote {\bibinfo {title} {{Chemical Bonding:
  The Orthogonal Valence-Bond View}},}\ }\href@noop {} {\bibfield  {journal}
  {\bibinfo  {journal} {International Journal of Molecular Sciences}\ }\textbf
  {\bibinfo {volume} {16}},\ \bibinfo {pages} {8896--8933} (\bibinfo {year}
  {2015})}\BibitemShut {NoStop}%
\bibitem [{\citenamefont {Weinbaum}(1933)}]{Weinbaum1933}%
  \BibitemOpen
  \bibfield  {author} {\bibinfo {author} {\bibfnamefont {S.}~\bibnamefont
  {Weinbaum}},\ }\bibfield  {title} {\enquote {\bibinfo {title} {{The Normal
  State of the Hydrogen Molecule}},}\ }\href@noop {} {\bibfield  {journal}
  {\bibinfo  {journal} {J. Chem.Phys.}\ }\textbf {\bibinfo {volume} {1}},\
  \bibinfo {pages} {593--596} (\bibinfo {year} {1933})}\BibitemShut {NoStop}%
\bibitem [{\citenamefont {Kollmar}\ and\ \citenamefont
  {Staemmler}(1979)}]{Kollmar1979}%
  \BibitemOpen
  \bibfield  {author} {\bibinfo {author} {\bibfnamefont {H.}~\bibnamefont
  {Kollmar}}\ and\ \bibinfo {author} {\bibfnamefont {V.}~\bibnamefont
  {Staemmler}},\ }\bibfield  {title} {\enquote {\bibinfo {title} {{Ab Initio
  Calculations of the Potential Energy Surface of the Reaction of Singlet
  Methylene with the Hydrogen Molecule.}}}\ }\href@noop {} {\bibfield
  {journal} {\bibinfo  {journal} {Theoret. Chim. Acta}\ }\textbf {\bibinfo
  {volume} {54}},\ \bibinfo {pages} {207--217} (\bibinfo {year}
  {1979})}\BibitemShut {NoStop}%
\bibitem [{\citenamefont {Gordon}\ and\ \citenamefont
  {Gano}(1984)}]{Gordon1984}%
  \BibitemOpen
  \bibfield  {author} {\bibinfo {author} {\bibfnamefont {M.~S.}\ \bibnamefont
  {Gordon}}\ and\ \bibinfo {author} {\bibfnamefont {D.~R.}\ \bibnamefont
  {Gano}},\ }\bibfield  {title} {\enquote {\bibinfo {title} {Ab initio study of
  the insertions of methylene and silylene into methane, silane, and
  hydrogen},}\ }\href@noop {} {\bibfield  {journal} {\bibinfo  {journal}
  {Journal of the American Chemical Society}\ }\textbf {\bibinfo {volume}
  {106}},\ \bibinfo {pages} {5421--5425} (\bibinfo {year} {1984})}\BibitemShut
  {NoStop}%
\bibitem [{\citenamefont {Sosa}\ and\ \citenamefont
  {Schlegel}(1984)}]{Sosa1984}%
  \BibitemOpen
  \bibfield  {author} {\bibinfo {author} {\bibfnamefont {C.}~\bibnamefont
  {Sosa}}\ and\ \bibinfo {author} {\bibfnamefont {H.~B.}\ \bibnamefont
  {Schlegel}},\ }\bibfield  {title} {\enquote {\bibinfo {title} {Carbene and
  silylene insertion reactions. ab initio calculations on the effects of
  fluorine substitution},}\ }\href@noop {} {\bibfield  {journal} {\bibinfo
  {journal} {Journal of the American Chemical Society}\ }\textbf {\bibinfo
  {volume} {106}},\ \bibinfo {pages} {5847--5852} (\bibinfo {year}
  {1984})}\BibitemShut {NoStop}%
\bibitem [{\citenamefont {Sax}\ and\ \citenamefont {Olbrich}(1985)}]{Sax1985}%
  \BibitemOpen
  \bibfield  {author} {\bibinfo {author} {\bibfnamefont {A.}~\bibnamefont
  {Sax}}\ and\ \bibinfo {author} {\bibfnamefont {G.}~\bibnamefont {Olbrich}},\
  }\bibfield  {title} {\enquote {\bibinfo {title} {{Ab Initio Study of the
  Insertion Reaction $\rm SiH_2(^1A_1)+H_2 \rightarrow SiH_4$}},}\ }\href@noop
  {} {\bibfield  {journal} {\bibinfo  {journal} {J. Am. Chem. Soc.}\ }\textbf
  {\bibinfo {volume} {107}},\ \bibinfo {pages} {4868--4874} (\bibinfo {year}
  {1985})}\BibitemShut {NoStop}%
\bibitem [{\citenamefont {Sax}(2017)}]{Sax2017}%
  \BibitemOpen
  \bibfield  {author} {\bibinfo {author} {\bibfnamefont {A.~F.}\ \bibnamefont
  {Sax}},\ }\bibfield  {title} {\enquote {\bibinfo {title} {{OVB analysis of
  symmetry allowed and symmetry forbidden chemical reactions}},}\ }\href
  {https://doi.org/10.1016/j.comptc.2017.02.009} {\bibfield  {journal}
  {\bibinfo  {journal} {Computational and Theoretical Chemistry}\ }\textbf
  {\bibinfo {volume} {1116}},\ \bibinfo {pages} {117--133} (\bibinfo {year}
  {2017})}\BibitemShut {NoStop}%
\bibitem [{\citenamefont {Sax}(2022)}]{Sax2022}%
  \BibitemOpen
  \bibfield  {author} {\bibinfo {author} {\bibfnamefont {A.~F.}\ \bibnamefont
  {Sax}},\ }\href@noop {} {\enquote {\bibinfo {title} {Chemical bonding in many
  electron molecules},}\ } (\bibinfo {year} {2022}),\ \Eprint
  {https://arxiv.org/abs/2205.01078} {arXiv:2205.01078 [physics.chem-ph]}
  \BibitemShut {NoStop}%
\bibitem [{Ten()}]{Tensor}%
  \BibitemOpen
  \href@noop {} {}\bibinfo {note} {Given a state composed of $N$ quantum
  particles, which can be in individual states $\ket{i}$, such as position
  states $\ket{\mbr_i}$, spin states $\ket{\sigma_i}$, energy states
  $\ket{\epsilon_i}$, or tensor products of them, e.g.,
  $\ket{\mbr}\otimes\ket{\sigma}$. Any state $\ket{\Psi}$ of the $N$-particle
  state can be represented by the quantum amplitude (projection amplitude,
  probability amplitude) $\overl{1,2,\cdots,N}{\Psi}$, the square of the
  modulus of the amplitude is called the transition probability or projection
  probability. The ket $\ket{1,2,\cdots,N}$ is the tensor product
  $\ket{1}\otimes\ket{2}\otimes\cdots\otimes\ket{N}$. If the individual states
  are position states, the position representation of $\ket{\Psi}$ is given by
  the quantum amplitude $\overl{\mbr_1,\mbr_2,\cdots,\mbr_N}{\Psi}$, which is
  called the wave function of state $\Psi$, it is written as
  $\Psi(\mbr_1,\mbr_2,\cdots,\mbr_N)$. Using sloppy language, the term ``wave
  function'' is used to describe quantum amplitudes irrespective of the
  representation. Individual energy states $\ket{\epsilon}$ are called (in
  chemistry) orbitals, the product $\ket{\epsilon}\otimes\ket{\sigma}$ is
  called a spin-orbital.}\BibitemShut {Stop}%
\bibitem [{PCb()}]{PCbook}%
  \BibitemOpen
  \href@noop {} {}\bibinfo {note} {See any good textbook on Physical Chemistry,
  for example "Physical Chemistry, A molecular approach", McQuarrie, D. A. and
  Simon, J. D., University Science Books, 1997}\BibitemShut {NoStop}%
\bibitem [{\citenamefont {Cohen-Tannoudji}, \citenamefont {Diu},\ and\
  \citenamefont {Lalo\"e}(1977)}]{Cohen1977}%
  \BibitemOpen
  \bibfield  {author} {\bibinfo {author} {\bibfnamefont {C.}~\bibnamefont
  {Cohen-Tannoudji}}, \bibinfo {author} {\bibfnamefont {B.}~\bibnamefont
  {Diu}},\ and\ \bibinfo {author} {\bibfnamefont {F.}~\bibnamefont {Lalo\"e}},\
  }\href@noop {} {\emph {\bibinfo {title} {{Quantum Mechanics, vol. 1}}}}\
  (\bibinfo  {publisher} {Hermann, John Wiley},\ \bibinfo {address} {Paris,
  France, New York, USA},\ \bibinfo {year} {1977})\BibitemShut {NoStop}%
\bibitem [{\citenamefont {Pauncz}(1979)}]{Pauncz1979}%
  \BibitemOpen
  \bibfield  {author} {\bibinfo {author} {\bibfnamefont {R.}~\bibnamefont
  {Pauncz}},\ }\href@noop {} {\emph {\bibinfo {title} {{Spin
  Eigenfunctions}}}}\ (\bibinfo  {publisher} {Plenum Press},\ \bibinfo
  {address} {New York, USA},\ \bibinfo {year} {1979})\BibitemShut {NoStop}%
\bibitem [{\citenamefont {Sax}(2012)}]{Sax2012}%
  \BibitemOpen
  \bibfield  {author} {\bibinfo {author} {\bibfnamefont {A.~F.}\ \bibnamefont
  {Sax}},\ }\bibfield  {title} {\enquote {\bibinfo {title} {{Localization of
  molecular orbitals on fragments}},}\ }\href@noop {} {\bibfield  {journal}
  {\bibinfo  {journal} {J. Comp. Chem.}\ }\textbf {\bibinfo {volume} {33}},\
  \bibinfo {pages} {495--1510} (\bibinfo {year} {2012})}\BibitemShut {NoStop}%
\bibitem [{\citenamefont {Sax}(2023)}]{Sax2023}%
  \BibitemOpen
  \bibfield  {author} {\bibinfo {author} {\bibfnamefont {A.~F.}\ \bibnamefont
  {Sax}},\ }\bibfield  {title} {\enquote {\bibinfo {title} {{Chemical Bonding
  in the C$_2$ Molecule}},}\ }\href@noop {} {\bibfield  {journal} {\bibinfo
  {journal} {Inorganics}\ }\textbf {\bibinfo {volume} {11}},\ \bibinfo {pages}
  {245} (\bibinfo {year} {2023})}\BibitemShut {NoStop}%
\bibitem [{\citenamefont {Sutcliffe}\ and\ \citenamefont
  {Woolley}(2012)}]{Sutcliffe2012}%
  \BibitemOpen
  \bibfield  {author} {\bibinfo {author} {\bibfnamefont {B.~T.}\ \bibnamefont
  {Sutcliffe}}\ and\ \bibinfo {author} {\bibfnamefont {R.~G.}\ \bibnamefont
  {Woolley}},\ }\bibfield  {title} {\enquote {\bibinfo {title} {{On the quantum
  theory of molecules}},}\ }\href {https://doi.org/10.1063/1.4755287}
  {\bibfield  {journal} {\bibinfo  {journal} {Journal of Chemical Physics}\
  }\textbf {\bibinfo {volume} {137}} (\bibinfo {year} {2012}),\
  10.1063/1.4755287},\ \Eprint {https://arxiv.org/abs/1206.4239}
  {arXiv:1206.4239} \BibitemShut {NoStop}%
\bibitem [{\citenamefont {Ruedenberg}\ and\ \citenamefont
  {Atchity}(1993)}]{Ruedenberg1993}%
  \BibitemOpen
  \bibfield  {author} {\bibinfo {author} {\bibfnamefont {K.}~\bibnamefont
  {Ruedenberg}}\ and\ \bibinfo {author} {\bibfnamefont {G.~J.}\ \bibnamefont
  {Atchity}},\ }\bibfield  {title} {\enquote {\bibinfo {title} {{A quantum
  chemical determination of diabatic states}},}\ }\href
  {https://doi.org/10.1063/1.466125} {\bibfield  {journal} {\bibinfo  {journal}
  {The Journal of Chemical Physics}\ }\textbf {\bibinfo {volume} {99}},\
  \bibinfo {pages} {3799--3803} (\bibinfo {year} {1993})}\BibitemShut {NoStop}%
\bibitem [{\citenamefont {Atchity}\ and\ \citenamefont
  {Ruedenberg}(1997)}]{Atchity1997}%
  \BibitemOpen
  \bibfield  {author} {\bibinfo {author} {\bibfnamefont {G.}~\bibnamefont
  {Atchity}}\ and\ \bibinfo {author} {\bibfnamefont {K.}~\bibnamefont
  {Ruedenberg}},\ }\bibfield  {title} {\enquote {\bibinfo {title}
  {{Determination of diabatic states through enforcement of configurational
  uniformity}},}\ }\href@noop {} {\bibfield  {journal} {\bibinfo  {journal}
  {Theor. Chem. Acc.}\ }\textbf {\bibinfo {volume} {97}},\ \bibinfo {pages}
  {47--58} (\bibinfo {year} {1997})}\BibitemShut {NoStop}%
\bibitem [{\citenamefont {Shu}\ \emph {et~al.}(2022)\citenamefont {Shu},
  \citenamefont {Varga}, \citenamefont {Kanchanakungwankul}, \citenamefont
  {Zhang},\ and\ \citenamefont {Truhlar}}]{Shu2022}%
  \BibitemOpen
  \bibfield  {author} {\bibinfo {author} {\bibfnamefont {Y.}~\bibnamefont
  {Shu}}, \bibinfo {author} {\bibfnamefont {Z.}~\bibnamefont {Varga}}, \bibinfo
  {author} {\bibfnamefont {S.}~\bibnamefont {Kanchanakungwankul}}, \bibinfo
  {author} {\bibfnamefont {L.}~\bibnamefont {Zhang}},\ and\ \bibinfo {author}
  {\bibfnamefont {D.~G.}\ \bibnamefont {Truhlar}},\ }\bibfield  {title}
  {\enquote {\bibinfo {title} {{Diabatic States of Molecules}},}\ }\href
  {https://doi.org/10.1021/acs.jpca.1c10583} {\bibfield  {journal} {\bibinfo
  {journal} {Journal of Physical Chemistry A}\ }\textbf {\bibinfo {volume}
  {126}},\ \bibinfo {pages} {992--1018} (\bibinfo {year} {2022})}\BibitemShut
  {NoStop}%
\bibitem [{\citenamefont {Nakamura}\ and\ \citenamefont
  {Truhlar}(2001)}]{Nakamura2001}%
  \BibitemOpen
  \bibfield  {author} {\bibinfo {author} {\bibfnamefont {H.}~\bibnamefont
  {Nakamura}}\ and\ \bibinfo {author} {\bibfnamefont {D.~G.}\ \bibnamefont
  {Truhlar}},\ }\bibfield  {title} {\enquote {\bibinfo {title} {{The direct
  calculation of diabatic states based on configurational uniformity.}}}\
  }\href@noop {} {\bibfield  {journal} {\bibinfo  {journal} {Journal of
  Chemical Physics}\ }\textbf {\bibinfo {volume} {115}},\ \bibinfo {pages}
  {10353} (\bibinfo {year} {2001})}\BibitemShut {NoStop}%
\bibitem [{\citenamefont {Ruedenberg}\ and\ \citenamefont
  {Schmidt}(2007)}]{Ruedenberg2007}%
  \BibitemOpen
  \bibfield  {author} {\bibinfo {author} {\bibfnamefont {K.}~\bibnamefont
  {Ruedenberg}}\ and\ \bibinfo {author} {\bibfnamefont {M.~W.}\ \bibnamefont
  {Schmidt}},\ }\bibfield  {title} {\enquote {\bibinfo {title} {{Why does
  electron sharing lead to covalent bonding? A variational analysis}},}\
  }\href@noop {} {\bibfield  {journal} {\bibinfo  {journal} {J. Comp. Chem.}\
  }\textbf {\bibinfo {volume} {28}},\ \bibinfo {pages} {391} (\bibinfo {year}
  {2007})}\BibitemShut {NoStop}%
\bibitem [{\citenamefont {Bitter}, \citenamefont {Ruedenberg},\ and\
  \citenamefont {Schwarz}(2007)}]{Bitter2007}%
  \BibitemOpen
  \bibfield  {author} {\bibinfo {author} {\bibfnamefont {T.}~\bibnamefont
  {Bitter}}, \bibinfo {author} {\bibfnamefont {K.}~\bibnamefont {Ruedenberg}},\
  and\ \bibinfo {author} {\bibfnamefont {W.~H.~E.}\ \bibnamefont {Schwarz}},\
  }\bibfield  {title} {\enquote {\bibinfo {title} {{Towards a physical
  understanding of electron-sharing two-center bonds. I. General aspects}},}\
  }\href@noop {} {\bibfield  {journal} {\bibinfo  {journal} {J. Comp. Chem.}\
  }\textbf {\bibinfo {volume} {28}},\ \bibinfo {pages} {411} (\bibinfo {year}
  {2007})}\BibitemShut {NoStop}%
\bibitem [{\citenamefont {Ruedenberg}\ and\ \citenamefont
  {Schmidt}(2009)}]{Ruedenberg2009}%
  \BibitemOpen
  \bibfield  {author} {\bibinfo {author} {\bibfnamefont {K.}~\bibnamefont
  {Ruedenberg}}\ and\ \bibinfo {author} {\bibfnamefont {M.~W.}\ \bibnamefont
  {Schmidt}},\ }\bibfield  {title} {\enquote {\bibinfo {title} {{Physical
  Understanding through Variational Reasoning: Electron Sharing and Covalent
  Bonding}},}\ }\href@noop {} {\bibfield  {journal} {\bibinfo  {journal} {J.
  Phys. Chem. A}\ }\textbf {\bibinfo {volume} {113}},\ \bibinfo {pages} {1954}
  (\bibinfo {year} {2009})}\BibitemShut {NoStop}%
\bibitem [{\citenamefont {Bitter}\ \emph {et~al.}(2010)\citenamefont {Bitter},
  \citenamefont {Wang}, \citenamefont {Ruedenberg},\ and\ \citenamefont
  {Schwarz}}]{Bitter2010}%
  \BibitemOpen
  \bibfield  {author} {\bibinfo {author} {\bibfnamefont {T.}~\bibnamefont
  {Bitter}}, \bibinfo {author} {\bibfnamefont {S.~G.}\ \bibnamefont {Wang}},
  \bibinfo {author} {\bibfnamefont {K.}~\bibnamefont {Ruedenberg}},\ and\
  \bibinfo {author} {\bibfnamefont {W.~H.~E.}\ \bibnamefont {Schwarz}},\
  }\bibfield  {title} {\enquote {\bibinfo {title} {{Towards a physical
  understanding of electron-sharing two-center bonds. II. Pseudo-potential
  based analysis of diatomic molecules}},}\ }\href@noop {} {\bibfield
  {journal} {\bibinfo  {journal} {Theor. Chem. Acc.}\ }\textbf {\bibinfo
  {volume} {127}},\ \bibinfo {pages} {237} (\bibinfo {year}
  {2010})}\BibitemShut {NoStop}%
\bibitem [{\citenamefont {Schmidt}, \citenamefont {Ivanic},\ and\ \citenamefont
  {Ruedenberg}(2014)}]{Schmidt2014}%
  \BibitemOpen
  \bibfield  {author} {\bibinfo {author} {\bibfnamefont {M.~W.}\ \bibnamefont
  {Schmidt}}, \bibinfo {author} {\bibfnamefont {J.}~\bibnamefont {Ivanic}},\
  and\ \bibinfo {author} {\bibfnamefont {K.}~\bibnamefont {Ruedenberg}},\
  }\bibfield  {title} {\enquote {\bibinfo {title} {{Covalent bonds are created
  by the drive of electron waves to lower their kinetic energy through
  expansion}},}\ }\href@noop {} {\bibfield  {journal} {\bibinfo  {journal} {J.
  Chem. Phys.}\ }\textbf {\bibinfo {volume} {140}},\ \bibinfo {pages} {1204104}
  (\bibinfo {year} {2014})}\BibitemShut {NoStop}%
\bibitem [{\citenamefont {Lennard-Jones}(1954)}]{Lennard1954}%
  \BibitemOpen
  \bibfield  {author} {\bibinfo {author} {\bibfnamefont {J.}~\bibnamefont
  {Lennard-Jones}},\ }\bibfield  {title} {\enquote {\bibinfo {title} {{New
  Ideas in Chemistry}},}\ }\href@noop {} {\bibfield  {journal} {\bibinfo
  {journal} {Adv. Sci.}\ }\textbf {\bibinfo {volume} {11}},\ \bibinfo {pages}
  {136--148} (\bibinfo {year} {1954})}\BibitemShut {NoStop}%
\bibitem [{\citenamefont {L\'evy-Leblond}\ and\ \citenamefont
  {Balibar}(1990)}]{Levy1990}%
  \BibitemOpen
  \bibfield  {author} {\bibinfo {author} {\bibfnamefont {J.-M.}\ \bibnamefont
  {L\'evy-Leblond}}\ and\ \bibinfo {author} {\bibfnamefont {F.}~\bibnamefont
  {Balibar}},\ }\href@noop {} {\emph {\bibinfo {title} {{Quantics, Rudiments of
  Quantum Physics}}}}\ (\bibinfo  {publisher} {North Holland, Elsevier Science
  Publishers B.V.},\ \bibinfo {address} {Amsterdam, The Netherlands},\ \bibinfo
  {year} {1990})\BibitemShut {NoStop}%
\bibitem [{\citenamefont {Daudel}\ \emph {et~al.}(1984)\citenamefont {Daudel},
  \citenamefont {Leroy}, \citenamefont {Peeters},\ and\ \citenamefont
  {Sana}}]{Daudel1984}%
  \BibitemOpen
  \bibfield  {author} {\bibinfo {author} {\bibfnamefont {R.}~\bibnamefont
  {Daudel}}, \bibinfo {author} {\bibfnamefont {G.}~\bibnamefont {Leroy}},
  \bibinfo {author} {\bibfnamefont {D.}~\bibnamefont {Peeters}},\ and\ \bibinfo
  {author} {\bibfnamefont {M.}~\bibnamefont {Sana}},\ }\href@noop {} {\emph
  {\bibinfo {title} {Quantum Chemistry}}}\ (\bibinfo  {publisher} {{John Wiley
  Sons}},\ \bibinfo {address} {Chichester, New York, Brisbane, Toronto,
  Singapore},\ \bibinfo {year} {1984})\BibitemShut {NoStop}%
\bibitem [{\citenamefont {Piela}(2014)}]{Piela2014}%
  \BibitemOpen
  \bibfield  {author} {\bibinfo {author} {\bibfnamefont {L.}~\bibnamefont
  {Piela}},\ }\href@noop {} {\emph {\bibinfo {title} {{Ideas of Quantum
  Chemistry}}}}\ (\bibinfo  {publisher} {Elsevier},\ \bibinfo {year}
  {2014})\BibitemShut {NoStop}%
\bibitem [{\citenamefont {Reuter}, \citenamefont {Engels},\ and\ \citenamefont
  {Peyerimhoff}(1991)}]{Reuter1991}%
  \BibitemOpen
  \bibfield  {author} {\bibinfo {author} {\bibfnamefont {W.}~\bibnamefont
  {Reuter}}, \bibinfo {author} {\bibfnamefont {B.}~\bibnamefont {Engels}},\
  and\ \bibinfo {author} {\bibfnamefont {S.~D.}\ \bibnamefont {Peyerimhoff}},\
  }\bibfield  {title} {\enquote {\bibinfo {title} {{Reactlon of Slnglet and
  Triplet Methylene wlth Ethane. A Multlreference Conflguratlon I nteractlon
  Study}},}\ }\href@noop {} {\bibfield  {journal} {\bibinfo  {journal} {J.
  Phys. Chem.}\ }\textbf {\bibinfo {volume} {96}},\ \bibinfo {pages}
  {6221--6232} (\bibinfo {year} {1991})}\BibitemShut {NoStop}%
\bibitem [{\citenamefont {Schmidt}\ \emph {et~al.}(1993)\citenamefont
  {Schmidt}, \citenamefont {Baldridge}, \citenamefont {Boatz}, \citenamefont
  {Elbert}, \citenamefont {Gordon}, \citenamefont {Jensen}, \citenamefont
  {Koseki}, \citenamefont {Matsunaga}, \citenamefont {Nguyen}, \citenamefont
  {Su}, \citenamefont {Windus}, \citenamefont {Dupuis},\ and\ \citenamefont
  {Montgomery}}]{Gamess}%
  \BibitemOpen
  \bibfield  {author} {\bibinfo {author} {\bibfnamefont {M.~W.}\ \bibnamefont
  {Schmidt}}, \bibinfo {author} {\bibfnamefont {K.~K.}\ \bibnamefont
  {Baldridge}}, \bibinfo {author} {\bibfnamefont {J.~A.}\ \bibnamefont
  {Boatz}}, \bibinfo {author} {\bibfnamefont {S.~T.}\ \bibnamefont {Elbert}},
  \bibinfo {author} {\bibfnamefont {M.~S.}\ \bibnamefont {Gordon}}, \bibinfo
  {author} {\bibfnamefont {J.~H.}\ \bibnamefont {Jensen}}, \bibinfo {author}
  {\bibfnamefont {S.}~\bibnamefont {Koseki}}, \bibinfo {author} {\bibfnamefont
  {N.}~\bibnamefont {Matsunaga}}, \bibinfo {author} {\bibfnamefont {K.~A.}\
  \bibnamefont {Nguyen}}, \bibinfo {author} {\bibfnamefont {S.~J.}\
  \bibnamefont {Su}}, \bibinfo {author} {\bibfnamefont {T.~L.}\ \bibnamefont
  {Windus}}, \bibinfo {author} {\bibfnamefont {M.}~\bibnamefont {Dupuis}},\
  and\ \bibinfo {author} {\bibfnamefont {J.~A.}\ \bibnamefont {Montgomery}},\
  }\bibfield  {title} {\enquote {\bibinfo {title} {General atomic and molecular
  electronic structure system},}\ }\href@noop {} {\bibfield  {journal}
  {\bibinfo  {journal} {J. Comp. Chem.}\ }\textbf {\bibinfo {volume} {14}},\
  \bibinfo {pages} {1347--1363} (\bibinfo {year} {1993})}\BibitemShut {NoStop}%
\end{thebibliography}%

\pagebreak
\setcounter{equation}{0}
\setcounter{figure}{0}
\setcounter{table}{0}
\setcounter{page}{1}
\makeatletter
\renewcommand{\theequation}{S\arabic{equation}}
\renewcommand{\thefigure}{S\arabic{figure}}
\renewcommand{\thetable}{S\arabic{table}}
\renewcommand{\bibnumfmt}[1]{[S#1]}
\renewcommand{\citenumfont}[1]{S#1}

\section{Supporting information}
\subsection{Parameters of fragments}
The geometry parameters of all fragments and all 3-rings are listed. Unsaturated fragments are calculated with the smallest CASSCF wave function that is necessary for a correct description, all 3-rings are calculated with closed-shell Hartree-Fock wave functions.
The 6-311G(2d) basis set was used throughout.

\subsubsection{Methylene}
The $^3B_1$ ground state and the excited $^1B_1$ state were calculated with single-configurational wave functions, the $^1A_1$ state was calculated with a two-configurational wave function $a|s\bar{s}|+b|p\bar{p}|$. The lowest singlet of linear methylene is the $^1\Delta_g$ state; in $C_{2v}$, one $\Delta$ component has $A_1$ symmetry, the other has $B$ symmetry, in detail: it is $B_1$, if the  molecule axis of linear methylene is the y-axis in $C_{2v}$,  or $B_2$ if the molecules axis is the y-axis. The $A_g$ component of the $^1\Delta_g$ state was calculated with the two-configurational wave function $|\pi_x\overline{\pi}_x|-|\pi_y\overline{\pi}_y|$, the $B$ component with a single-configurational wave function. The geometry parameters of the methylene cation and anion were taken from the literature.

\setlength{\tabcolsep}{2mm}
\begin{table}[h]
\begin{tabular}{ll|l|l|l|l|ll}
System & State & Energy & C-H & HCH & $a^2$ & $b^2$ \\
\toprule
CH$_2$ & $^3B_1$      & -38.9251067 &	1.070 &	128.6 &    \\
CH$_2$ & 1-$^1A_1$    & -38.9086023 &	1.098 &	102.3 & 0.96 & 0.04 \\
CH$_2$ & 2-$^1A_1$    & -38.8512556 &	1.058 &	169.8 & 0.73 & -0.68 \\
CH$_2$ & $^1B_1$      & -38.8591646 &	1.066 &	139.0 &    \\
CH$_2$ & $^1\Delta_g$ & -38.8513871 &   1.056 & 180\\
\midrule
CH$_2^-$ & $^2B_1$    &             &   1.12  & 104\\
\midrule
CH$_2^+$ & $^2A_1$    &             &   1.09  & 139\\
\bottomrule
\end{tabular}
\caption{Methylene in different states, methylene  cation and anion. Energies are in hartrees, lengths in \AA{} and angles in degrees.}
\end{table}

The singlet-triplet splitting $\Delta E_{ST} = E(1-^1A_1)-E(^3B_1) = -0.0165044$ or $-43.3$\,kJ/mol.

\subsubsection{Silylene}
The excited $^3B_1$ state  and the excited $^1B_1$ state were calculated with single-configurational wave functions, the $^1A_1$ ground state was calculated with a two-configurational wave function $a|s\bar{s}|+b|p\bar{p}|$. The geometry parameters of the silylene cation and anion were taken from the literature.

\begin{table}[h]
\begin{tabular}{l|l|l|l|l|ll}
System &State & Energy & Si-H & HSiH & $a^2$ & $b^2$ \\
\toprule
SiH$_2$   & $^1A_1$ & -290.0418089 & 1.513 &	94.2 & 0.95 & 0.05 \\
SiH$_2$   & $^3B_1$ & -290.0117125 & 1.474 &	117.2 \\
SiH$_2$   & $^1B_1$ & -289.9565196 & 1.472 &	122.8 \\
\midrule
SiH$_2^-$ & $^2B_1$ &              & 1.55  &	93.5 \\
SiH$_2^-$ & $^2B_1$ &              & 1.488 & 119.4\\
\midrule
SiH$_2^+$ & $^2B_1$ &              & 1.456 & 180\\
\bottomrule
\end{tabular}
\caption{Silylene in different states, silylene  cation and anion.  Energies are in hartrees, lengths in \AA{} and angles in degrees.}
\end{table}

The singlet-triplet splitting $\Delta E_{ST} = E(^1A_1)-E(^3B_1) = 0.0300964$ or $79.0$\,kJ/mol.

\subsubsection{Ethene}
The singlet ground state has $D_{2h}$ symmetry, the triplet ground state has $D_{2d}$ symmetry; in 3-rings the triplet fragment has $C_{2v}$ symmetry. The singlet state $^1A_{1g}$ is calculated with the CAS(2,2) wave function $a|\pi\overline{\pi}|+b|\pi^*\overline{\pi^*}|$. Angle oop in the Table is the out-of-plane angle of the CH$_2$ group.

\begin{table}[h]
\begin{tabular}{r|r|r|r|r|r|rr}
State & Energy & C=C & C-H & HCH & oop & $a^2$ & $b^2$ \\
\toprule
$^1A_{1g}$ ($D_{2h}$) & -78.0772454 & 1.331 & 1.074 & 116.4 &      & 0.96 & 0.04  \\
$^3B_1$   ($C_{2v}$)  & -77.9529294 & 1.532 & 1.075 & 114.3 & 32.0 \\
\bottomrule
\end{tabular}
\caption{Ethene, ground and excited state. Energies are in hartrees, lengths in \AA, and angles in degrees.}
\end{table}

The singlet-triplet splitting $\Delta E_{ST} = E(^1A_{1g})-E(^3B_1) = 0.124316$ or $326.4$\,kJ/mol.

\subsubsection{Disilene}
The singlet ground state has $C_{2h}$ symmetry, the triplet ground state has $C_2$ symmetry; in 3-rings the triplet fragment has $C_{2v}$ symmetry. The singlet state $^1A_{1g}$ is calculated with the CAS(2,2) wave function $a|\pi\overline{\pi}|+b|\pi^*\overline{\pi^*}|$. Angle oop in the Table is the out-of-plane angle of the SiH$_2$ group.

\begin{threeparttable}[h]
\caption{Disilene, ground and excited state.  Energies are in hartrees, lengths in \AA, and angles in degrees.}
\begin{tabular}{r|r|r|r|r|r|rrr}
State & Energy & Si=Si & Si-H & HSiH & oop & $a^2$ & $b^2$ \\
\toprule
$^1A_{g}$ ($C_{2h}$)               & -580.1458985	& 2.236 & 1.480 & 109.5	& 41.2 & 0.90 & 0.10   \\
$^1A_{1}$ ($C_{2v}$)               & -580.1378171	& 2.160 & 1.469 & 115.0	& 7.6  & 0.95 &	0.05  \\
$^3B_{2}$ ($C_{2v}$)\tn{1}~\qquad  &-580.1127347	& 2.383 & 1.484 & 108.3 & 50.8 \\
\bottomrule
\end{tabular}
 \begin{tablenotes}
 \item[1]Molecule plane is the y-z plane, accordingly has the $\pi^*$ MO $b_2$ symmetry.
 \end{tablenotes}
\end{threeparttable}

The singlet-triplet splitting $\Delta E_{ST} = E(^1A_{g})-E(^3B_2) = 0.0331638$ or $87.1$\,kJ/mol.

The singlet-triplet splitting $\Delta E_{ST} = E(^1A_{1})-E(^3B_2) = 0.0250824$ or $65.9$\,kJ/mol.

\subsubsection{Silaethene}
Only $C_s$ structures are considered. The singlet ground state has a trans-bent structure; the triplet fragment in the 3-rings is cis-bent. The singlet state $^1A$ is calculated with a CAS(2,2) wave function, which is a linear combination of three CSFs, $a|\pi\overline{\pi}|+b|\pi\overline{\pi^*}| + c|\pi^*\overline{\pi^*}|$. Different signs of the oop angles indicate the trans-bent structure.

\begin{table}[h]
\begin{tabular}{r|r|r|r|r|r|r|r|r|rrrr}
State & Energy & C=Si & C-H & Si-H & HCH & HSiH & oop(C) & oop(Si) & $a^2$ & $b^2$ & $c^2$ \\
\toprule
$^1A'$  & -329.0942429 &	1.724 &	1.074 &	1.471 &	114.8 &	113.9 &	11.4 & -20.0 & 0.78 & 0.19 & 0.03\\
$^3A''$  & -329.0402377 &	1.881 &	1.075 &	1.485 &	114.4 &	108.2 & 16.9 &	54.2 \\
\bottomrule
\end{tabular}
\caption{Silaethene, ground and excited state. Energies are in hartrees, lengths in \AA, and angles in degrees.}
\end{table}

The singlet-triplet splitting $\Delta E_{ST} = E(^1A)-E(^3B) = 0.0540052$ or $141.8$\,kJ/mol.

\newpage
\subsection{Comparison of CAS(4,4) and CAS(6,6) results}
The reactions \textbf{R1v} and \textbf{R2v} were studied with CAS(6,6) wave function including the $\sigma$ and $\sigma^*$ MOs of ethene and disilene, respectively. The juxtaposition of the energies and geometry parameters show that CAS(4,4) wave functions are sufficient for a description of these reactions, and, therefore, all reactions studied were investigated with CAS(4,4) wave functions only.

\begin{figure}
\includegraphics[width=0.8\textwidth]{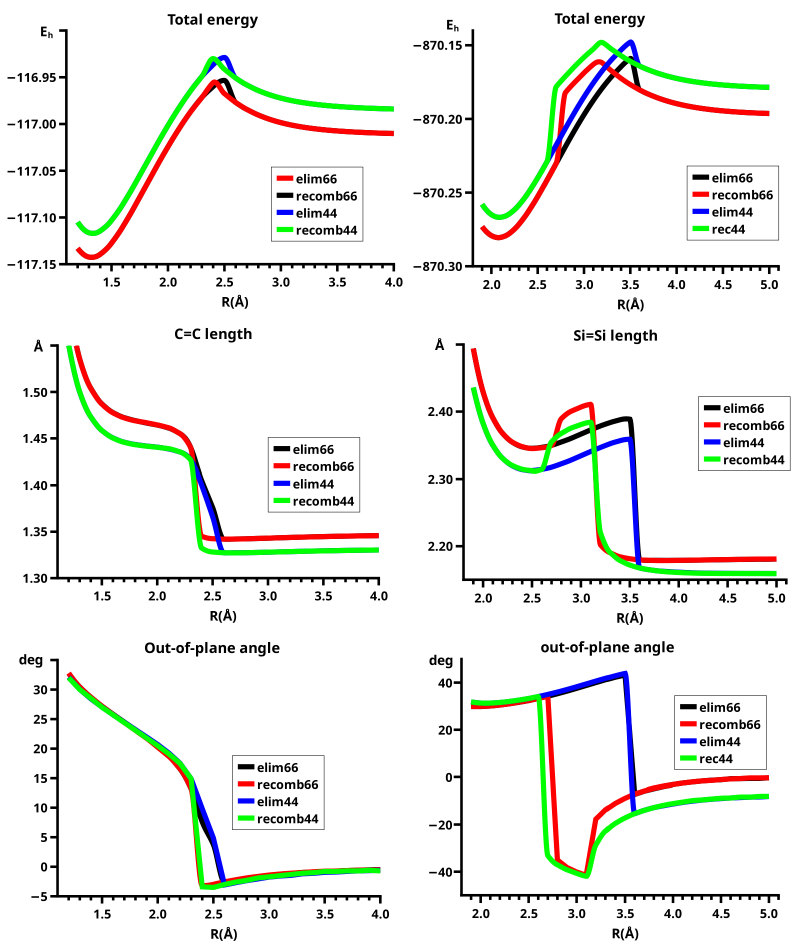}
\end{figure}

\begin{figure}
\includegraphics[width=0.8\textwidth]{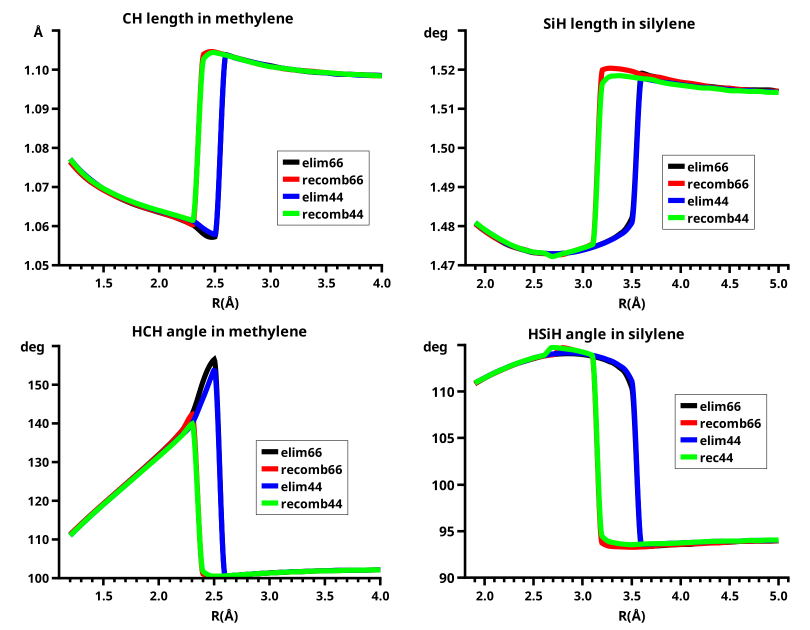}

\includegraphics[width=0.8\textwidth]{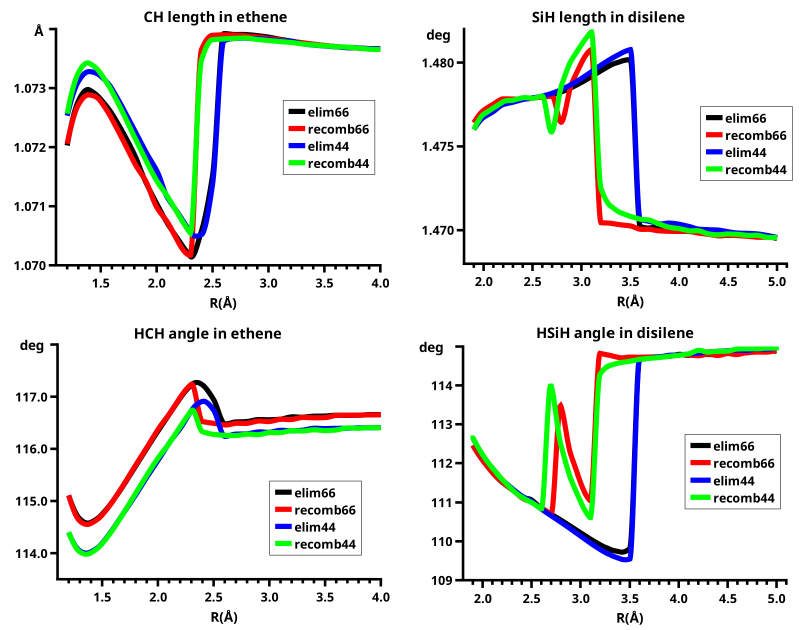}
\caption{Comparison of total energies and geometry parameters of CAS(4,4) and CAS(6,6) calculations of \ce{C3H6} and \ce{Si3H6}.}
\end{figure}

\newpage
\subsection{Graphical representation of physical quantities}
As explained in section IV, curves representing physical quantities along different MEPs should be represented by discontinuous curves. The more convenient and less cumbersome representation by continuous curves is indeed not correct.
As an example, the corresponding curves for the recombination reaction of CSi$_2$H$_6$ in $C_{2v}$ symmetry are shown.

\begin{figure}
\includegraphics[width=0.32\textwidth]{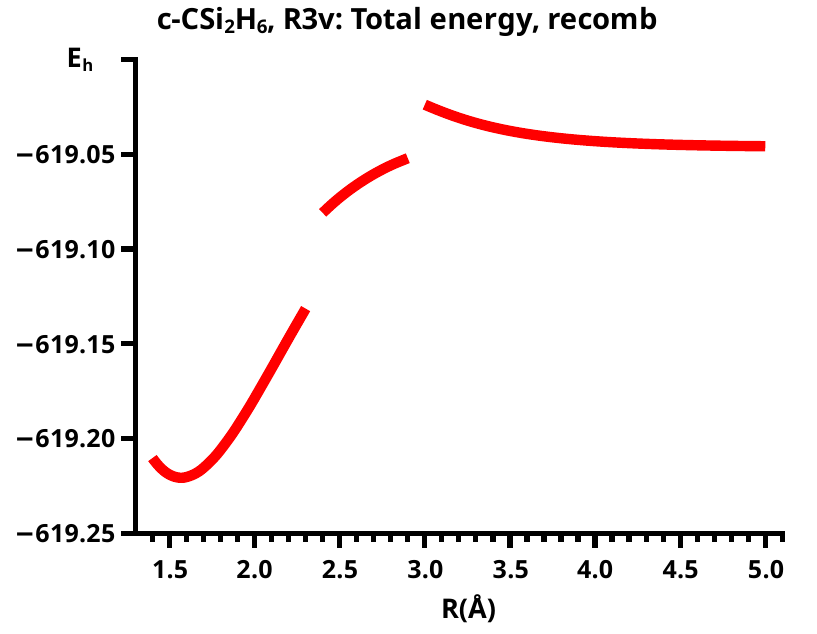}
\includegraphics[width=0.32\textwidth]{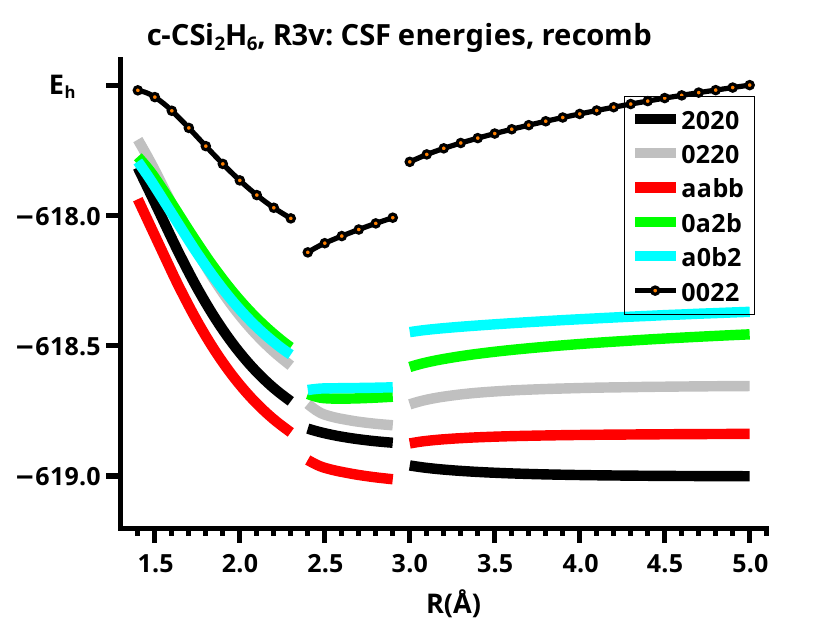}
\includegraphics[width=0.32\textwidth]{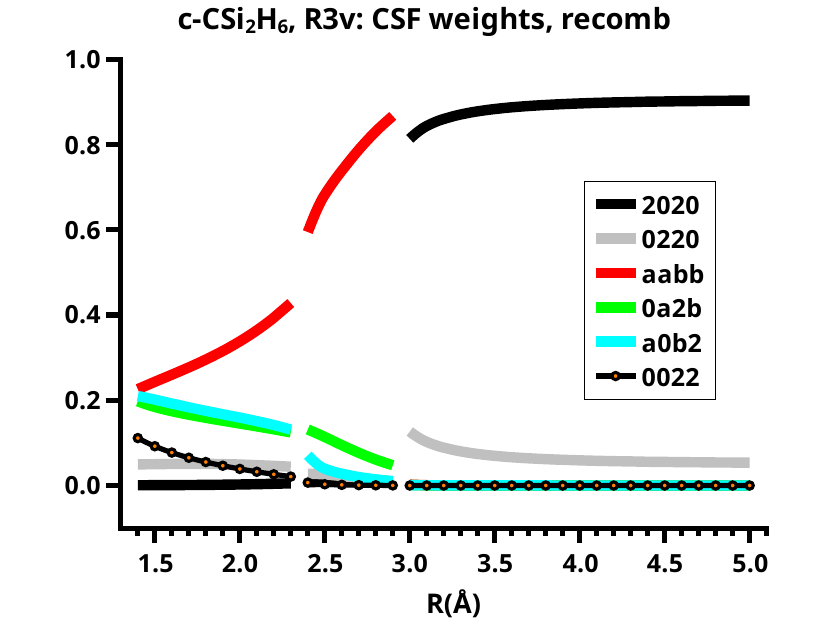}

\includegraphics[width=0.32\textwidth]{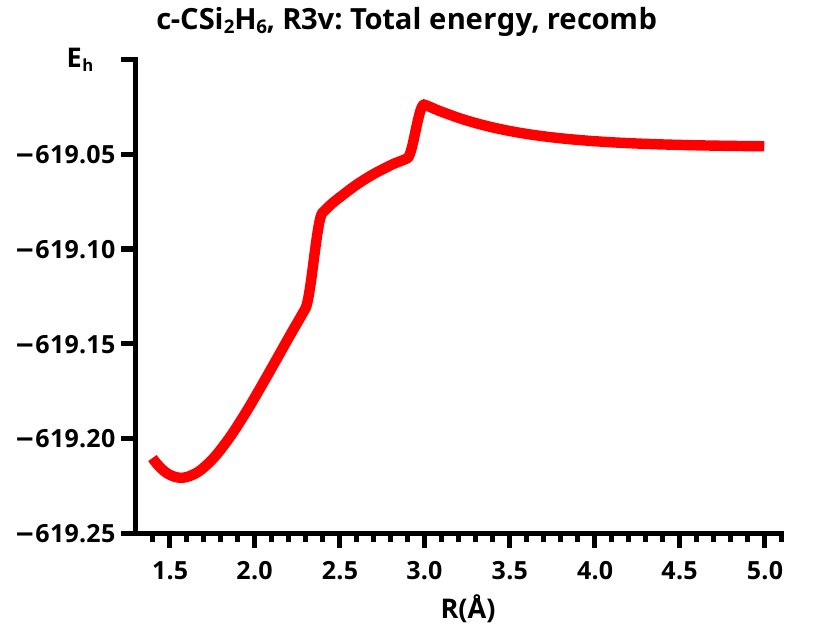}
\includegraphics[width=0.32\textwidth]{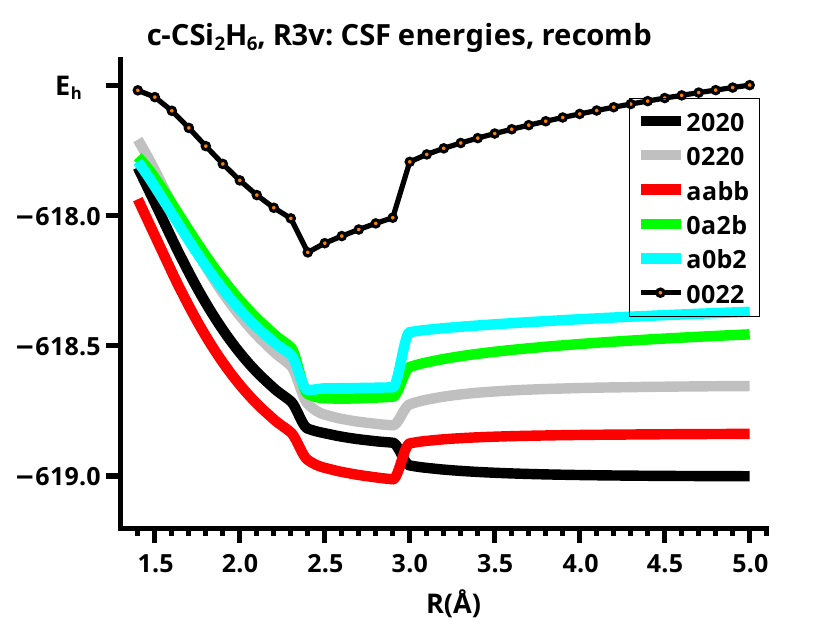}
\includegraphics[width=0.32\textwidth]{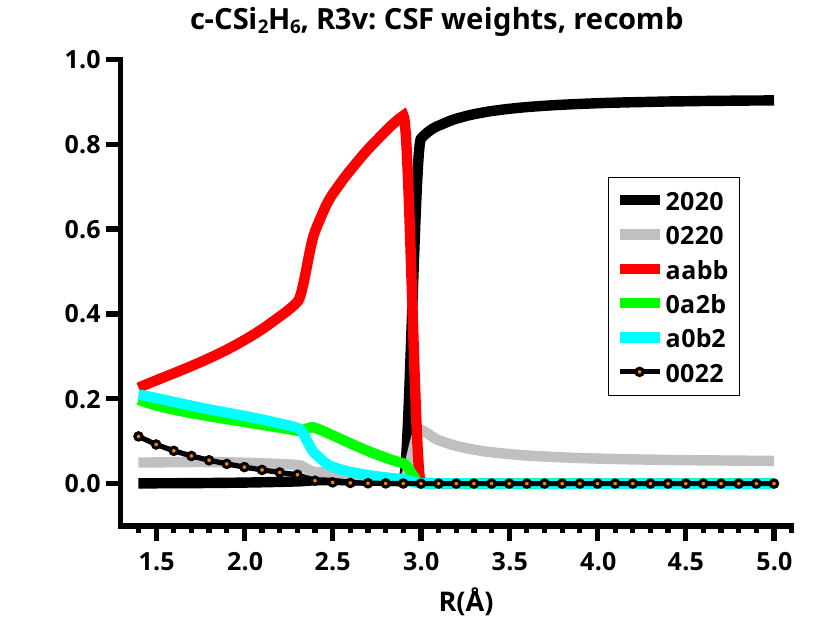}
\caption{Graphs of the total energy, the CSF energies, and the CSF weights. Top: Discontinuous representation. Bottom: Continuous representation.}
\end{figure}

\newpage
\subsection{Energetics}
$E_{\rm min}$ are the lowest energies of the respective PECs; $E_{\rm max}$ are the energies at the crossing points. $E_{\rm diss}$ are the energies of the dissociated systems at $R=5.0$\,\AA.
$\Delta E_{\rm reac} = E_{\rm diss}-E_{\rm min}$,\; $\Delta E^{\ddagger}_e = E_{\rm max}-E_{\rm min}$,  $\Delta E^{\ddagger}_r = E_{\rm max}-E_{\rm diss}$. The values can be different, if elimination and recombination reactions have different crossing points.  $\Delta E_{\rm reac}^{\rm corr}$ are the $\Delta E_{\rm reac}$ values corrected by the respective singlet-triplet splittings . $\Delta E_{ST}(\ce{CH2}) = -43.3$\,kJ/mol;  $\Delta E_{ST}(\ce{Si2H4}) = 65.9$\,kJ/mol, in $C_{2v}$ geometry calculated.

\ce{C3H6}: In the elimination reaction, the \ce{CH2} product is in the excited singlet state, the ground state energy is by 43.3\,kJ/mol lower.\newline
 In the recombination reaction, the \ce{CH2} educt is in the excited singlet state, the ground state energy is by 43.3\,kJ/mol lower, and $\Delta E^{\ddagger}_r$ is by that amount higher.

\ce{CSi2H6}: In the elimination reaction, both products are in  excited triplet states, the ground state energy is lower by $\Delta E_{ST}$(\ce{Si2H4})= 65.9\,kJ/mol.\newline
 In the recombination reaction, the \ce{CH2} educt is in the excited singlet state, the ground state energy is lower by $\Delta E_{ST}$(\ce{CH2}) = 43.3\,kJ/mol; and $\Delta E^{\ddagger}_r$ is by that amount higher.\newline

\begin{threeparttable}[h]
\caption{Energy differences for the  elimination reactions in $C_{2v}$ in kJ/mol.}
\begin{tabular}{r|r|r|r|rrrr}
Energy differences & \ce{c-C3H6} & \ce{c-Si3H6} & \ce{c-CSi2H6} &  \ce{c-SiC2H6} \\
\toprule
$\Delta E_{\rm reac}$              & 344.2\tn{1} & 231.4 & 478.6\tn{2} & 103.2 \\
$\Delta E_{\rm reac}^{\rm corr}$   & 300.9       & 231.4 & 412.7       & 103.2 \\
$\Delta E^{\ddagger}_e$            & 492.4       & 311.2 &             & 273.4 \\
\bottomrule
\end{tabular}
 \begin{tablenotes}
 \item[1]\ce{CH2} in excited singlet state.
 \item[2]\ce{CH2} and disilene in excited triplet states.
 \end{tablenotes}
 \end{threeparttable}

\vspace{5mm}
\begin{threeparttable}[h]
\caption{Energy differences for the  recombination reactions in $C_{2v}$ in kJ/mol.}
\begin{tabular}{r|r|r|r|rrrr}
Energy differences & \ce{c-C3H6} & \ce{c-Si3H6} & \ce{c-CSi2H6} &  \ce{c-SiC2H6} \\
\toprule
$\Delta E_{\rm reac}$              & 344.2\tn{1} & 231.4 & 458.1\tn{1} & 103.2 \\
$\Delta E_{\rm reac}^{\rm corr}$   & 300.9       & 231.4 & 414.8       & 103.2 \\
$\Delta E^{\ddagger}_r$            & 145.0       &  79.8 &  57.6       & 170.2 \\
$\Delta E^{\ddagger}_r{}^{\rm corr}$ & 188.3       &  79.8 & 101.9       & 170.2 \\
\bottomrule
\end{tabular}
 \begin{tablenotes}
 \item[1]\ce{CH2} in excited singlet state.
 \end{tablenotes}
 \end{threeparttable}

\newpage
\subsection{Electron densities}
Going from high $C_{2v}$ symmetry to low symmetry $C_s$ symmetry of the molecular system the geometry, the characteristics, and the electron density change. Only in the elimination reaction of \ce{c-CSi2H6} the characteristics does not change.
The contour values of the molecular densities were 0.065 for all systems but \ce{c-C3H6} for which the contour value 0.1 was chosen.

\begin{figure}
\includegraphics[width=0.5\textwidth]{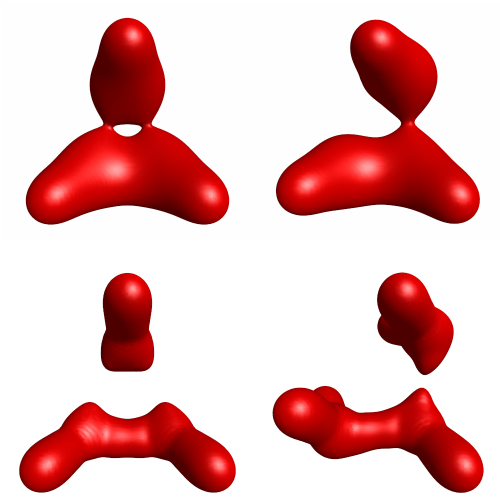}
\caption{Electron densities of \ce{c-C3H6} (top) and \ce{c-Si3H6} (bottom) for elimination reaction. Top, left: $R=1.8$\,\AA, right: $R=1.9$\,\AA. Bottom, left: $R=2.5$\,\AA, right: $R=2.6$\,\AA. }
\end{figure}
\begin{figure}
\includegraphics[width=0.5\textwidth]{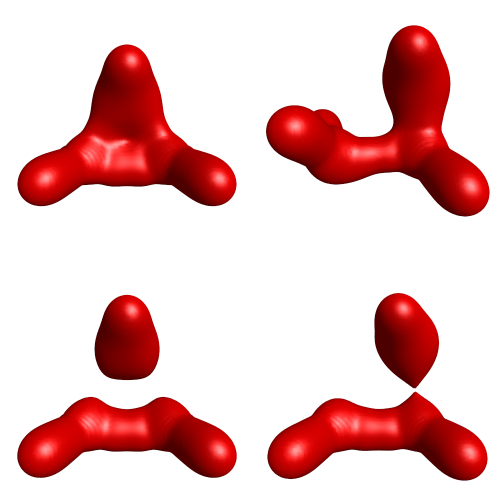}
\caption{Electron densities of \ce{c-CSi2H6}, recombination reaction (top), and elimination reaction (bottom). Top, left: $R=1.8$\,\AA, right: $R=1.9$\,\AA. Bottom, left: $R=2.2$\,\AA, right: $R=2.3$\,\AA. }
\end{figure}
\begin{figure}
\includegraphics[width=0.5\textwidth]{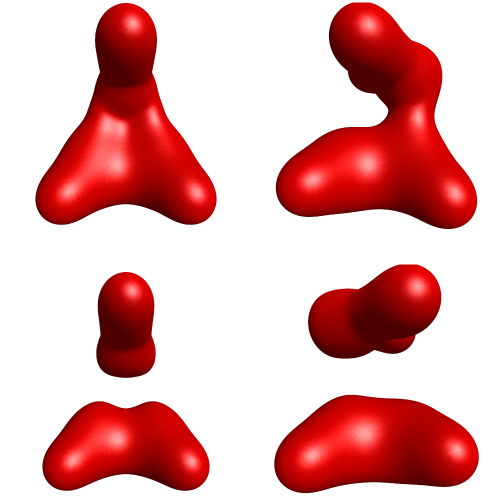}
\caption{Electron densities of \ce{c-SiC2H6}, recombination reaction (top), and elimination reaction (bottom). Top, left: $R=1.9$\,\AA, right: $R=2.0$\,\AA. Bottom, left: $R=2.3$\,\AA, right: $R=2.4$\,\AA. }
\end{figure}

\end{document}